\documentclass[fleqn,10pt]{wlscirep}
\usepackage[utf8]{inputenc}
\usepackage[T1]{fontenc}

\usepackage[T1]{fontenc}
\usepackage{amsmath,amsfonts,amssymb}

\usepackage{cases}

\usepackage{graphicx}

\usepackage{booktabs}

\usepackage{subcaption}
\DeclareCaptionLabelFormat{boldparens}{\color{black}\textbf{(#2)}}
\usepackage[per-mode=symbol, range-phrase=\textendash, range-units=single, detect-all]{siunitx}
\DeclareSIUnit\bit{b}
\DeclareSIUnit\fiber{fiber}
\DeclareSIUnit\channel{channel}
\DeclareSIUnit\group{group}
\DeclareSIUnit\facet{facet}
\DeclareSIUnit\bps{bps}
\DeclareSIUnit\bm{Bm}

\usepackage{chemformula}

\usepackage{pifont}

\newif\ifcoffee
\input{random.tex}

\def\coffeescale{1}

\def\coffeeA{\pdfliteral{%
        q \coffeescale\space 0 0 \coffeescale\space 0 0 cm
        .5 0 0 .5 0 -440 cm
        0.80 0.68 0.60 rg
        375 410 m
        373 411 368 413 364 415 c
        350 421 356 424 385 426 c
        408 427 410 428 420 434 c
        425 438 432 442 435 443 c
        438 444 445 447 450 451 c
        456 454 463 457 466 458 c
        474 460 495 476 516 497 c
        526 507 533 517 538 527 c
        542 535 547 544 549 546 c
        559 557 563 567 563 581 c
        563 593 564 596 570 601 c
        578 610 583 611 588 606 c
        594 601 593 589 586 562 c
        574 521 559 496 525 461 c
        499 434 478 422 443 413 c
        427 409 379 407 375 410 c
        h f
        142 605 m
        138 609 138 614 139 642 c
        140 674 142 686 152 707 c
        154 712 158 722 160 728 c
        165 746 174 762 179 763 c
        181 764 183 768 184 771 c
        186 778 213 807 230 818 c
        261 838 278 848 297 853 c
        332 864 364 866 401 860 c
        424 857 438 852 462 841 c
        496 824 518 810 530 796 c
        533 792 541 783 549 776 c
        564 760 572 747 575 730 c
        577 724 580 714 583 708 c
        591 692 594 677 594 658 c
        594 643 593 640 588 631 c
        581 619 577 618 577 627 c
        577 632 576 639 574 644 c
        573 650 569 662 566 671 c
        564 681 561 690 560 692 c
        559 694 558 698 558 700 c
        558 702 553 710 547 716 c
        540 724 535 732 533 739 c
        531 746 527 754 523 757 c
        520 761 517 764 517 766 c
        517 769 490 795 478 803 c
        450 822 415 832 368 835 c
        336 836 330 836 302 828 c
        286 823 251 806 243 799 c
        239 796 231 789 225 784 c
        219 779 212 772 208 769 c
        204 765 199 762 196 762 c
        189 761 189 761 193 756 c
        196 753 196 752 190 744 c
        175 724 167 699 164 664 c
        163 649 161 635 160 632 c
        158 628 158 622 159 616 c
        162 604 159 599 151 599 c
        147 599 144 601 142 605 c
        h f
        533 839 m
        532 841 532 843 534 845 c
        537 848 543 845 542 841 c
        541 836 534 834 533 839 c
        h f
        599 856 m
        600 861 604 861 605 857 c
        605 854 604 853 602 853 c
        600 853 599 854 599 856 c
        h f
        0.77 0.58 0.43 rg
        387 409 m
        378 410 366 416 365 419 c
        365 421 379 424 401 426 c
        408 426 413 428 419 433 c
        424 437 432 441 437 443 c
        442 445 448 448 450 449 c
        452 451 459 454 465 456 c
        486 464 528 502 538 525 c
        542 532 549 544 554 552 c
        563 566 564 568 564 580 c
        564 592 565 594 572 600 c
        580 609 584 609 589 603 c
        593 599 593 598 590 584 c
        580 533 560 496 522 459 c
        499 437 480 425 453 417 c
        434 412 403 408 387 409 c
        h f
        141 609 m
        138 613 138 617 139 632 c
        140 642 141 655 141 662 c
        142 677 145 688 156 711 c
        161 721 164 730 164 733 c
        164 736 165 739 166 740 c
        167 741 169 746 170 750 c
        174 761 177 764 185 759 c
        189 758 192 755 193 754 c
        194 753 190 746 185 739 c
        179 732 175 724 175 722 c
        175 720 173 715 172 712 c
        167 704 165 689 162 660 c
        161 647 159 634 158 630 c
        157 627 157 620 158 615 c
        160 605 159 603 151 603 c
        147 603 144 605 141 609 c
        h f
        580 638 m
        577 641 573 653 568 671 c
        565 678 562 688 561 693 c
        559 697 558 703 558 704 c
        558 706 553 711 548 716 c
        541 723 538 729 535 738 c
        527 765 486 805 449 820 c
        415 834 352 841 323 835 c
        300 829 285 824 268 815 c
        260 811 252 808 251 808 c
        249 808 247 806 246 804 c
        244 802 237 795 229 789 c
        221 783 212 775 208 772 c
        201 764 194 763 189 769 c
        186 773 186 773 196 786 c
        207 798 227 816 231 816 c
        232 816 236 820 241 823 c
        246 827 253 832 257 834 c
        261 835 267 839 271 841 c
        288 853 327 861 362 862 c
        406 862 429 856 472 834 c
        505 817 517 808 543 781 c
        562 760 575 739 575 728 c
        575 725 578 716 582 707 c
        590 690 594 672 593 660 c
        593 656 593 650 592 648 c
        592 642 587 636 584 636 c
        583 636 581 637 580 638 c
        h f
        534 841 m
        534 843 536 844 537 844 c
        539 844 541 843 541 841 c
        541 839 539 838 537 838 c
        536 838 534 839 534 841 c
        h f
        0.71 0.54 0.40 rg
        384 411 m
        381 412 376 414 373 416 c
        367 420 l
        373 420 l
        375 420 380 419 382 417 c
        385 414 386 414 386 418 c
        386 422 393 425 404 425 c
        407 425 414 428 419 432 c
        425 436 433 441 437 442 c
        441 444 448 447 452 450 c
        456 453 461 455 463 455 c
        467 455 479 461 490 471 c
        495 476 505 483 510 488 c
        523 499 533 512 539 527 c
        542 533 547 541 550 544 c
        559 556 564 567 565 582 c
        566 593 566 596 572 600 c
        575 603 580 605 583 605 c
        592 605 593 599 588 578 c
        577 529 560 500 522 460 c
        491 429 457 413 410 411 c
        399 411 386 411 384 411 c
        h f
        142 609 m
        139 613 139 642 143 649 c
        145 652 145 654 143 658 c
        141 662 141 664 142 667 c
        144 670 145 674 145 677 c
        145 680 146 683 147 684 c
        148 685 149 689 149 692 c
        150 696 151 697 157 698 c
        164 698 164 698 164 691 c
        164 677 159 639 156 631 c
        155 626 154 622 156 620 c
        157 619 158 615 158 611 c
        158 606 157 605 151 605 c
        148 605 144 607 142 609 c
        h f
        579 645 m
        575 649 573 654 573 655 c
        573 658 571 666 561 697 c
        559 703 554 711 549 716 c
        540 726 534 736 534 743 c
        534 745 531 751 527 756 c
        522 760 519 765 519 766 c
        519 771 492 797 478 805 c
        443 827 410 836 358 836 c
        318 837 305 834 275 820 c
        267 815 256 811 252 809 c
        247 807 244 805 244 804 c
        244 801 218 781 212 779 c
        210 778 206 775 204 772 c
        200 766 194 765 190 770 c
        187 773 190 780 194 782 c
        195 783 198 786 201 790 c
        209 799 229 816 233 816 c
        234 816 238 819 241 822 c
        244 825 251 830 257 833 c
        263 836 271 840 274 842 c
        288 852 334 861 364 861 c
        400 861 428 855 463 838 c
        498 821 514 810 539 783 c
        562 758 575 738 575 726 c
        575 723 578 716 581 710 c
        593 686 596 652 588 642 c
        584 637 l
        579 645 l
        h f
        166 733 m
        166 738 167 739 171 739 c
        178 739 180 737 176 732 c
        171 726 166 727 166 733 c
        h f
        176 744 m
        171 747 171 749 173 754 c
        175 761 180 762 187 757 c
        193 753 193 749 188 745 c
        183 740 182 740 176 744 c
        h f
        534 841 m
        534 843 536 844 537 844 c
        539 844 541 843 541 841 c
        541 839 539 838 537 838 c
        536 838 534 839 534 841 c
        h f
        0.67 0.44 0.30 rg
        391 411 m
        386 413 387 420 392 420 c
        394 420 397 421 397 423 c
        398 424 402 425 406 425 c
        410 425 414 426 417 429 c
        420 433 452 450 470 457 c
        480 460 518 493 527 505 c
        532 511 538 521 540 527 c
        543 532 548 540 551 545 c
        563 560 567 569 566 578 c
        564 589 568 598 576 602 c
        593 612 594 593 579 547 c
        570 519 549 488 521 459 c
        498 436 471 421 438 414 c
        426 412 394 410 391 411 c
        h f
        142 610 m
        139 614 141 643 144 648 c
        147 652 147 654 144 658 c
        141 663 141 664 143 667 c
        144 669 145 672 145 675 c
        145 677 146 681 148 683 c
        150 686 151 690 151 692 c
        151 695 153 696 157 695 c
        163 695 163 694 163 685 c
        162 661 160 642 157 633 c
        155 628 154 623 154 621 c
        155 619 156 615 156 612 c
        158 608 157 608 150 608 c
        146 608 142 609 142 610 c
        h f
        578 649 m
        574 653 571 662 566 683 c
        565 688 563 694 562 695 c
        561 696 560 699 560 702 c
        560 705 556 711 549 717 c
        542 725 538 730 536 739 c
        534 745 530 753 526 758 c
        522 763 519 767 519 768 c
        519 773 497 793 481 804 c
        463 816 443 825 423 830 c
        398 837 340 841 324 837 c
        298 830 281 824 265 815 c
        262 813 257 812 254 812 c
        249 812 247 810 243 804 c
        241 800 238 797 236 797 c
        235 797 230 793 226 789 c
        221 784 218 782 214 783 c
        211 783 207 782 203 777 c
        199 774 196 771 194 771 c
        189 771 190 776 196 783 c
        199 786 206 793 210 798 c
        215 803 223 810 230 813 c
        236 816 243 821 244 823 c
        246 826 251 829 256 831 c
        260 833 267 836 271 839 c
        284 849 335 861 362 861 c
        399 861 435 852 471 833 c
        504 816 517 806 541 781 c
        562 758 573 741 573 729 c
        573 727 576 719 579 711 c
        587 694 592 676 592 663 c
        592 645 586 640 578 649 c
        h f
        167 732 m
        167 734 167 735 169 736 c
        174 737 176 736 174 733 c
        172 730 169 730 167 732 c
        h f
        176 746 m
        172 750 172 752 176 757 c
        178 761 179 761 185 757 c
        193 753 194 752 188 747 c
        183 742 181 742 176 746 c
        h f
        535 840 m
        534 841 535 842 537 842 c
        539 842 540 841 540 840 c
        539 839 538 838 537 838 c
        537 838 536 839 535 840 c
        h f
        0.63 0.33 0.10 rg
        391 414 m
        390 415 391 416 393 416 c
        395 416 397 418 398 419 c
        399 421 402 423 406 423 c
        411 423 415 424 419 429 c
        422 433 430 437 435 439 c
        440 441 447 444 450 447 c
        453 449 461 452 468 455 c
        475 457 481 461 483 462 c
        484 464 493 472 503 479 c
        522 495 534 509 541 526 c
        544 533 549 541 552 544 c
        563 558 566 566 566 580 c
        566 591 567 593 572 598 c
        579 604 584 605 588 600 c
        591 597 590 592 584 569 c
        573 526 556 497 520 460 c
        501 441 484 430 459 421 c
        436 413 394 409 391 414 c
        h f
        142 619 m
        141 624 141 629 142 631 c
        143 633 144 637 143 640 c
        142 644 143 647 147 649 c
        152 653 l
        147 658 l
        141 665 143 668 153 668 c
        160 668 l
        159 654 l
        158 647 156 636 154 631 c
        151 623 151 620 153 618 c
        157 613 156 610 149 610 c
        144 610 143 611 142 619 c
        h f
        578 651 m
        577 653 575 656 574 659 c
        571 674 565 693 563 697 c
        561 700 560 704 560 706 c
        560 708 556 713 552 717 c
        544 723 538 734 535 745 c
        535 748 532 753 529 756 c
        527 759 522 766 518 772 c
        515 777 511 783 509 784 c
        507 786 504 789 504 791 c
        503 793 499 796 494 798 c
        489 801 480 806 474 810 c
        458 820 442 826 421 832 c
        386 841 329 843 311 836 c
        309 835 304 833 301 832 c
        298 830 293 829 290 829 c
        287 829 284 828 283 827 c
        282 825 279 825 276 826 c
        271 827 269 826 269 823 c
        269 817 262 814 254 815 c
        248 815 247 816 247 821 c
        247 825 249 827 257 830 c
        263 832 269 835 272 838 c
        274 840 279 843 282 844 c
        285 845 291 847 295 848 c
        333 862 390 863 429 850 c
        447 845 486 825 503 814 c
        522 801 559 761 567 745 c
        570 737 573 729 573 727 c
        573 724 576 714 581 704 c
        588 688 589 684 589 668 c
        589 651 589 650 584 649 c
        582 649 579 650 578 651 c
        h f
        151 674 m
        146 675 146 679 150 683 c
        152 684 153 687 153 690 c
        153 695 159 695 161 688 c
        164 676 160 671 151 674 c
        h f
        179 748 m
        175 750 174 755 177 757 c
        179 758 190 754 190 752 c
        190 750 185 745 184 745 c
        183 745 181 746 179 748 c
        h f
        212 787 m
        205 790 206 793 214 800 c
        224 810 231 813 231 809 c
        231 807 233 805 235 804 c
        241 803 240 800 233 798 c
        229 798 226 795 225 793 c
        223 787 218 784 212 787 c
        h f
        237 813 m
        238 814 240 814 241 813 c
        242 813 240 812 238 812 c
        236 812 235 813 237 813 c
        h f
        0.57 0.16 0.01 rg
        403 416 m
        404 420 411 420 412 417 c
        412 416 410 415 407 414 c
        404 414 403 414 403 416 c
        h f
        429 418 m
        435 422 434 425 427 423 c
        422 422 420 423 420 425 c
        420 429 433 438 438 438 c
        441 438 446 440 450 443 c
        454 446 460 450 464 451 c
        468 452 478 458 485 464 c
        492 469 500 474 503 474 c
        507 474 508 475 508 478 c
        508 480 511 484 513 486 c
        518 489 519 489 521 486 c
        524 483 523 482 520 480 c
        515 476 517 473 524 472 c
        528 471 527 469 516 458 c
        499 441 478 429 454 422 c
        435 416 422 414 429 418 c
        h f
        523 493 m
        523 494 525 496 526 498 c
        530 501 533 497 531 494 c
        529 491 523 491 523 493 c
        h f
        550 505 m
        550 507 548 510 546 512 c
        543 514 542 513 539 509 c
        537 507 535 504 534 504 c
        532 504 533 506 535 508 c
        537 510 539 516 540 521 c
        542 526 546 534 550 539 c
        563 556 568 568 568 581 c
        568 594 570 598 580 600 c
        589 602 589 597 582 567 c
        575 541 567 522 557 507 c
        552 499 550 498 550 505 c
        h f
        144 625 m
        142 628 144 631 148 631 c
        152 631 152 628 148 625 c
        146 623 145 623 144 625 c
        h f
        147 643 m
        147 646 152 649 154 647 c
        157 644 155 640 151 640 c
        149 640 147 641 147 643 c
        h f
        581 653 m
        579 654 577 657 577 659 c
        577 661 576 666 574 669 c
        572 673 571 680 572 687 c
        572 695 572 698 570 696 c
        566 693 562 697 562 703 c
        562 705 558 711 553 716 c
        544 725 536 739 536 746 c
        536 748 533 753 530 757 c
        527 760 522 767 520 772 c
        512 792 500 804 489 804 c
        487 804 486 806 486 808 c
        486 813 483 815 479 812 c
        477 810 475 811 470 816 c
        466 820 462 822 460 821 c
        458 821 453 822 448 825 c
        437 831 406 838 384 839 c
        371 840 369 841 368 844 c
        366 850 360 852 358 848 c
        356 846 344 844 329 845 c
        326 845 319 845 313 845 c
        305 844 302 845 302 847 c
        303 848 306 850 309 850 c
        313 851 321 852 327 854 c
        342 858 385 858 406 854 c
        436 849 461 838 465 830 c
        467 826 468 825 471 827 c
        476 829 476 829 498 816 c
        514 806 560 759 565 746 c
        567 741 570 732 572 725 c
        574 719 577 708 580 703 c
        587 690 591 670 589 660 c
        587 651 586 650 581 653 c
        h f
        148 659 m
        146 662 149 666 153 666 c
        156 666 158 664 158 661 c
        158 657 150 655 148 659 c
        h f
        155 678 m
        157 679 158 678 159 678 c
        160 677 159 677 156 677 c
        154 677 153 677 155 678 c
        h f
        212 791 m
        208 794 208 795 212 796 c
        217 798 222 794 219 790 c
        217 789 215 789 212 791 c
        h f
        251 819 m
        249 822 251 824 256 822 c
        257 821 258 820 258 818 c
        256 816 253 816 251 819 c
        h f
        275 831 m
        273 833 274 836 277 836 c
        279 836 280 834 280 832 c
        280 829 278 828 275 831 c
        h f
        286 836 m
        283 840 282 841 285 842 c
        291 846 295 845 295 838 c
        295 830 292 830 286 836 c
        h f
        0.34 0.05 0.01 rg
        491 455 m
        482 460 492 471 503 469 c
        516 466 519 459 508 455 c
        500 452 495 452 491 455 c
        h f
        548 517 m
        541 520 541 522 548 534 c
        554 543 561 547 567 544 c
        570 542 567 526 562 520 c
        557 515 555 514 548 517 c
        h f
        568 564 m
        571 570 578 570 577 565 c
        576 563 574 561 571 561 c
        567 560 566 560 568 564 c
        h f
        575 584 m
        575 588 578 593 581 593 c
        584 593 584 590 582 585 c
        581 582 575 580 575 584 c
        h f
        581 669 m
        576 674 575 676 576 683 c
        578 694 579 696 581 696 c
        584 696 589 674 588 668 c
        587 664 587 664 581 669 c
        h f
        554 716 m
        544 726 542 730 537 744 c
        534 754 530 762 528 764 c
        526 765 526 767 526 769 c
        529 773 534 772 535 767 c
        537 762 545 761 545 765 c
        545 772 558 760 565 746 c
        569 739 575 714 575 707 c
        575 701 565 706 554 716 c
        h f
        514 792 m
        513 793 513 796 513 799 c
        513 804 513 804 517 800 c
        519 798 521 795 521 793 c
        521 790 517 789 514 792 c
        h f
        495 811 m
        493 814 491 816 492 816 c
        495 816 506 809 506 807 c
        506 804 500 806 495 811 c
        h f
        476 824 m
        476 826 477 827 479 827 c
        483 825 483 821 479 821 c
        478 821 476 822 476 824 c
        h f
        445 836 m
        441 841 443 841 449 837 c
        451 835 452 834 451 833 c
        450 832 447 833 445 836 c
        h f
        409 845 m
        409 847 411 849 413 849 c
        414 849 416 848 416 847 c
        416 846 414 844 413 843 c
        411 843 409 843 409 845 c
        h f
        Q
    }}

\def\coffeeB{\pdfliteral{%
        q \coffeescale\space 0 0 \coffeescale\space 0 0 cm
        .5 0 0 .5 0 -500 cm
        0.86 0.77 0.70 rg
        119 537 m
        108 544 101 556 96.79 573 c
        94.66 582 93.17 589 93.48 589 c
        94.91 591 121 563 127 554 c
        134 543 134 541 130 536 c
        128 532 127 532 119 537 c
        h f
        62.17 668 m
        57.50 672 55.08 705 58.27 720 c
        59.63 727 61.08 736 61.49 741 c
        63.32 762 67.92 791 71.71 806 c
        79.42 836 111 887 134 909 c
        160 932 214 959 214 949 c
        214 945 207 930 202 924 c
        200 921 194 917 188 915 c
        175 909 173 907 165 892 c
        161 885 152 874 145 867 c
        139 860 131 850 128 844 c
        125 839 120 827 115 819 c
        111 810 104 798 101 792 c
        94.41 781 91.12 771 88.69 753 c
        87.87 747 85.95 741 84.42 739 c
        82.86 737 82.39 734 83.35 733 c
        86.09 728 84.33 673 81.37 670 c
        76.77 665 65.75 664 62.17 668 c
        h f
        587 815 m
        587 819 585 824 584 827 c
        581 833 585 833 589 826 c
        592 822 592 819 591 815 c
        589 809 l
        587 815 l
        h f
        0.80 0.68 0.59 rg
        112 545 m
        110 549 103 561 103 563 c
        103 564 106 564 109 564 c
        114 564 117 562 121 557 c
        127 549 128 545 123 545 c
        122 545 119 544 117 544 c
        116 543 113 544 112 545 c
        h f
        60.54 678 m
        57.41 689 58.92 725 62.73 730 c
        64.21 732 64.87 735 64.21 738 c
        63.54 740 64.09 744 65.43 747 c
        67.85 751 68.37 755 70.06 782 c
        71.69 808 75.87 820 96.71 857 c
        107 876 118 891 131 903 c
        142 914 169 931 179 933 c
        182 934 188 936 190 939 c
        200 946 206 942 201 932 c
        197 923 192 919 188 919 c
        186 919 180 917 176 914 c
        169 910 167 906 163 896 c
        158 883 149 870 138 861 c
        135 858 130 851 126 845 c
        110 813 104 802 100 794 c
        90.36 778 85.66 765 85.42 753 c
        85.31 748 83.94 743 82.36 741 c
        80.21 738 80.03 736 81.61 730 c
        84.21 722 83.31 676 80.49 673 c
        79.36 672 74.99 670 70.76 670 c
        63.19 668 63.04 669 60.54 678 c
        h f
        0.76 0.58 0.42 rg
        63.74 675 m
        58.40 680 60.11 724 65.91 731 c
        70.74 736 73.92 736 78.29 731 c
        80.97 728 81.61 722 81.49 701 c
        81.34 676 81.23 675 76.20 673 c
        69.17 670 67.91 671 63.74 675 c
        h f
        72.81 746 m
        75.68 749 78.74 750 81.02 749 c
        86.31 747 80.99 742 73.80 742 c
        68.24 742 l
        72.81 746 l
        h f
        72.89 755 m
        68.05 757 68.20 763 73.34 772 c
        77.15 778 77.22 779 74.25 785 c
        71.56 790 71.42 793 73.43 802 c
        76.24 815 83.19 832 89.46 841 c
        91.85 845 97.16 854 101 862 c
        106 872 115 883 126 896 c
        145 915 l
        156 915 l
        167 914 l
        164 909 l
        162 906 161 902 161 900 c
        161 889 155 879 141 865 c
        133 857 126 848 125 845 c
        125 842 122 836 119 833 c
        116 829 112 821 109 814 c
        107 808 103 801 102 799 c
        100 797 99.13 795 99.13 794 c
        99.13 793 96.72 789 93.78 787 c
        90.77 784 88.43 779 88.43 776 c
        88.42 773 85.76 766 82.50 762 c
        77.69 754 75.90 753 72.89 755 c
        h f
        171 918 m
        169 920 173 926 176 926 c
        177 926 181 928 184 930 c
        189 934 191 934 193 933 c
        196 929 194 923 189 923 c
        187 923 183 922 181 920 c
        176 917 173 916 171 918 c
        h f
        0.71 0.55 0.40 rg
        66.28 676 m
        60.78 680 60.69 723 66.16 729 c
        70.88 734 73.97 734 78.88 729 c
        82.36 725 82.56 723 80.65 711 c
        79.22 703 79.20 695 80.59 689 c
        83.70 675 76.26 669 66.28 676 c
        h f
        71.33 763 m
        71.33 765 73.88 769 77.00 771 c
        82.66 775 l
        77.94 784 l
        73.69 791 73.43 793 75.35 805 c
        76.52 812 80.18 823 83.49 829 c
        86.80 835 94.07 848 99.66 859 c
        110 878 133 906 144 912 c
        148 914 151 914 155 912 c
        160 910 160 909 159 901 c
        157 885 154 880 139 864 c
        129 854 124 847 124 844 c
        124 841 122 837 120 835 c
        118 833 114 826 111 819 c
        109 812 105 805 103 803 c
        102 801 99.51 797 98.20 795 c
        96.82 792 93.45 789 90.14 789 c
        84.78 788 84.51 788 85.35 779 c
        86.00 772 85.40 769 83.28 768 c
        81.64 768 79.82 765 79.22 763 c
        77.82 757 71.33 757 71.33 763 c
        h f
        173 921 m
        173 922 175 923 176 923 c
        177 923 177 922 177 921 c
        176 920 175 919 174 919 c
        174 919 173 920 173 921 c
        h f
        183 925 m
        182 928 189 932 191 930 c
        193 929 189 923 186 923 c
        185 923 184 924 183 925 c
        h f
        0.67 0.44 0.30 rg
        67.59 675 m
        63.24 677 62.77 679 62.77 692 c
        62.77 701 63.64 707 64.70 707 c
        65.76 707 66.23 712 65.75 717 c
        64.77 727 68.34 732 75.31 730 c
        81.24 728 82.09 724 79.80 710 c
        78.51 702 78.55 695 79.91 690 c
        82.64 681 82.62 681 78.66 677 c
        74.87 673 74.29 673 67.59 675 c
        h f
        74.62 791 m
        71.17 796 81.37 828 89.85 838 c
        91.41 840 96.15 848 100 857 c
        108 874 123 894 137 905 c
        147 913 149 914 155 910 c
        158 908 159 906 157 903 c
        155 900 155 895 156 891 c
        157 884 156 882 141 867 c
        131 857 124 848 124 846 c
        124 843 122 838 119 835 c
        116 831 112 822 109 816 c
        107 810 103 803 101 800 c
        98.84 798 96.99 795 96.99 793 c
        96.99 792 96.27 791 95.39 791 c
        94.51 791 89.70 790 84.71 790 c
        79.72 789 75.18 790 74.62 791 c
        h f
        0.62 0.35 0.13 rg
        66.51 679 m
        62.59 683 63.58 686 71.08 690 c
        77.21 693 80.26 690 79.37 682 c
        78.68 675 71.47 674 66.51 679 c
        h f
        65.79 700 m
        65.17 701 67.61 706 71.20 710 c
        77.74 717 l
        77.74 711 l
        77.74 700 68.94 691 65.79 700 c
        h f
        76.78 793 m
        74.90 796 79.53 817 83.85 826 c
        86.17 830 90.56 837 93.60 841 c
        96.64 845 99.13 849 99.13 851 c
        99.13 855 108 871 117 883 c
        130 900 145 911 150 909 c
        153 908 154 907 153 905 c
        151 900 151 897 153 889 c
        155 883 153 881 138 866 c
        129 856 122 848 122 845 c
        122 842 120 838 117 834 c
        114 831 110 823 108 818 c
        103 803 95.60 794 85.80 792 c
        81.41 792 77.35 792 76.78 793 c
        h f
        0.54 0.18 0.03 rg
        69.41 680 m
        67.48 683 67.96 684 71.73 685 c
        77.43 686 78.70 685 77.01 681 c
        75.44 677 72.62 677 69.41 680 c
        h f
        82.02 797 m
        82.02 798 82.98 799 84.16 799 c
        85.34 799 86.30 798 86.30 797 c
        86.30 796 85.34 795 84.16 795 c
        82.98 795 82.02 796 82.02 797 c
        h f
        87.37 803 m
        79.15 805 77.31 814 84.03 820 c
        86.45 822 88.44 825 88.44 828 c
        88.44 832 101 846 106 846 c
        110 846 110 847 106 851 c
        103 855 103 855 106 864 c
        108 869 114 877 118 883 c
        133 900 144 903 151 892 c
        154 886 154 885 150 879 c
        148 876 140 868 133 862 c
        123 853 120 849 120 845 c
        120 842 118 837 115 834 c
        112 831 109 824 107 818 c
        104 807 100 801 96.30 802 c
        94.92 802 90.90 803 87.37 803 c
        h f
        0.40 0.05 0.00 rg
        89.00 807 m
        84.54 809 82.48 811 83.23 813 c
        85.02 818 93.83 821 99.00 820 c
        102 819 103 817 102 812 c
        102 804 98.82 803 89.00 807 c
        h f
        97.95 834 m
        95.97 837 98.92 839 107 841 c
        112 842 114 841 114 839 c
        114 833 100 829 97.95 834 c
        h f
        114 864 m
        107 866 108 869 120 881 c
        127 888 131 891 134 890 c
        136 889 139 890 140 891 c
        143 893 144 893 147 891 c
        152 887 150 881 143 873 c
        131 862 125 859 114 864 c
        h f
        Q
    }}

\def\coffeeC{\pdfliteral{%
        q \coffeescale\space 0 0 \coffeescale\space 0 0 cm
        .5 0 0 .5 0 -220 cm
        0.98 0.85 0.77 rg
        240 148 m
        236 151 220 154 220 150 c
        220 149 219 148 217 148 c
        215 148 214 149 214 150 c
        214 154 198 157 192 154 c
        185 151 184 152 183 158 c
        181 166 178 168 171 168 c
        168 168 166 169 166 171 c
        166 175 160 180 155 181 c
        152 181 150 184 149 189 c
        149 195 148 196 139 197 c
        130 198 l
        136 201 l
        143 206 143 211 136 217 c
        132 220 131 222 133 228 c
        134 234 133 236 127 241 c
        121 246 121 247 123 253 c
        125 256 127 259 128 260 c
        133 262 136 276 134 281 c
        132 284 132 285 135 287 c
        141 290 141 299 135 305 c
        130 311 132 314 141 314 c
        152 314 156 320 150 328 c
        145 334 l
        154 333 l
        163 331 163 331 165 337 c
        166 340 165 343 163 345 c
        159 349 158 354 161 354 c
        162 354 165 352 168 349 c
        170 346 174 344 175 344 c
        180 344 190 354 190 359 c
        190 364 195 369 198 367 c
        199 366 200 364 200 362 c
        200 356 209 355 218 359 c
        222 361 228 363 232 363 c
        236 364 241 364 243 364 c
        245 365 248 363 249 361 c
        251 357 252 357 258 358 c
        263 359 265 359 266 357 c
        266 355 269 354 274 354 c
        277 354 282 354 283 352 c
        284 351 287 350 290 350 c
        292 350 295 348 295 347 c
        296 345 297 344 299 344 c
        300 344 304 341 307 339 c
        310 336 314 334 316 334 c
        317 334 320 330 323 326 c
        325 321 329 317 331 317 c
        333 316 335 313 335 310 c
        336 307 338 304 340 302 c
        343 301 345 299 345 297 c
        345 292 351 279 355 277 c
        357 275 357 269 353 251 c
        351 242 351 241 354 237 c
        358 233 358 232 354 229 c
        348 225 345 218 345 213 c
        345 210 343 208 341 207 c
        339 207 337 205 337 204 c
        337 199 320 185 307 178 c
        300 175 294 170 293 169 c
        293 168 291 166 289 166 c
        288 166 284 164 282 162 c
        278 158 276 157 267 158 c
        257 159 256 158 256 154 c
        256 151 254 148 251 147 c
        244 144 244 144 240 148 c
        h f
        542 301 m
        533 304 522 314 516 325 c
        511 335 509 339 509 349 c
        509 356 511 363 512 365 c
        513 366 514 369 513 371 c
        512 376 529 393 540 398 c
        544 400 555 402 563 403 c
        579 405 579 405 589 394 c
        591 393 594 390 596 388 c
        599 385 601 381 601 379 c
        601 376 603 371 606 367 c
        611 360 611 358 609 352 c
        607 348 606 341 607 336 c
        607 330 606 327 603 326 c
        601 325 599 322 598 320 c
        597 317 595 316 594 316 c
        593 316 589 313 585 309 c
        580 304 577 302 573 303 c
        570 303 566 303 566 302 c
        564 299 550 299 542 301 c
        h f
        590 363 m
        585 366 583 367 583 364 c
        583 363 585 362 587 361 c
        590 361 592 360 593 360 c
        593 360 592 361 590 363 c
        h f
        0.97 0.80 0.68 rg
        216 163 m
        215 165 212 166 211 166 c
        209 165 208 165 208 166 c
        208 168 214 169 221 169 c
        226 168 230 165 229 161 c
        227 157 218 158 216 163 c
        h f
        191 164 m
        191 165 193 166 194 166 c
        195 166 196 165 196 164 c
        196 163 194 162 193 162 c
        191 162 190 163 191 164 c
        h f
        260 166 m
        260 168 267 172 271 172 c
        273 172 274 171 274 170 c
        273 169 271 167 269 167 c
        267 167 264 167 263 166 c
        261 165 260 165 260 166 c
        h f
        198 179 m
        200 183 204 183 204 180 c
        204 178 202 176 201 176 c
        199 176 198 178 198 179 c
        h f
        181 180 m
        177 183 180 190 186 193 c
        192 195 193 197 190 200 c
        188 201 186 200 185 198 c
        183 193 176 194 175 199 c
        174 201 172 203 169 203 c
        166 203 163 204 163 205 c
        162 206 159 206 156 206 c
        151 205 150 206 145 213 c
        142 218 139 226 138 230 c
        137 236 136 239 131 241 c
        124 245 124 251 129 256 c
        134 261 138 272 137 278 c
        137 280 138 284 140 287 c
        142 290 143 293 141 299 c
        139 309 139 309 144 309 c
        147 309 152 314 159 321 c
        171 332 177 334 184 328 c
        187 326 188 326 195 333 c
        203 341 207 342 210 335 c
        212 329 214 329 220 333 c
        223 335 225 336 230 334 c
        239 331 245 331 248 335 c
        250 336 253 338 256 338 c
        260 338 264 339 267 340 c
        273 343 274 342 276 333 c
        278 326 282 323 282 329 c
        282 335 292 339 296 336 c
        298 335 302 333 305 331 c
        308 330 311 328 311 326 c
        312 325 314 324 315 324 c
        318 324 333 311 333 308 c
        333 307 335 303 338 301 c
        341 299 343 296 343 295 c
        343 294 345 289 348 283 c
        353 271 354 263 351 250 c
        349 244 349 240 351 237 c
        353 233 352 232 348 227 c
        345 223 343 219 343 216 c
        343 214 341 210 338 208 c
        336 205 333 203 333 202 c
        332 200 329 197 326 194 c
        320 189 317 188 301 187 c
        289 187 282 187 282 188 c
        280 191 268 189 258 184 c
        248 179 245 179 241 187 c
        237 198 225 203 224 194 c
        223 190 211 185 201 184 c
        198 184 193 183 190 182 c
        184 178 183 178 181 180 c
        h f
        226 182 m
        226 184 227 185 227 183 c
        228 181 228 180 227 179 c
        227 178 226 180 226 182 c
        h f
        542 303 m
        534 306 525 314 519 323 c
        513 333 512 336 511 347 c
        511 358 512 360 519 370 c
        523 377 528 383 529 384 c
        531 385 532 388 532 389 c
        532 391 536 394 541 396 c
        549 400 551 400 554 398 c
        558 396 560 396 568 398 c
        576 401 577 401 578 399 c
        579 397 581 396 583 396 c
        588 396 591 390 591 379 c
        592 372 591 371 586 369 c
        574 365 575 361 588 356 c
        595 353 596 354 599 358 c
        603 363 610 362 608 356 c
        607 354 606 348 606 341 c
        605 330 605 330 592 317 c
        580 305 577 304 568 303 c
        556 301 547 301 542 303 c
        h f
        576 378 m
        575 379 573 380 571 380 c
        570 380 569 379 570 378 c
        570 377 572 376 574 376 c
        576 376 576 377 576 378 c
        h f
        551 390 m
        552 391 551 392 550 392 c
        548 392 546 391 546 390 c
        546 389 547 388 548 388 c
        549 388 551 389 551 390 c
        h f
        Q
    }}

\def\coffeeD{\pdfliteral{%
        q \coffeescale\space 0 0 \coffeescale\space 0 0 cm
        .5 0 0 .5 0 -530 cm
        0.98 0.90 0.81 rg
        495 591 m
        492 593 492 593 496 596 c
        498 597 499 599 498 600 c
        497 601 485 599 479 596 c
        473 593 471 594 470 601 c
        470 606 468 608 462 611 c
        458 614 453 619 451 622 c
        449 625 445 629 442 630 c
        439 632 436 635 436 637 c
        435 639 432 641 429 643 c
        425 645 422 646 417 644 c
        413 643 410 641 410 640 c
        410 637 386 610 382 609 c
        373 606 358 614 353 622 c
        351 626 348 628 343 630 c
        339 631 335 632 334 633 c
        331 636 318 635 313 632 c
        309 630 308 629 310 624 c
        311 621 311 620 308 620 c
        306 620 302 619 299 618 c
        297 616 292 615 288 615 c
        283 615 l
        283 628 l
        283 638 282 642 279 647 c
        276 652 275 652 258 654 c
        231 656 228 656 223 651 c
        212 643 198 649 198 662 c
        198 667 201 678 204 683 c
        205 685 205 690 205 694 c
        205 702 205 703 200 704 c
        197 705 195 707 195 709 c
        195 714 185 720 175 722 c
        161 723 157 726 157 735 c
        157 741 158 743 162 745 c
        164 747 166 750 166 752 c
        166 756 159 764 154 764 c
        151 764 146 771 146 776 c
        146 778 149 783 154 787 c
        160 792 161 794 161 800 c
        161 814 159 820 153 824 c
        148 828 147 832 153 833 c
        155 834 157 835 158 836 c
        159 839 157 882 155 885 c
        155 887 154 890 154 893 c
        154 895 154 902 154 909 c
        154 916 154 921 156 922 c
        157 922 164 923 172 923 c
        185 922 186 922 192 928 c
        198 934 198 935 191 951 c
        189 956 189 957 192 961 c
        195 966 197 966 203 965 c
        225 962 225 962 233 966 c
        243 971 247 975 248 981 c
        249 989 249 999 246 1003 c
        244 1009 246 1013 253 1014 c
        258 1015 268 1007 270 1000 c
        271 993 275 992 285 995 c
        292 998 296 1002 297 1013 c
        300 1028 307 1029 316 1016 c
        320 1011 324 1007 326 1006 c
        331 1003 338 1005 345 1012 c
        351 1018 356 1020 357 1016 c
        358 1014 360 1012 363 1012 c
        366 1011 369 1009 369 1007 c
        370 1005 373 1003 377 1001 c
        381 999 387 997 390 995 c
        396 992 417 991 420 993 c
        420 994 421 999 421 1004 c
        421 1012 422 1014 425 1014 c
        433 1015 437 1012 437 1004 c
        437 1000 436 996 435 996 c
        432 994 433 984 437 981 c
        441 976 442 976 451 979 c
        456 980 458 982 459 985 c
        459 988 461 989 466 989 c
        475 990 477 987 474 972 c
        472 959 l
        482 950 l
        490 942 492 941 496 942 c
        498 943 505 944 511 944 c
        521 945 523 945 525 941 c
        529 935 529 933 521 924 c
        514 917 514 916 515 908 c
        516 903 518 897 522 892 c
        529 883 530 883 544 888 c
        550 890 553 891 559 889 c
        565 887 566 887 566 880 c
        566 872 559 866 551 866 c
        544 866 537 861 537 857 c
        537 855 536 853 535 852 c
        534 851 534 849 535 846 c
        536 844 537 841 536 838 c
        535 836 536 832 537 831 c
        538 830 539 826 539 823 c
        538 818 539 817 541 816 c
        551 814 552 808 544 800 c
        541 797 539 794 539 790 c
        539 786 540 785 547 782 c
        563 775 561 762 544 764 c
        533 766 522 760 522 752 c
        522 748 523 746 529 743 c
        535 741 537 738 540 732 c
        546 721 545 717 539 715 c
        537 714 533 714 531 714 c
        530 715 527 714 526 714 c
        525 713 528 710 535 705 c
        550 694 551 693 551 686 c
        551 683 554 677 557 672 c
        567 657 567 653 558 640 c
        556 638 555 633 555 629 c
        555 624 554 622 552 621 c
        550 621 547 618 546 615 c
        544 612 540 610 535 608 c
        530 607 524 604 521 601 c
        516 596 515 596 513 597 c
        509 600 505 598 505 594 c
        505 590 499 588 495 591 c
        h f
        383 665 m
        383 666 382 667 381 667 c
        380 667 380 666 380 665 c
        381 664 382 663 382 663 c
        383 663 383 664 383 665 c
        h f
        409 673 m
        417 678 416 680 406 680 c
        397 681 392 676 396 671 c
        399 668 402 668 409 673 c
        h f
        387 680 m
        387 681 385 683 383 684 c
        381 685 379 686 379 687 c
        379 691 355 693 348 689 c
        346 688 343 688 340 690 c
        336 692 334 693 331 691 c
        328 689 327 684 330 684 c
        332 685 338 684 346 683 c
        356 682 360 682 363 684 c
        369 688 372 687 378 681 c
        384 674 387 674 387 680 c
        h f
        437 679 m
        437 684 432 684 432 678 c
        431 675 432 674 434 674 c
        436 674 437 676 437 679 c
        h f
        295 697 m
        295 701 295 701 293 700 c
        292 699 292 696 292 695 c
        294 691 295 692 295 697 c
        h f
        430 941 m
        430 944 426 945 424 942 c
        422 940 423 939 426 939 c
        428 939 430 940 430 941 c
        h f
        393 945 m
        393 949 386 948 385 944 c
        385 942 385 941 389 941 c
        391 942 393 943 393 945 c
        h f
        359 944 m
        362 945 362 951 360 953 c
        359 953 355 954 351 955 c
        344 956 344 956 344 952 c
        344 950 345 947 348 946 c
        354 943 355 943 359 944 c
        h f
        374 598 m
        375 599 376 600 378 600 c
        379 600 380 599 381 598 c
        381 596 380 595 378 595 c
        375 595 374 596 374 598 c
        h f
        36.00 1005 m
        36.09 1007 36.51 1007 37.08 1006 c
        37.60 1005 37.54 1003 36.94 1003 c
        36.35 1002 35.92 1003 36.00 1005 c
        h f
        0.92 0.83 0.80 rg
        482 602 m
        480 602 478 604 477 605 c
        476 606 471 609 467 611 c
        462 614 457 617 456 619 c
        454 621 448 628 442 634 c
        436 641 431 646 431 647 c
        431 652 408 647 407 642 c
        406 640 402 635 398 631 c
        394 626 390 621 390 620 c
        390 616 383 611 378 611 c
        370 611 358 619 355 625 c
        353 630 351 631 342 634 c
        325 638 304 637 304 630 c
        304 626 297 619 292 619 c
        287 619 286 621 285 633 c
        285 642 284 645 280 649 c
        277 652 273 654 271 654 c
        269 654 260 656 251 658 c
        236 661 233 661 228 659 c
        225 658 222 656 221 655 c
        219 651 209 650 205 654 c
        201 658 201 664 206 677 c
        212 693 210 700 197 713 c
        187 723 181 726 170 726 c
        166 726 164 727 162 730 c
        158 735 159 739 166 746 c
        175 754 173 759 162 764 c
        153 769 149 775 152 780 c
        153 782 156 786 158 787 c
        161 790 162 793 163 799 c
        165 814 163 820 156 825 c
        149 830 149 831 154 832 c
        158 833 158 834 159 847 c
        160 855 160 868 159 877 c
        158 885 157 897 156 903 c
        155 918 157 920 174 920 c
        186 920 188 921 192 925 c
        200 932 201 935 197 945 c
        192 959 197 967 208 962 c
        211 960 216 959 221 960 c
        230 961 250 971 252 977 c
        254 982 255 1001 252 1003 c
        251 1004 250 1006 250 1008 c
        250 1015 262 1009 266 998 c
        270 989 275 988 288 994 c
        298 999 298 999 299 1007 c
        301 1019 304 1021 311 1016 c
        318 1011 319 1008 314 1005 c
        309 1001 312 998 319 1000 c
        322 1001 327 1001 330 1001 c
        335 1000 337 1001 343 1006 c
        350 1013 350 1013 358 1011 c
        362 1009 366 1007 367 1005 c
        368 1003 374 999 381 996 c
        394 989 406 988 418 990 c
        423 991 424 991 424 989 c
        424 987 425 986 427 987 c
        428 988 430 986 430 984 c
        433 978 437 975 446 975 c
        453 975 456 976 460 981 c
        466 986 471 988 471 983 c
        471 982 470 981 469 981 c
        468 982 467 982 467 981 c
        467 980 468 979 469 979 c
        471 979 471 976 470 969 c
        469 958 l
        480 948 l
        490 937 l
        504 940 l
        517 942 518 942 521 939 c
        526 936 525 931 517 926 c
        508 920 507 916 512 903 c
        517 891 528 879 533 881 c
        535 881 539 883 543 884 c
        552 888 558 887 559 882 c
        561 876 558 873 549 870 c
        544 869 538 866 537 865 c
        532 860 531 850 534 839 c
        535 834 536 826 536 823 c
        537 819 539 816 541 815 c
        547 812 547 807 541 800 c
        532 792 534 786 545 780 c
        557 774 558 766 546 768 c
        537 769 522 762 519 755 c
        517 750 520 738 523 740 c
        529 743 541 730 540 722 c
        540 718 539 718 528 719 c
        514 719 512 717 520 712 c
        532 704 546 693 548 690 c
        549 688 550 685 550 683 c
        550 681 552 675 555 669 c
        561 657 562 654 556 644 c
        553 639 549 632 547 627 c
        542 617 540 614 533 612 c
        531 612 526 609 522 607 c
        516 603 514 602 506 603 c
        501 604 494 604 491 603 c
        488 602 484 602 482 602 c
        h f
        367 660 m
        367 661 366 662 364 662 c
        363 662 361 661 361 660 c
        361 659 363 658 364 658 c
        366 658 367 659 367 660 c
        h f
        385 665 m
        385 669 381 670 379 668 c
        377 666 381 662 383 662 c
        384 662 385 663 385 665 c
        h f
        409 672 m
        418 677 418 677 414 682 c
        412 685 411 685 410 684 c
        408 682 405 681 403 681 c
        400 681 396 680 394 678 c
        391 674 391 674 395 670 c
        400 666 400 666 409 672 c
        h f
        460 675 m
        458 677 454 674 456 672 c
        456 671 457 671 459 672 c
        460 673 460 675 460 675 c
        h f
        438 680 m
        437 683 431 684 430 681 c
        430 679 430 676 431 675 c
        434 672 434 672 436 675 c
        437 677 438 679 438 680 c
        h f
        387 680 m
        387 686 371 695 365 692 c
        362 691 358 691 355 692 c
        351 692 349 691 347 690 c
        345 687 344 687 340 690 c
        335 695 332 695 328 690 c
        324 686 326 682 332 684 c
        335 685 338 684 339 683 c
        342 681 358 680 363 682 c
        368 685 375 683 380 678 c
        384 674 387 675 387 680 c
        h f
        297 698 m
        297 701 296 703 294 703 c
        291 703 291 697 293 691 c
        294 687 297 691 297 698 c
        h f
        498 927 m
        498 927 496 929 495 930 c
        492 933 492 932 493 928 c
        494 925 498 924 498 927 c
        h f
        430 942 m
        430 945 429 946 425 946 c
        418 947 417 944 421 940 c
        424 936 430 938 430 942 c
        h f
        394 944 m
        394 946 392 948 390 950 c
        387 952 386 951 384 947 c
        383 944 383 942 385 941 c
        388 939 394 941 394 944 c
        h f
        374 951 m
        372 953 369 951 368 945 c
        366 940 l
        371 945 l
        374 948 375 950 374 951 c
        h f
        359 944 m
        360 945 365 954 365 956 c
        365 956 360 956 353 957 c
        342 957 342 957 342 952 c
        343 946 354 941 359 944 c
        h f
        313 956 m
        313 959 308 961 307 959 c
        306 958 307 956 308 955 c
        311 953 313 953 313 956 c
        h f
        361 998 m
        362 1001 352 1004 350 1002 c
        349 1001 349 1000 350 999 c
        351 996 359 996 361 998 c
        h f
        425 992 m
        423 993 422 995 422 997 c
        422 1002 428 1001 429 996 c
        430 992 427 989 425 992 c
        h f
        425 1007 m
        424 1008 424 1010 425 1011 c
        426 1013 431 1011 431 1008 c
        431 1005 427 1005 425 1007 c
        h f
        0.96 0.80 0.66 rg
        479 605 m
        478 605 474 608 470 610 c
        462 614 443 633 435 643 c
        432 648 430 649 425 649 c
        417 649 407 645 405 640 c
        404 638 400 633 397 630 c
        393 626 390 621 389 619 c
        386 607 365 610 356 624 c
        351 633 335 639 322 637 c
        317 637 309 636 305 636 c
        300 636 298 635 298 633 c
        299 632 300 631 301 631 c
        304 631 301 623 297 621 c
        289 616 286 619 286 632 c
        286 642 286 644 281 649 c
        277 654 274 655 264 656 c
        257 657 248 659 245 660 c
        235 663 232 662 225 657 c
        215 651 210 650 205 654 c
        204 655 202 658 202 660 c
        202 665 206 678 208 680 c
        208 681 209 686 209 691 c
        209 700 209 701 199 712 c
        188 723 181 726 170 726 c
        159 726 157 736 166 745 c
        174 754 173 759 162 764 c
        156 767 150 773 150 777 c
        150 777 153 782 157 786 c
        165 795 165 798 163 815 c
        163 819 161 823 160 823 c
        159 823 157 824 155 826 c
        153 828 153 829 157 832 c
        161 834 161 836 161 852 c
        161 862 160 878 158 888 c
        156 901 156 908 157 913 c
        158 919 l
        171 919 l
        184 919 190 921 196 928 c
        201 934 201 936 197 946 c
        196 950 195 955 196 958 c
        197 963 197 963 203 962 c
        216 959 230 960 234 963 c
        235 965 240 967 244 969 c
        248 970 251 973 252 976 c
        255 982 254 1000 252 1003 c
        250 1005 250 1011 253 1011 c
        257 1011 264 1004 266 998 c
        268 989 276 987 290 994 c
        297 998 299 998 305 997 c
        313 995 324 996 324 999 c
        324 1000 327 1000 330 1000 c
        335 1000 338 1002 344 1007 c
        350 1013 351 1013 354 1011 c
        356 1010 359 1009 361 1009 c
        364 1009 366 1008 367 1005 c
        368 1002 372 999 378 997 c
        383 995 389 992 391 991 c
        393 990 400 989 406 989 c
        413 989 420 987 421 986 c
        423 985 425 984 427 985 c
        428 985 430 985 430 984 c
        430 979 437 975 446 975 c
        451 975 455 974 455 973 c
        455 972 456 973 459 976 c
        462 979 464 980 467 979 c
        471 978 471 978 470 968 c
        469 958 l
        480 948 l
        490 937 l
        504 940 l
        517 942 518 942 521 939 c
        526 936 525 931 517 926 c
        509 920 508 916 511 905 c
        514 895 524 883 530 881 c
        532 880 536 881 539 883 c
        546 887 556 887 559 884 c
        563 877 558 872 548 870 c
        544 870 539 867 537 865 c
        533 860 530 850 531 846 c
        534 842 536 829 535 821 c
        535 815 536 814 538 814 c
        542 814 546 811 546 808 c
        546 807 544 804 540 801 c
        533 792 534 786 543 780 c
        558 772 559 768 544 768 c
        533 768 522 762 519 755 c
        517 750 520 738 523 740 c
        529 743 541 730 540 722 c
        540 718 539 718 528 719 c
        514 719 512 717 520 712 c
        532 704 546 693 548 690 c
        549 688 550 685 550 683 c
        550 681 552 675 555 669 c
        561 657 562 654 556 644 c
        553 639 549 632 547 627 c
        542 617 540 614 533 612 c
        531 612 526 609 523 607 c
        517 603 514 603 498 603 c
        488 603 479 604 479 605 c
        h f
        370 627 m
        370 628 370 629 369 628 c
        368 628 367 627 367 626 c
        367 625 368 624 369 624 c
        370 624 370 625 370 627 c
        h f
        375 633 m
        376 635 365 641 362 639 c
        358 638 359 635 364 636 c
        367 637 369 636 370 634 c
        371 631 373 630 375 633 c
        h f
        378 654 m
        379 655 379 656 378 656 c
        377 656 376 655 376 654 c
        376 653 376 653 377 653 c
        377 653 378 653 378 654 c
        h f
        369 658 m
        371 662 372 662 376 660 c
        378 659 381 658 384 658 c
        388 658 388 659 388 663 c
        388 669 390 670 397 666 c
        401 664 402 664 409 668 c
        415 672 418 672 424 671 c
        429 671 434 669 437 668 c
        441 666 443 666 447 667 c
        449 668 454 669 457 669 c
        463 669 464 669 464 675 c
        464 683 461 684 457 679 c
        452 674 444 675 442 681 c
        440 685 440 685 429 684 c
        420 684 417 684 414 686 c
        412 689 411 689 408 687 c
        403 684 402 684 404 688 c
        404 691 404 692 400 694 c
        393 696 391 696 393 694 c
        396 691 396 684 394 682 c
        393 682 391 682 391 684 c
        390 685 388 687 386 688 c
        384 688 379 691 375 693 c
        370 697 368 697 366 696 c
        365 695 357 694 347 694 c
        338 694 328 694 326 693 c
        323 692 322 693 322 694 c
        322 696 321 697 320 698 c
        317 699 313 696 313 693 c
        313 691 312 690 310 689 c
        306 687 307 685 314 684 c
        323 682 333 676 333 672 c
        333 671 335 668 338 667 c
        341 666 345 663 347 660 c
        349 657 351 656 354 657 c
        357 657 360 657 362 656 c
        365 654 366 654 369 658 c
        h f
        355 665 m
        352 665 351 667 350 669 c
        350 672 348 673 345 674 c
        339 675 336 678 342 678 c
        349 678 351 678 354 675 c
        356 673 359 671 361 670 c
        364 668 363 663 360 664 c
        359 664 357 664 355 665 c
        h f
        362 678 m
        363 679 365 680 366 680 c
        367 680 369 679 369 678 c
        369 677 367 676 365 676 c
        363 676 362 677 362 678 c
        h f
        308 659 m
        308 659 307 660 306 661 c
        305 661 304 661 304 660 c
        304 659 305 658 306 658 c
        307 658 308 658 308 659 c
        h f
        323 663 m
        319 668 317 668 317 664 c
        317 662 319 660 324 660 c
        326 660 325 661 323 663 c
        h f
        298 688 m
        303 701 300 708 291 703 c
        289 702 286 701 284 702 c
        279 702 278 696 282 694 c
        285 694 286 692 286 689 c
        286 683 289 679 293 680 c
        294 680 297 684 298 688 c
        h f
        498 929 m
        494 933 490 933 492 929 c
        493 925 496 923 498 925 c
        499 926 499 928 498 929 c
        h f
        424 930 m
        425 935 431 935 432 930 c
        433 926 433 926 436 929 c
        439 932 439 932 436 935 c
        431 938 433 943 439 943 c
        446 943 446 946 441 950 c
        435 955 431 955 432 950 c
        433 946 433 946 431 948 c
        428 951 420 951 416 947 c
        413 943 413 943 416 939 c
        418 936 419 933 418 932 c
        417 931 417 930 418 929 c
        420 926 423 927 424 930 c
        h f
        406 934 m
        411 940 407 941 401 936 c
        397 932 397 932 400 932 c
        402 932 405 933 406 934 c
        h f
        372 941 m
        375 946 379 946 381 941 c
        383 937 391 936 394 940 c
        398 944 398 945 392 951 c
        387 956 386 957 381 956 c
        378 955 375 956 374 957 c
        370 960 361 961 359 959 c
        357 957 355 957 350 959 c
        345 961 343 961 339 958 c
        337 956 333 956 330 956 c
        327 957 326 957 324 952 c
        320 945 317 945 317 952 c
        317 955 315 958 313 960 c
        307 964 302 963 302 958 c
        302 955 304 953 309 952 c
        314 950 315 949 314 945 c
        313 939 317 936 321 939 c
        323 941 330 941 344 941 c
        352 942 360 939 362 936 c
        364 933 369 936 372 941 c
        h f
        329 947 m
        329 949 331 951 333 952 c
        336 953 337 953 337 951 c
        338 950 338 948 337 947 c
        336 944 329 944 329 947 c
        h f
        256 943 m
        256 945 255 946 252 946 c
        248 946 248 945 251 941 c
        254 938 256 939 256 943 c
        h f
        295 953 m
        295 954 294 955 292 955 c
        288 955 287 953 290 951 c
        294 950 295 950 295 953 c
        h f
        308 976 m
        308 980 306 979 301 974 c
        298 972 297 969 298 967 c
        299 964 300 964 304 969 c
        306 971 308 975 308 976 c
        h f
        318 973 m
        318 974 317 975 316 975 c
        314 975 313 974 313 973 c
        313 972 314 972 315 972 c
        316 972 317 972 318 973 c
        h f
        381 974 m
        379 976 377 977 375 976 c
        371 974 373 972 379 972 c
        383 972 383 972 381 974 c
        h f
        397 974 m
        397 976 396 977 395 976 c
        394 975 394 974 394 973 c
        394 972 394 972 395 972 c
        396 972 397 973 397 974 c
        h f
        408 982 m
        408 985 405 986 403 983 c
        401 981 401 981 404 981 c
        406 981 408 981 408 982 c
        h f
        352 991 m
        354 993 356 995 357 995 c
        358 995 360 996 361 997 c
        364 1000 360 1004 353 1004 c
        347 1004 347 1003 348 999 c
        348 995 346 992 342 995 c
        338 998 335 997 336 994 c
        337 991 342 988 347 988 c
        348 988 350 989 352 991 c
        h f
        467 982 m
        462 983 463 986 467 986 c
        469 986 471 985 471 983 c
        471 982 471 981 470 981 c
        470 981 468 981 467 982 c
        h f
        425 995 m
        424 996 424 997 425 998 c
        426 999 430 997 430 995 c
        430 992 426 993 425 995 c
        h f
        300 1009 m
        301 1019 304 1021 311 1016 c
        318 1011 319 1005 312 1006 c
        309 1007 308 1007 308 1005 c
        308 1003 306 1002 303 1002 c
        299 1002 298 1002 300 1009 c
        h f
        424 1009 m
        424 1010 426 1011 428 1011 c
        430 1011 431 1010 431 1009 c
        431 1008 430 1007 428 1007 c
        426 1007 424 1008 424 1009 c
        h f
        0.88 0.67 0.59 rg
        476 609 m
        465 615 436 642 434 649 c
        433 653 432 653 424 652 c
        412 650 403 646 403 643 c
        403 642 399 637 394 633 c
        387 628 385 624 386 622 c
        386 620 385 618 384 617 c
        382 615 380 615 373 617 c
        362 620 358 624 358 629 c
        358 632 359 634 363 635 c
        369 637 372 635 368 631 c
        364 627 364 622 369 622 c
        371 622 372 623 372 625 c
        372 627 373 630 374 631 c
        376 633 376 634 373 636 c
        367 640 362 641 358 638 c
        354 634 346 634 339 638 c
        335 640 330 640 317 639 c
        307 639 300 639 300 640 c
        299 641 295 644 290 646 c
        285 649 280 652 279 653 c
        278 655 273 656 268 657 c
        263 658 257 660 254 661 c
        252 662 245 664 239 665 c
        228 666 227 666 223 662 c
        219 658 211 655 209 658 c
        206 661 207 672 210 678 c
        212 682 213 686 212 694 c
        211 704 210 705 201 714 c
        189 726 184 730 178 730 c
        174 730 170 731 168 733 c
        164 736 l
        169 744 l
        173 748 175 754 175 756 c
        175 759 167 769 164 769 c
        162 769 157 775 157 778 c
        157 780 159 785 162 788 c
        166 793 167 794 166 807 c
        165 820 164 822 160 825 c
        157 827 157 828 159 830 c
        160 832 161 837 161 842 c
        161 847 162 853 162 855 c
        164 858 161 887 158 902 c
        156 910 164 916 175 916 c
        186 916 195 921 201 930 c
        205 936 207 937 213 938 c
        217 939 220 941 221 943 c
        222 947 221 950 217 950 c
        216 950 215 951 215 952 c
        215 954 221 954 222 951 c
        223 950 224 950 227 953 c
        232 960 240 964 247 964 c
        258 964 261 966 261 973 c
        261 976 262 981 264 983 c
        266 987 267 988 273 987 c
        279 986 281 986 287 990 c
        294 995 295 995 313 994 c
        327 993 332 993 335 990 c
        339 986 350 986 354 990 c
        357 993 359 993 368 992 c
        383 990 385 990 383 987 c
        380 984 382 981 387 981 c
        390 981 390 981 390 984 c
        389 987 389 988 391 987 c
        393 986 394 986 395 986 c
        395 986 394 983 392 979 c
        388 968 391 966 399 973 c
        404 977 406 979 409 978 c
        412 977 413 978 414 982 c
        414 984 415 987 416 987 c
        417 987 422 984 428 979 c
        436 971 438 970 442 970 c
        445 971 449 971 451 970 c
        454 968 455 969 458 972 c
        459 974 462 975 464 975 c
        467 975 467 974 467 966 c
        467 959 468 956 473 950 c
        477 947 481 940 484 935 c
        489 925 493 922 500 921 c
        504 921 505 921 505 916 c
        505 913 506 909 507 906 c
        509 903 511 898 513 894 c
        515 889 520 884 534 876 c
        538 874 537 873 534 868 c
        529 861 528 850 532 836 c
        533 831 534 825 534 822 c
        535 819 535 816 535 815 c
        535 815 537 814 539 813 c
        544 811 543 807 538 801 c
        535 798 533 794 533 791 c
        533 784 534 782 543 778 c
        546 776 550 774 550 773 c
        550 772 546 771 541 771 c
        523 769 513 757 517 743 c
        518 739 519 738 524 738 c
        531 738 533 736 535 729 c
        538 720 536 719 525 721 c
        519 723 517 723 515 721 c
        511 717 511 712 516 710 c
        523 708 542 694 545 690 c
        547 688 548 684 548 682 c
        548 680 550 674 552 669 c
        555 663 557 657 557 655 c
        557 650 551 637 548 636 c
        547 635 544 631 542 627 c
        540 620 537 618 528 613 c
        518 608 514 607 504 607 c
        497 607 490 606 489 606 c
        487 605 481 607 476 609 c
        h f
        379 654 m
        379 655 381 656 383 657 c
        386 657 388 658 388 663 c
        389 669 390 670 397 666 c
        401 664 402 664 409 668 c
        415 672 418 672 424 671 c
        429 671 434 669 437 668 c
        441 666 443 666 447 667 c
        449 668 454 669 457 669 c
        463 669 464 669 464 675 c
        464 683 462 684 457 679 c
        450 674 444 675 442 681 c
        440 685 440 685 429 684 c
        420 684 417 684 414 686 c
        412 689 411 689 408 687 c
        403 684 402 684 404 688 c
        404 691 403 692 398 695 c
        391 698 388 697 393 692 c
        396 689 396 684 394 682 c
        393 682 391 682 391 684 c
        388 688 368 698 366 695 c
        364 694 357 694 345 695 c
        335 695 327 695 326 694 c
        325 691 323 692 322 696 c
        320 702 314 698 308 686 c
        308 686 310 685 314 684 c
        322 682 333 676 333 672 c
        333 671 336 668 339 666 c
        343 665 346 661 347 659 c
        350 654 350 654 354 656 c
        356 657 358 657 361 656 c
        365 655 366 655 369 658 c
        372 663 378 662 377 656 c
        376 654 376 653 377 653 c
        378 653 379 653 379 654 c
        h f
        350 669 m
        350 672 348 673 345 674 c
        339 675 336 678 342 678 c
        348 678 352 678 352 676 c
        353 675 355 673 358 671 c
        365 668 365 665 357 665 c
        352 665 351 666 350 669 c
        h f
        309 659 m
        307 660 305 661 304 661 c
        303 659 308 656 310 656 c
        311 656 311 658 309 659 c
        h f
        323 663 m
        320 667 313 669 313 667 c
        313 664 319 660 322 660 c
        326 660 326 660 323 663 c
        h f
        298 688 m
        301 696 301 701 299 705 c
        297 707 296 707 292 705 c
        289 703 285 702 284 702 c
        281 702 280 701 280 697 c
        280 694 281 693 283 693 c
        286 694 286 693 286 688 c
        286 677 294 677 298 688 c
        h f
        243 708 m
        243 709 242 710 241 710 c
        240 710 240 709 240 707 c
        240 706 240 705 241 706 c
        242 706 243 707 243 708 c
        h f
        424 930 m
        425 935 431 935 432 930 c
        433 926 433 926 436 929 c
        439 932 439 932 436 935 c
        431 938 433 943 439 943 c
        446 943 446 946 440 951 c
        434 956 430 957 429 954 c
        428 953 425 952 423 951 c
        420 951 416 949 414 947 c
        412 944 412 944 415 941 c
        417 939 418 935 417 931 c
        417 926 417 925 420 925 c
        422 926 424 928 424 930 c
        h f
        406 934 m
        411 940 407 941 401 936 c
        397 932 397 932 400 932 c
        402 932 405 933 406 934 c
        h f
        253 937 m
        256 937 257 938 257 943 c
        258 946 259 950 260 951 c
        261 953 261 954 260 954 c
        258 954 258 953 258 952 c
        258 950 255 950 251 951 c
        244 952 243 951 246 945 c
        247 942 248 938 247 936 c
        247 934 247 934 248 935 c
        249 936 251 937 253 937 c
        h f
        372 941 m
        375 946 379 946 381 941 c
        383 937 391 936 394 940 c
        398 944 398 945 392 952 c
        388 957 386 957 381 956 c
        376 955 374 956 372 958 c
        370 961 359 962 357 959 c
        357 958 353 959 349 960 c
        343 961 341 961 338 958 c
        336 956 334 956 330 956 c
        327 957 325 957 323 953 c
        319 947 317 947 317 953 c
        317 959 320 963 324 963 c
        328 963 328 966 325 967 c
        323 968 321 970 320 972 c
        318 978 305 980 300 975 c
        298 973 296 972 295 972 c
        294 972 295 969 296 967 c
        297 964 299 960 300 957 c
        302 954 304 952 307 952 c
        311 951 312 950 312 946 c
        312 939 316 936 320 939 c
        323 941 326 941 331 939 c
        338 938 339 938 341 941 c
        343 943 344 943 349 942 c
        353 940 357 939 358 939 c
        360 939 362 938 362 937 c
        363 933 369 936 372 941 c
        h f
        329 947 m
        329 949 331 951 333 952 c
        336 953 337 953 337 951 c
        338 950 338 948 337 947 c
        336 944 329 944 329 947 c
        h f
        307 966 m
        305 967 308 970 311 970 c
        313 970 315 969 315 967 c
        315 964 309 963 307 966 c
        h f
        268 939 m
        268 940 267 941 265 941 c
        263 941 261 940 261 939 c
        261 938 263 938 265 938 c
        267 938 268 938 268 939 c
        h f
        241 944 m
        241 949 240 949 235 947 c
        228 943 228 943 230 941 c
        230 940 233 939 236 939 c
        240 939 241 940 241 944 c
        h f
        275 948 m
        275 949 275 950 274 950 c
        273 950 272 949 272 948 c
        272 947 272 946 273 946 c
        273 946 274 947 275 948 c
        h f
        297 952 m
        296 955 290 958 289 957 c
        288 956 288 954 289 953 c
        290 950 298 949 297 952 c
        h f
        355 966 m
        355 967 353 968 352 968 c
        351 968 351 967 352 966 c
        352 965 353 964 354 964 c
        355 964 356 965 355 966 c
        h f
        383 972 m
        383 976 380 977 375 976 c
        372 975 372 975 375 972 c
        379 969 383 969 383 972 c
        h f
        333 979 m
        332 981 330 982 328 982 c
        326 982 326 981 329 978 c
        334 973 336 973 333 979 c
        h f
        349 981 m
        349 982 348 982 347 982 c
        346 982 345 982 345 981 c
        345 980 346 979 347 979 c
        348 979 349 980 349 981 c
        h f
        512 933 m
        513 935 514 937 516 937 c
        519 937 519 931 515 931 c
        513 930 512 931 512 933 c
        h f
        204 948 m
        204 949 205 950 205 950 c
        206 950 208 949 208 948 c
        209 947 208 946 207 946 c
        205 946 204 947 204 948 c
        h f
        0.96 0.67 0.44 rg
        474 611 m
        464 616 444 635 438 646 c
        433 653 431 654 419 651 c
        414 650 404 648 396 647 c
        385 646 380 645 379 643 c
        377 641 374 641 368 641 c
        362 642 359 641 355 639 c
        353 638 350 637 348 637 c
        346 638 346 639 350 645 c
        352 649 356 652 357 653 c
        358 654 363 653 369 651 c
        378 647 378 647 381 650 c
        383 652 386 655 389 657 c
        393 659 394 659 397 658 c
        400 656 402 656 408 662 c
        415 670 421 671 425 665 c
        427 662 428 662 430 663 c
        431 664 435 664 439 665 c
        443 665 450 665 455 666 c
        460 666 464 667 465 668 c
        466 669 466 673 466 677 c
        465 685 461 688 456 682 c
        453 678 448 679 445 683 c
        443 686 441 687 430 687 c
        421 688 415 689 408 692 c
        393 700 391 700 388 696 c
        386 691 383 691 376 697 c
        370 702 368 702 365 698 c
        363 696 359 696 344 697 c
        333 697 324 698 323 699 c
        322 700 320 701 317 701 c
        315 701 313 702 312 703 c
        312 704 309 705 306 705 c
        303 705 300 706 299 707 c
        298 708 296 708 293 707 c
        291 706 286 705 284 704 c
        276 704 276 697 281 687 c
        286 679 295 673 296 677 c
        298 682 302 686 304 684 c
        306 683 310 681 314 680 c
        318 679 323 677 324 675 c
        327 671 321 670 313 673 c
        305 676 300 675 305 671 c
        309 667 308 665 300 665 c
        293 664 292 664 292 660 c
        291 654 296 649 299 653 c
        300 655 301 655 305 653 c
        309 651 310 648 310 645 c
        310 637 305 637 299 644 c
        296 647 292 649 290 649 c
        288 649 284 651 281 654 c
        277 656 272 658 268 658 c
        265 658 260 659 257 661 c
        254 663 246 664 239 665 c
        227 667 226 666 223 662 c
        221 660 217 658 214 658 c
        209 657 208 658 208 664 c
        207 668 208 673 211 678 c
        214 684 214 686 212 695 c
        211 703 209 707 201 714 c
        189 726 184 730 178 730 c
        174 730 170 731 168 733 c
        164 736 l
        169 744 l
        173 748 175 754 175 756 c
        175 759 167 769 164 769 c
        162 769 157 775 157 778 c
        157 780 159 785 162 788 c
        166 793 167 794 166 807 c
        165 820 164 822 161 825 c
        157 828 157 828 159 831 c
        161 833 163 839 163 845 c
        165 860 165 856 160 899 c
        159 908 159 908 164 911 c
        167 913 172 914 175 914 c
        183 914 195 921 201 929 c
        204 933 207 936 211 937 c
        215 937 218 939 219 940 c
        223 943 222 948 218 950 c
        216 950 215 951 215 952 c
        215 954 221 954 222 951 c
        223 950 226 952 229 956 c
        237 963 238 964 255 965 c
        265 966 267 965 268 963 c
        271 957 270 955 267 958 c
        264 959 262 959 258 955 c
        254 951 252 950 246 952 c
        241 953 239 953 236 950 c
        234 948 231 946 229 946 c
        228 946 227 945 227 943 c
        227 941 229 939 235 938 c
        239 937 244 935 245 933 c
        247 931 248 931 250 932 c
        251 934 257 935 262 935 c
        270 936 277 936 283 937 c
        286 938 282 941 279 941 c
        276 941 276 942 277 945 c
        279 948 279 950 276 955 c
        272 962 l
        276 965 l
        282 969 280 972 273 971 c
        265 969 262 973 264 980 c
        265 986 265 986 275 985 c
        284 984 285 984 290 989 c
        295 993 296 993 303 992 c
        309 991 311 989 311 986 c
        312 981 317 981 317 987 c
        317 991 317 991 324 991 c
        329 991 333 990 334 989 c
        336 987 338 986 341 986 c
        343 985 344 984 344 981 c
        344 978 344 976 346 976 c
        349 975 353 978 351 981 c
        350 984 362 993 365 992 c
        367 991 372 991 376 990 c
        380 990 384 989 384 989 c
        385 988 370 977 368 977 c
        366 977 364 975 363 973 c
        362 970 362 970 375 969 c
        383 969 391 968 393 967 c
        397 966 398 967 400 971 c
        403 977 404 977 411 976 c
        419 975 420 975 417 981 c
        416 983 415 984 416 984 c
        418 984 426 980 426 979 c
        426 975 439 969 443 971 c
        446 971 449 971 451 970 c
        454 968 456 968 459 971 c
        466 975 467 974 467 965 c
        467 959 468 956 473 951 c
        479 944 487 932 487 930 c
        487 929 485 929 483 928 c
        481 928 479 927 478 925 c
        478 923 479 923 484 924 c
        488 925 491 924 491 923 c
        492 922 495 921 499 921 c
        504 921 505 921 505 915 c
        505 912 506 909 506 908 c
        507 907 509 903 511 898 c
        514 889 522 881 530 878 c
        536 876 536 874 532 865 c
        528 859 528 856 529 845 c
        530 838 531 832 532 830 c
        533 829 533 825 533 822 c
        532 817 533 815 536 813 c
        541 809 542 805 538 801 c
        529 794 531 784 542 778 c
        546 776 550 774 550 773 c
        550 771 543 770 533 770 c
        533 770 531 771 529 772 c
        525 773 524 772 524 769 c
        524 766 523 764 521 762 c
        517 760 515 751 517 743 c
        518 739 519 738 524 738 c
        531 738 533 736 535 729 c
        538 720 536 719 525 721 c
        519 723 517 723 515 721 c
        511 717 511 712 516 710 c
        523 708 542 694 545 690 c
        547 688 548 684 548 682 c
        548 680 550 674 552 669 c
        555 663 557 657 557 655 c
        557 650 551 637 548 636 c
        547 635 544 631 542 627 c
        540 620 537 618 528 613 c
        518 608 515 607 499 607 c
        484 607 481 607 474 611 c
        h f
        482 670 m
        482 671 481 672 479 672 c
        477 672 476 672 476 670 c
        476 667 481 667 482 670 c
        h f
        244 704 m
        248 709 247 712 242 712 c
        238 712 235 708 237 704 c
        239 700 241 700 244 704 c
        h f
        234 728 m
        234 729 233 730 231 729 c
        230 729 228 728 227 726 c
        227 725 228 724 230 725 c
        232 725 233 726 234 728 c
        h f
        198 730 m
        198 731 198 731 197 731 c
        196 731 195 730 195 729 c
        195 727 196 726 197 727 c
        198 728 198 729 198 730 c
        h f
        425 923 m
        426 924 428 925 431 925 c
        434 925 438 927 439 928 c
        441 930 444 932 446 932 c
        450 933 450 934 451 942 c
        451 951 451 952 447 952 c
        445 952 441 953 439 955 c
        435 957 434 957 425 954 c
        414 950 l
        412 955 l
        410 959 406 961 406 957 c
        406 956 407 954 408 952 c
        410 949 410 948 406 945 c
        402 941 402 941 400 945 c
        399 948 396 952 392 954 c
        388 958 385 959 381 958 c
        378 958 376 958 375 959 c
        374 960 372 961 369 962 c
        366 962 363 964 361 965 c
        358 969 349 971 349 968 c
        349 967 346 965 342 964 c
        334 960 335 960 334 964 c
        334 966 331 968 328 968 c
        325 969 323 970 323 972 c
        323 974 325 975 332 972 c
        335 971 336 972 336 974 c
        336 981 333 984 326 984 c
        322 984 318 983 318 982 c
        318 981 316 981 313 981 c
        309 982 305 981 302 980 c
        300 978 296 976 294 975 c
        292 974 289 972 288 971 c
        286 968 286 968 291 966 c
        296 964 296 960 290 959 c
        286 959 282 955 284 952 c
        286 950 296 945 299 945 c
        301 945 303 946 304 947 c
        307 951 309 949 309 944 c
        309 941 310 939 312 938 c
        315 936 344 936 344 938 c
        344 940 356 938 360 935 c
        364 931 367 932 372 937 c
        377 941 378 941 379 939 c
        379 937 382 936 385 936 c
        387 936 391 934 393 932 c
        397 928 400 928 408 932 c
        412 935 413 935 413 933 c
        413 925 422 918 425 923 c
        h f
        468 931 m
        470 933 465 936 463 934 c
        461 932 463 929 465 929 c
        466 929 467 929 468 931 c
        h f
        424 963 m
        424 965 421 967 419 966 c
        418 964 420 961 422 961 c
        423 961 424 962 424 963 c
        h f
        376 618 m
        374 619 374 620 376 621 c
        377 621 378 624 378 626 c
        378 629 380 630 382 631 c
        384 631 385 630 384 625 c
        382 614 380 613 376 618 c
        h f
        390 635 m
        390 639 395 642 397 641 c
        399 640 398 639 396 636 c
        393 632 390 632 390 635 c
        h f
        337 647 m
        336 648 332 650 330 650 c
        322 650 323 654 332 657 c
        340 660 341 660 343 657 c
        345 654 344 644 342 644 c
        341 644 339 645 337 647 c
        h f
        304 694 m
        304 695 305 696 307 696 c
        308 696 309 695 309 694 c
        308 693 307 692 306 692 c
        305 692 304 693 304 694 c
        h f
        512 933 m
        513 935 514 937 516 937 c
        519 937 519 931 515 931 c
        513 930 512 931 512 933 c
        h f
        386 975 m
        384 977 385 978 388 979 c
        391 980 392 979 392 977 c
        391 974 388 973 386 975 c
        h f
        389 985 m
        389 987 389 988 391 987 c
        393 986 394 986 395 986 c
        395 986 395 985 395 984 c
        393 981 391 982 389 985 c
        h f
        0.92 0.57 0.30 rg
        476 612 m
        472 614 465 618 462 622 c
        458 625 452 632 448 636 c
        443 639 439 645 437 648 c
        434 655 431 657 428 655 c
        423 654 408 651 401 650 c
        387 649 379 647 378 644 c
        378 641 376 641 368 641 c
        362 642 357 642 355 640 c
        348 637 347 640 351 646 c
        356 654 358 655 369 651 c
        378 647 l
        385 653 l
        392 659 393 659 396 657 c
        399 654 399 654 404 658 c
        406 660 410 663 411 665 c
        414 670 421 670 425 665 c
        426 662 428 661 430 663 c
        431 664 435 664 439 665 c
        443 665 450 665 455 666 c
        460 666 464 667 465 668 c
        467 671 465 685 463 685 c
        461 685 458 683 455 682 c
        449 679 449 679 446 683 c
        443 687 441 688 430 688 c
        420 688 416 689 407 693 c
        394 700 388 701 387 696 c
        386 691 383 691 376 697 c
        370 702 368 702 365 698 c
        361 694 320 698 313 704 c
        311 705 308 706 306 706 c
        303 706 300 707 299 709 c
        298 710 296 710 292 708 c
        290 707 285 706 283 706 c
        276 706 274 697 279 690 c
        281 687 283 684 283 682 c
        283 681 284 679 286 678 c
        290 676 297 675 297 677 c
        297 678 298 680 299 682 c
        301 685 303 685 304 684 c
        306 683 310 681 314 680 c
        318 679 323 677 324 675 c
        327 671 321 670 313 673 c
        307 675 305 675 303 674 c
        302 673 302 672 305 670 c
        310 667 308 665 300 665 c
        293 664 292 664 292 658 c
        292 650 288 649 282 654 c
        279 657 274 659 269 659 c
        263 660 258 662 256 664 c
        254 666 251 667 244 667 c
        234 667 230 669 228 676 c
        227 680 l
        223 676 l
        221 674 220 671 220 669 c
        220 665 215 661 213 662 c
        209 665 211 673 219 685 c
        228 698 229 703 222 704 c
        219 705 213 708 208 712 c
        200 719 199 721 200 727 c
        200 733 200 733 194 734 c
        190 735 183 735 179 735 c
        173 735 172 735 172 739 c
        172 741 173 744 174 746 c
        179 755 179 760 173 766 c
        170 770 167 773 165 773 c
        161 773 160 779 164 786 c
        168 794 169 808 166 820 c
        163 831 162 846 164 852 c
        166 858 166 861 162 896 c
        160 906 l
        167 909 l
        171 911 175 912 178 912 c
        185 912 198 919 202 926 c
        204 930 208 933 212 934 c
        219 936 237 935 243 931 c
        248 928 249 928 253 931 c
        257 933 261 934 276 933 c
        294 932 295 932 296 936 c
        298 941 306 950 308 949 c
        309 948 309 946 309 944 c
        309 938 314 935 320 936 c
        324 937 329 936 333 936 c
        338 935 341 935 343 937 c
        346 940 357 939 360 935 c
        363 931 367 932 372 937 c
        377 941 378 941 379 939 c
        379 937 382 936 384 936 c
        386 936 390 934 391 932 c
        395 928 400 928 407 932 c
        413 935 l
        414 929 l
        416 922 419 920 424 923 c
        426 924 430 925 432 925 c
        434 925 438 927 439 929 c
        441 931 445 932 448 932 c
        453 932 453 933 452 936 c
        452 938 451 943 451 946 c
        451 951 450 952 447 952 c
        445 952 441 953 439 954 c
        437 956 434 957 431 957 c
        427 957 426 958 426 962 c
        425 966 424 967 421 968 c
        419 968 417 967 417 966 c
        417 963 419 959 422 959 c
        425 959 425 956 420 954 c
        417 952 416 952 412 957 c
        409 961 408 961 406 960 c
        404 958 404 956 406 953 c
        408 948 408 947 405 944 c
        401 941 401 941 400 945 c
        399 947 396 952 393 955 c
        388 960 386 961 381 960 c
        376 959 374 959 372 962 c
        371 963 369 964 367 964 c
        365 964 364 967 367 969 c
        368 969 375 969 383 968 c
        398 966 398 966 401 970 c
        404 973 406 974 410 973 c
        414 972 416 973 418 975 c
        421 978 421 978 428 972 c
        433 968 436 966 442 966 c
        445 966 449 965 451 964 c
        452 962 454 963 457 966 c
        463 971 465 971 465 965 c
        465 957 468 950 474 945 c
        476 943 478 940 477 940 c
        477 940 473 939 469 938 c
        459 936 458 932 466 927 c
        470 924 474 922 480 922 c
        488 922 498 918 498 915 c
        498 914 500 911 503 907 c
        506 904 508 899 508 898 c
        508 892 511 888 522 880 c
        531 873 532 873 529 868 c
        525 860 524 852 526 848 c
        528 844 532 828 532 820 c
        533 816 534 813 536 812 c
        540 810 540 807 534 799 c
        529 793 529 788 535 779 c
        540 773 539 772 531 772 c
        526 773 524 772 523 770 c
        523 768 520 764 518 761 c
        514 757 514 754 514 748 c
        515 739 518 735 525 735 c
        528 735 530 734 532 731 c
        535 725 533 723 523 724 c
        515 725 514 724 511 721 c
        507 715 510 712 524 703 c
        539 694 544 689 546 681 c
        547 678 549 670 551 665 c
        555 653 554 646 546 637 c
        543 634 541 631 541 630 c
        541 626 533 618 524 614 c
        516 610 512 609 500 609 c
        489 609 482 610 476 612 c
        h f
        483 670 m
        483 671 482 672 480 672 c
        478 672 476 671 476 670 c
        476 666 482 667 483 670 c
        h f
        494 674 m
        494 676 493 678 492 678 c
        491 678 490 676 490 674 c
        490 672 491 671 492 671 c
        493 671 494 672 494 674 c
        h f
        444 696 m
        444 697 443 697 442 697 c
        441 697 440 697 440 696 c
        440 695 441 694 442 694 c
        443 694 444 695 444 696 c
        h f
        246 706 m
        249 713 248 718 243 715 c
        238 712 235 705 238 702 c
        241 699 242 700 246 706 c
        h f
        235 725 m
        238 726 240 728 240 729 c
        240 732 231 732 226 729 c
        218 723 226 720 235 725 c
        h f
        203 760 m
        203 763 200 764 197 762 c
        194 760 196 757 200 757 c
        203 757 204 758 203 760 c
        h f
        495 909 m
        496 910 495 911 494 911 c
        493 911 492 910 492 909 c
        492 908 493 907 493 907 c
        493 907 494 908 495 909 c
        h f
        301 643 m
        298 646 298 654 301 654 c
        304 654 309 649 309 645 c
        309 641 305 640 301 643 c
        h f
        338 647 m
        336 649 332 650 330 650 c
        322 650 323 654 332 657 c
        337 658 341 659 342 659 c
        344 658 344 644 342 644 c
        341 644 339 645 338 647 c
        h f
        304 694 m
        304 695 305 696 307 696 c
        308 696 309 695 309 694 c
        308 693 307 692 306 692 c
        305 692 304 693 304 694 c
        h f
        278 947 m
        279 950 279 954 278 957 c
        276 961 276 963 279 965 c
        284 970 284 971 277 972 c
        272 974 272 974 275 978 c
        276 980 280 982 284 983 c
        287 984 291 986 292 988 c
        295 992 299 992 305 989 c
        313 984 311 982 297 980 c
        296 980 292 977 290 974 c
        285 969 l
        289 967 l
        294 963 295 961 290 961 c
        288 961 285 959 284 956 c
        281 952 281 952 287 948 c
        291 946 293 944 293 944 c
        293 943 290 943 285 943 c
        278 943 277 943 278 947 c
        h f
        335 963 m
        335 965 335 966 336 967 c
        337 968 338 971 338 975 c
        338 983 l
        343 977 l
        348 971 348 968 340 964 c
        335 961 335 961 335 963 c
        h f
        353 972 m
        350 975 352 983 357 987 c
        363 991 375 989 373 985 c
        373 983 372 981 372 980 c
        372 979 371 979 370 979 c
        369 979 366 977 363 974 c
        359 969 356 969 353 972 c
        h f
        319 988 m
        319 989 320 990 321 990 c
        321 990 324 989 326 989 c
        328 989 329 989 329 988 c
        329 987 327 986 324 986 c
        321 986 318 987 319 988 c
        h f
        0.95 0.56 0.14 rg
        476 614 m
        470 616 466 619 465 622 c
        464 624 460 628 457 630 c
        449 634 438 646 436 652 c
        434 657 431 658 422 656 c
        418 655 415 654 414 655 c
        411 660 410 665 412 667 c
        415 670 422 669 422 666 c
        422 663 427 661 430 663 c
        431 664 435 664 439 665 c
        443 665 450 665 455 666 c
        460 666 464 667 465 668 c
        466 669 466 673 466 678 c
        464 687 460 689 453 683 c
        449 679 l
        446 683 l
        443 687 441 688 430 688 c
        418 688 413 689 400 697 c
        393 702 387 701 387 696 c
        387 691 382 691 376 697 c
        370 702 368 702 365 698 c
        362 695 337 696 330 700 c
        326 702 322 703 319 703 c
        316 703 313 704 311 706 c
        310 707 308 708 307 708 c
        305 708 302 709 300 710 c
        297 712 296 712 292 710 c
        289 708 285 707 283 708 c
        280 708 278 708 277 706 c
        274 702 275 693 279 689 c
        281 688 283 684 283 681 c
        283 676 283 676 290 676 c
        294 676 297 677 297 677 c
        297 678 298 680 299 682 c
        301 685 303 685 304 684 c
        306 683 310 681 314 680 c
        318 679 323 677 324 675 c
        327 671 321 670 313 673 c
        307 675 305 675 303 674 c
        302 673 302 672 305 670 c
        310 667 308 665 300 665 c
        293 664 292 664 292 658 c
        292 654 292 653 289 653 c
        287 653 283 654 281 656 c
        278 658 273 660 269 660 c
        263 661 259 663 256 666 c
        253 670 251 671 246 670 c
        239 669 232 671 232 675 c
        232 676 234 679 237 681 c
        243 687 243 688 235 688 c
        230 688 229 689 229 692 c
        229 696 237 702 239 699 c
        241 696 243 697 244 702 c
        245 704 247 707 249 708 c
        252 710 252 710 249 715 c
        247 721 247 721 242 718 c
        240 716 238 713 238 712 c
        238 711 236 708 235 705 c
        232 701 232 701 229 703 c
        227 705 224 707 222 707 c
        216 709 203 718 201 721 c
        201 722 201 725 202 726 c
        206 733 200 737 183 737 c
        179 738 175 738 174 739 c
        174 740 175 743 177 747 c
        180 752 180 756 179 759 c
        178 764 171 773 169 773 c
        168 773 166 774 164 776 c
        162 778 162 779 165 784 c
        167 787 168 794 169 800 c
        170 811 167 835 164 839 c
        163 840 163 843 163 847 c
        165 855 165 870 164 888 c
        162 904 l
        169 908 l
        172 909 177 911 179 911 c
        186 911 202 919 203 923 c
        205 930 213 933 226 931 c
        232 931 239 929 243 928 c
        247 927 249 927 255 930 c
        263 934 l
        268 929 l
        274 924 l
        286 927 l
        293 929 300 930 300 929 c
        303 928 304 931 302 938 c
        301 942 301 944 303 947 c
        307 951 309 949 309 943 c
        309 939 311 937 315 936 c
        317 935 321 935 322 936 c
        323 936 328 936 332 935 c
        339 933 341 934 343 936 c
        344 939 346 939 351 938 c
        355 938 359 936 360 935 c
        363 931 367 932 372 937 c
        377 941 378 941 379 939 c
        379 937 382 936 384 936 c
        386 936 389 934 390 932 c
        393 928 400 927 407 932 c
        413 936 l
        414 928 l
        415 923 416 921 420 921 c
        422 921 424 922 425 923 c
        426 924 428 925 431 925 c
        433 925 437 927 439 928 c
        445 934 456 932 465 924 c
        471 918 472 918 479 919 c
        489 921 493 919 491 911 c
        490 907 491 905 493 905 c
        494 905 496 906 497 907 c
        499 911 506 902 507 895 c
        508 890 510 887 518 881 c
        528 873 530 870 525 866 c
        522 863 522 853 526 844 c
        528 839 530 831 531 826 c
        532 821 533 815 535 812 c
        537 808 537 807 533 800 c
        528 793 528 788 533 778 c
        535 775 535 775 530 774 c
        525 772 516 764 514 757 c
        512 752 513 739 517 736 c
        518 734 521 733 523 733 c
        525 733 527 732 528 730 c
        530 726 529 726 522 726 c
        517 726 513 725 511 723 c
        507 719 508 711 513 709 c
        516 708 519 705 521 703 c
        524 701 528 698 531 697 c
        539 693 544 687 545 679 c
        546 675 548 669 550 665 c
        553 658 553 657 551 651 c
        550 647 547 641 544 637 c
        542 634 538 629 537 626 c
        535 623 530 619 522 615 c
        512 610 510 609 506 611 c
        503 613 501 613 489 611 c
        487 610 482 612 476 614 c
        h f
        284 669 m
        284 670 284 671 283 671 c
        282 672 281 671 281 670 c
        281 668 282 667 283 667 c
        284 667 284 668 284 669 c
        h f
        483 671 m
        483 674 483 674 480 674 c
        479 674 477 672 476 670 c
        476 668 477 667 479 667 c
        481 668 483 669 483 671 c
        h f
        496 671 m
        496 674 493 678 491 678 c
        488 678 486 673 489 670 c
        491 667 496 668 496 671 c
        h f
        444 697 m
        444 698 443 699 441 699 c
        440 699 438 698 438 697 c
        438 695 440 694 441 694 c
        443 694 444 695 444 697 c
        h f
        329 712 m
        329 713 329 714 328 714 c
        328 714 327 713 327 712 c
        326 711 326 710 327 710 c
        328 710 329 711 329 712 c
        h f
        284 714 m
        284 715 283 715 282 715 c
        280 715 279 715 279 714 c
        279 713 280 712 282 712 c
        283 712 284 713 284 714 c
        h f
        223 717 m
        225 719 227 721 229 721 c
        234 721 243 726 243 729 c
        243 733 239 735 231 733 c
        222 731 219 727 218 719 c
        216 713 220 711 223 717 c
        h f
        207 757 m
        207 764 203 767 197 764 c
        189 760 193 755 206 753 c
        207 753 207 755 207 757 c
        h f
        185 771 m
        184 775 180 775 180 772 c
        180 770 182 769 183 769 c
        185 769 186 770 185 771 c
        h f
        303 916 m
        303 917 301 918 300 918 c
        299 918 299 917 299 916 c
        299 915 300 914 302 914 c
        303 914 304 915 303 916 c
        h f
        351 646 m
        356 654 358 655 368 651 c
        373 649 379 648 380 648 c
        382 648 383 648 382 648 c
        380 647 379 645 378 644 c
        377 641 376 640 371 641 c
        363 643 362 643 354 641 c
        347 640 l
        351 646 l
        h f
        301 648 m
        298 654 298 654 301 654 c
        304 654 309 649 308 646 c
        307 642 303 643 301 648 c
        h f
        338 647 m
        336 650 333 651 330 650 c
        322 650 323 654 332 657 c
        337 658 341 659 342 659 c
        344 657 344 644 342 644 c
        341 644 339 645 338 647 c
        h f
        388 653 m
        392 655 394 655 393 653 c
        392 652 390 651 389 651 c
        386 651 386 651 388 653 c
        h f
        400 652 m
        399 654 404 659 406 657 c
        406 657 406 655 404 653 c
        402 652 401 651 400 652 c
        h f
        304 694 m
        304 695 305 696 307 696 c
        308 696 309 695 309 694 c
        308 693 307 692 306 692 c
        305 692 304 693 304 694 c
        h f
        452 940 m
        452 942 451 945 451 948 c
        451 951 450 952 446 952 c
        444 952 440 954 438 956 c
        435 959 433 961 432 961 c
        430 961 430 962 430 964 c
        430 966 431 966 434 965 c
        437 964 441 964 443 964 c
        445 964 448 964 448 963 c
        450 960 457 960 460 963 c
        461 965 462 964 463 962 c
        469 944 469 944 465 941 c
        460 937 454 937 452 940 c
        h f
        400 944 m
        398 949 399 952 402 952 c
        405 952 407 946 404 943 c
        402 941 401 941 400 944 c
        h f
        278 944 m
        277 947 280 950 283 949 c
        286 948 286 947 284 945 c
        282 943 280 942 278 944 c
        h f
        393 953 m
        392 954 392 956 392 957 c
        392 961 388 963 382 962 c
        377 960 370 964 371 966 c
        372 968 384 967 392 964 c
        394 963 397 963 398 965 c
        408 974 408 973 415 963 c
        418 956 419 955 416 958 c
        411 962 406 962 401 956 c
        396 952 395 951 393 953 c
        h f
        279 960 m
        279 963 280 966 281 967 c
        284 969 283 973 280 973 c
        277 973 276 980 279 981 c
        280 982 283 982 285 981 c
        288 980 288 979 287 975 c
        287 971 287 968 289 966 c
        292 963 292 963 289 963 c
        287 963 285 962 284 959 c
        283 952 279 952 279 960 c
        h f
        335 963 m
        335 965 335 966 336 966 c
        337 966 338 965 338 965 c
        338 964 337 962 336 962 c
        335 961 335 962 335 963 c
        h f
        354 975 m
        351 977 353 982 357 985 c
        363 989 366 988 368 984 c
        372 978 358 970 354 975 c
        h f
        296 984 m
        294 988 298 991 301 987 c
        303 985 304 983 303 982 c
        301 979 297 980 296 984 c
        h f
        307 984 m
        306 985 306 986 307 986 c
        308 986 310 985 310 984 c
        311 983 311 982 310 982 c
        309 982 307 983 307 984 c
        h f
        322 988 m
        322 989 323 989 324 989 c
        325 989 326 989 326 988 c
        326 987 325 986 324 986 c
        323 986 322 987 322 988 c
        h f
        0.82 0.44 0.32 rg
        473 615 m
        471 616 468 619 467 621 c
        466 624 462 628 458 630 c
        450 634 438 646 436 652 c
        434 657 431 658 422 656 c
        418 655 415 654 414 655 c
        411 660 410 665 412 667 c
        415 670 422 669 422 666 c
        422 663 427 661 430 663 c
        431 664 435 664 439 665 c
        468 666 467 666 466 676 c
        465 686 461 689 455 686 c
        453 685 452 684 452 683 c
        452 682 452 681 451 680 c
        450 680 447 681 446 683 c
        443 687 441 688 430 688 c
        418 688 413 689 400 697 c
        393 702 387 701 387 696 c
        387 691 382 691 376 697 c
        370 702 368 702 365 698 c
        363 696 356 696 336 697 c
        334 697 332 699 330 700 c
        328 702 324 703 320 703 c
        317 703 313 704 311 705 c
        304 711 300 712 293 711 c
        286 710 l
        286 717 l
        285 723 285 723 280 724 c
        278 724 274 725 273 726 c
        272 727 267 728 262 728 c
        254 728 253 729 247 734 c
        240 742 238 742 230 736 c
        222 731 218 730 215 734 c
        214 735 209 736 204 736 c
        199 736 193 738 190 739 c
        188 741 185 742 183 742 c
        180 742 180 749 183 754 c
        185 757 186 757 189 755 c
        191 753 196 751 202 749 c
        211 747 211 747 216 752 c
        219 755 222 758 222 759 c
        222 762 206 769 200 769 c
        196 769 193 771 189 775 c
        185 780 182 782 178 782 c
        173 782 172 782 170 792 c
        169 798 169 803 170 806 c
        171 809 171 813 171 816 c
        170 819 169 824 169 827 c
        169 831 167 835 165 837 c
        162 841 162 846 165 852 c
        166 856 166 862 165 876 c
        164 887 163 897 164 899 c
        165 905 175 910 185 912 c
        197 913 206 917 206 920 c
        206 923 213 930 217 930 c
        218 930 220 929 222 928 c
        225 924 229 924 234 928 c
        237 930 239 930 241 929 c
        246 926 255 928 259 931 c
        262 935 266 933 268 928 c
        270 923 274 922 282 925 c
        290 928 295 926 295 918 c
        296 913 297 911 300 911 c
        305 910 309 915 306 921 c
        305 924 305 926 306 929 c
        308 931 308 933 304 938 c
        300 943 300 943 303 947 c
        307 951 309 950 309 943 c
        309 939 311 937 315 936 c
        317 935 321 935 322 936 c
        323 936 328 936 332 935 c
        339 933 341 934 343 936 c
        344 939 346 939 351 938 c
        355 938 359 936 360 935 c
        363 931 367 932 372 937 c
        377 941 378 941 379 939 c
        379 937 382 936 384 936 c
        386 936 389 934 390 932 c
        393 928 400 927 407 932 c
        413 936 l
        414 928 l
        415 923 416 921 420 921 c
        422 921 424 922 425 923 c
        426 924 428 925 431 925 c
        434 925 438 927 439 929 c
        443 932 449 934 447 930 c
        446 929 448 929 451 929 c
        455 929 458 927 462 923 c
        467 918 468 917 474 918 c
        482 920 487 917 486 912 c
        484 907 488 903 495 903 c
        502 903 505 901 505 894 c
        505 889 513 880 518 880 c
        522 880 526 874 524 870 c
        524 868 522 866 521 865 c
        518 863 519 854 523 847 c
        528 840 532 830 532 822 c
        532 819 533 815 535 812 c
        538 807 538 807 533 802 c
        529 798 528 796 529 792 c
        532 775 532 775 528 773 c
        519 769 512 759 511 751 c
        510 740 515 731 523 731 c
        526 731 528 731 527 730 c
        526 728 524 728 522 729 c
        518 730 508 724 506 720 c
        503 713 513 703 532 693 c
        539 690 540 688 544 677 c
        548 664 550 646 548 643 c
        547 642 545 640 543 639 c
        541 638 538 634 535 629 c
        528 617 522 614 498 614 c
        486 614 475 614 473 615 c
        h f
        483 672 m
        484 676 482 677 478 674 c
        475 671 476 667 479 667 c
        481 668 483 670 483 672 c
        h f
        495 669 m
        497 671 494 678 492 678 c
        490 678 488 677 487 676 c
        486 674 486 673 488 670 c
        491 667 493 666 495 669 c
        h f
        444 697 m
        444 699 443 701 441 701 c
        438 701 437 699 438 696 c
        439 693 444 693 444 697 c
        h f
        328 710 m
        330 713 328 714 326 711 c
        324 709 324 708 325 708 c
        326 708 328 709 328 710 c
        h f
        351 646 m
        356 654 358 655 368 651 c
        373 649 379 648 380 648 c
        382 648 383 648 382 648 c
        380 647 379 645 378 644 c
        377 641 376 640 371 641 c
        363 643 362 643 354 641 c
        347 640 l
        351 646 l
        h f
        301 648 m
        298 654 298 654 301 654 c
        304 654 309 649 308 646 c
        307 642 303 643 301 648 c
        h f
        338 647 m
        336 650 333 651 330 650 c
        322 650 323 654 332 657 c
        337 658 341 659 342 659 c
        344 657 344 644 342 644 c
        341 644 339 645 338 647 c
        h f
        400 652 m
        399 654 404 659 406 657 c
        406 657 406 655 404 653 c
        402 652 401 651 400 652 c
        h f
        280 656 m
        278 658 274 660 271 660 c
        264 660 258 663 258 667 c
        258 668 256 670 254 670 c
        231 672 229 673 237 679 c
        239 680 241 682 241 683 c
        239 684 244 688 247 688 c
        251 688 251 691 248 693 c
        245 695 238 696 239 694 c
        239 693 238 692 236 692 c
        234 692 232 693 232 694 c
        232 695 233 696 234 696 c
        235 696 236 696 235 697 c
        235 698 235 699 236 700 c
        237 701 238 700 239 699 c
        241 696 243 697 244 702 c
        245 704 247 708 250 709 c
        252 711 254 713 253 714 c
        252 715 252 717 253 719 c
        254 722 256 723 260 723 c
        264 723 267 721 271 716 c
        275 710 276 708 276 701 c
        275 695 276 692 279 689 c
        281 688 283 684 283 681 c
        283 676 283 676 290 676 c
        294 676 297 677 297 677 c
        297 678 298 680 299 682 c
        301 685 303 685 304 684 c
        306 683 310 681 314 680 c
        318 679 323 677 324 675 c
        327 671 321 670 313 673 c
        307 675 305 675 303 674 c
        302 673 302 672 305 670 c
        310 667 308 665 300 665 c
        293 664 292 664 292 658 c
        292 654 292 653 289 653 c
        287 653 283 654 280 656 c
        h f
        284 669 m
        284 670 284 671 283 671 c
        282 672 281 671 281 670 c
        281 668 282 667 283 667 c
        284 667 284 668 284 669 c
        h f
        304 694 m
        304 695 305 696 307 696 c
        308 696 309 695 309 694 c
        308 693 307 692 306 692 c
        305 692 304 693 304 694 c
        h f
        229 709 m
        228 714 230 718 234 719 c
        238 719 238 719 237 713 c
        237 707 231 704 229 709 c
        h f
        209 723 m
        209 724 210 724 211 724 c
        212 724 213 724 213 723 c
        213 722 212 721 211 721 c
        210 721 209 722 209 723 c
        h f
        452 940 m
        452 942 451 945 451 948 c
        451 951 450 952 447 952 c
        442 952 437 955 437 958 c
        437 961 438 961 448 960 c
        459 959 464 956 463 950 c
        463 949 461 947 459 946 c
        457 945 455 942 454 940 c
        454 937 454 937 452 940 c
        h f
        400 944 m
        398 949 399 952 402 952 c
        405 952 407 946 404 943 c
        402 941 401 941 400 944 c
        h f
        278 944 m
        277 947 280 950 283 949 c
        286 948 286 947 284 945 c
        282 943 280 942 278 944 c
        h f
        393 954 m
        391 958 397 966 403 968 c
        407 969 408 969 410 966 c
        412 962 412 961 408 961 c
        407 961 403 959 400 956 c
        395 951 394 951 393 954 c
        h f
        279 960 m
        279 963 280 966 281 967 c
        284 969 283 973 280 973 c
        276 973 276 980 280 982 c
        284 983 288 977 287 973 c
        286 971 287 968 289 966 c
        292 963 292 963 289 963 c
        287 963 285 962 284 959 c
        283 952 279 952 279 960 c
        h f
        335 963 m
        335 965 335 966 336 966 c
        337 966 338 965 338 965 c
        338 964 337 962 336 962 c
        335 961 335 962 335 963 c
        h f
        377 962 m
        375 964 376 966 380 966 c
        385 966 386 964 382 962 c
        378 961 378 961 377 962 c
        h f
        356 975 m
        353 975 352 976 353 979 c
        354 984 364 988 368 984 c
        370 982 370 981 367 978 c
        364 975 362 973 362 974 c
        361 974 359 974 356 975 c
        h f
        296 984 m
        294 988 298 991 301 987 c
        303 985 304 983 303 982 c
        301 979 297 980 296 984 c
        h f
        307 984 m
        306 985 306 986 307 986 c
        308 986 310 985 310 984 c
        311 983 311 982 310 982 c
        309 982 307 983 307 984 c
        h f
        322 988 m
        322 989 323 989 324 989 c
        325 989 326 989 326 988 c
        326 987 325 986 324 986 c
        323 986 322 987 322 988 c
        h f
        0.93 0.42 0.05 rg
        475 616 m
        471 617 468 620 466 622 c
        465 624 462 627 461 628 c
        453 632 448 638 441 649 c
        436 655 433 660 433 661 c
        433 662 449 663 457 663 c
        461 663 465 664 467 665 c
        470 667 472 667 474 665 c
        477 664 480 664 485 665 c
        495 669 497 670 497 674 c
        497 679 489 682 488 677 c
        487 675 485 675 482 675 c
        479 676 477 675 474 673 c
        470 670 469 671 469 681 c
        468 687 468 688 463 687 c
        460 687 456 687 454 686 c
        451 686 450 687 448 692 c
        447 695 445 698 444 699 c
        441 703 437 700 437 695 c
        437 691 436 691 428 690 c
        421 689 418 690 411 695 c
        399 703 393 704 387 700 c
        382 697 379 697 375 701 c
        371 705 369 705 364 701 c
        361 699 357 698 349 699 c
        343 699 337 700 335 701 c
        333 702 328 703 323 704 c
        319 704 314 706 314 707 c
        313 707 311 708 309 708 c
        307 708 304 710 301 711 c
        298 714 296 714 293 712 c
        288 709 286 710 286 717 c
        285 723 285 723 280 724 c
        278 724 274 725 273 726 c
        272 727 267 728 262 728 c
        254 728 253 729 247 734 c
        240 742 238 742 230 736 c
        222 731 218 730 215 734 c
        214 735 209 736 204 736 c
        199 736 193 738 190 739 c
        188 741 185 742 183 742 c
        180 742 180 749 183 754 c
        185 757 186 757 189 755 c
        191 753 196 751 202 749 c
        211 747 211 747 216 752 c
        219 755 222 758 222 759 c
        222 762 206 769 200 769 c
        196 769 193 771 189 775 c
        185 780 182 782 178 782 c
        172 782 172 782 171 791 c
        170 796 170 804 170 809 c
        170 813 170 821 169 826 c
        168 830 167 845 167 858 c
        167 871 166 886 165 892 c
        164 903 l
        169 906 l
        172 908 176 909 178 909 c
        184 909 204 917 205 919 c
        207 924 213 930 216 930 c
        218 930 220 929 222 927 c
        225 925 226 925 231 926 c
        233 927 237 927 238 927 c
        242 924 251 925 255 928 c
        260 931 266 931 269 926 c
        271 923 272 922 280 925 c
        292 928 295 926 295 918 c
        296 913 297 911 300 911 c
        306 910 309 915 306 921 c
        305 924 305 926 306 928 c
        307 929 307 932 307 934 c
        305 938 l
        310 934 l
        314 931 316 931 321 932 c
        324 933 330 934 335 933 c
        340 932 344 933 344 934 c
        346 937 352 936 358 932 c
        363 928 363 928 369 931 c
        374 934 376 934 383 932 c
        387 931 390 929 390 928 c
        390 927 394 927 400 927 c
        410 928 411 927 411 924 c
        412 922 413 919 415 919 c
        419 917 436 922 441 926 c
        448 931 456 929 464 921 c
        471 914 476 913 479 916 c
        482 919 487 917 486 912 c
        484 907 488 903 495 903 c
        502 903 505 901 505 894 c
        505 890 509 885 518 880 c
        524 876 526 870 522 866 c
        518 862 518 856 522 848 c
        524 845 527 836 528 829 c
        529 822 531 814 533 812 c
        535 807 535 806 532 801 c
        528 796 528 791 531 780 c
        532 777 531 776 529 776 c
        525 776 519 771 519 768 c
        519 767 517 765 515 763 c
        507 753 511 731 522 731 c
        524 731 526 731 526 729 c
        526 728 525 728 522 729 c
        518 730 508 724 506 720 c
        503 713 513 703 532 693 c
        539 690 540 688 543 679 c
        548 665 550 650 548 646 c
        547 644 544 640 541 637 c
        538 634 535 629 533 626 c
        528 617 521 614 500 614 c
        490 614 479 615 475 616 c
        h f
        332 712 m
        333 714 332 716 330 717 c
        327 718 321 713 323 710 c
        325 707 332 709 332 712 c
        h f
        357 647 m
        356 648 357 651 360 651 c
        361 651 361 650 361 648 c
        361 645 359 645 357 647 c
        h f
        280 660 m
        274 663 274 666 280 664 c
        287 663 289 660 287 658 c
        286 657 283 658 280 660 c
        h f
        418 660 m
        417 661 418 662 419 662 c
        420 663 421 662 421 661 c
        421 658 419 657 418 660 c
        h f
        260 667 m
        258 669 256 671 256 673 c
        256 674 253 674 246 673 c
        238 672 236 673 236 675 c
        236 678 244 687 247 687 c
        250 687 251 690 249 693 c
        246 696 248 705 252 705 c
        257 705 258 708 255 712 c
        250 717 253 723 260 722 c
        264 722 267 721 271 716 c
        275 711 276 708 275 706 c
        273 703 273 694 276 687 c
        278 682 277 681 275 678 c
        273 676 270 674 268 674 c
        266 674 265 673 264 668 c
        264 663 l
        260 667 l
        h f
        291 671 m
        291 672 293 672 293 671 c
        296 670 296 669 292 669 c
        291 669 290 670 291 671 c
        h f
        232 694 m
        232 695 234 696 235 696 c
        236 696 238 695 239 694 c
        239 693 238 692 236 692 c
        234 692 232 693 232 694 c
        h f
        230 709 m
        228 714 230 718 234 719 c
        239 719 239 719 235 712 c
        232 706 231 705 230 709 c
        h f
        209 723 m
        209 724 210 724 211 724 c
        212 724 213 724 213 723 c
        213 722 212 721 211 721 c
        210 721 209 722 209 723 c
        h f
        455 951 m
        453 955 451 956 448 956 c
        446 955 442 956 440 958 c
        436 961 436 961 440 961 c
        452 961 460 958 462 954 c
        464 951 463 949 462 947 c
        459 944 458 944 455 951 c
        h f
        395 957 m
        393 960 395 963 398 963 c
        401 963 401 962 400 958 c
        399 955 396 955 395 957 c
        h f
        377 962 m
        375 964 376 966 380 966 c
        385 966 386 964 382 962 c
        378 961 378 961 377 962 c
        h f
        405 966 m
        405 967 406 968 407 968 c
        408 968 409 967 409 966 c
        410 965 409 964 407 964 c
        405 964 404 965 405 966 c
        h f
        283 977 m
        281 978 281 978 283 979 c
        285 980 286 979 286 978 c
        286 975 286 975 283 977 c
        h f
        357 981 m
        360 985 365 988 367 985 c
        369 983 361 975 357 975 c
        354 975 354 976 357 981 c
        h f
        0.91 0.31 0.05 rg
        473 617 m
        471 618 469 620 469 621 c
        469 623 464 627 459 631 c
        452 635 446 642 441 648 c
        437 654 433 659 433 660 c
        433 661 441 662 459 664 c
        468 665 471 665 478 665 c
        490 665 503 667 505 669 c
        511 674 526 669 526 661 c
        526 659 528 658 529 658 c
        531 658 534 657 536 655 c
        537 653 541 651 543 650 c
        548 648 548 643 543 642 c
        538 641 534 637 532 629 c
        529 619 524 616 504 617 c
        495 617 485 617 482 616 c
        479 616 475 616 473 617 c
        h f
        357 647 m
        356 648 357 651 360 651 c
        361 651 361 650 361 648 c
        361 645 359 645 357 647 c
        h f
        280 660 m
        274 663 274 666 280 664 c
        287 663 289 660 287 658 c
        286 657 283 658 280 660 c
        h f
        531 662 m
        529 663 528 667 527 670 c
        526 674 525 677 525 678 c
        524 679 511 675 504 672 c
        500 670 498 670 498 674 c
        498 680 493 682 488 679 c
        485 677 483 676 482 677 c
        481 677 478 676 476 674 c
        470 670 469 671 469 680 c
        469 686 468 687 465 688 c
        462 689 459 692 457 693 c
        455 695 452 697 450 699 c
        447 700 447 701 449 703 c
        452 707 451 707 443 707 c
        435 707 434 707 434 703 c
        434 700 435 697 436 695 c
        437 691 436 691 428 690 c
        421 689 418 690 413 693 c
        406 697 406 698 409 700 c
        414 703 411 706 405 703 c
        403 702 399 702 396 702 c
        394 703 390 702 387 700 c
        382 697 379 697 375 701 c
        371 705 369 705 364 701 c
        361 699 357 698 349 699 c
        334 699 331 701 332 709 c
        333 717 329 719 324 715 c
        320 711 318 711 318 716 c
        318 719 317 721 316 721 c
        315 721 313 722 311 723 c
        308 726 308 726 302 722 c
        298 720 295 717 295 716 c
        295 710 292 714 291 721 c
        290 731 292 732 299 730 c
        304 728 305 728 307 731 c
        310 736 313 736 316 731 c
        321 724 336 729 336 738 c
        336 741 335 741 328 740 c
        319 739 313 741 308 747 c
        306 750 297 749 297 745 c
        297 744 296 742 295 741 c
        294 739 293 739 288 742 c
        285 745 282 749 281 751 c
        280 753 278 755 276 755 c
        273 755 270 747 271 742 c
        272 739 271 739 263 740 c
        257 740 250 741 247 742 c
        237 745 228 744 226 739 c
        224 734 221 734 219 739 c
        217 742 217 743 221 747 c
        224 749 229 752 232 752 c
        241 754 245 762 236 762 c
        234 762 231 764 228 766 c
        226 769 222 770 215 770 c
        204 771 202 772 190 781 c
        186 784 181 787 180 787 c
        178 787 176 789 174 791 c
        171 794 171 798 171 814 c
        171 825 170 834 168 837 c
        167 840 166 845 167 855 c
        168 862 167 875 166 882 c
        165 890 165 897 165 899 c
        168 905 177 909 190 910 c
        204 911 208 913 211 923 c
        213 928 216 931 216 927 c
        216 926 219 924 222 922 c
        228 919 236 920 236 924 c
        236 925 239 926 247 926 c
        253 926 258 927 260 928 c
        262 930 263 930 265 926 c
        268 920 278 918 283 922 c
        287 926 292 925 292 918 c
        292 916 293 911 295 909 c
        298 905 299 905 302 906 c
        306 909 313 920 312 922 c
        311 923 311 926 310 929 c
        310 933 310 934 313 932 c
        315 931 317 931 317 932 c
        317 932 318 931 319 929 c
        322 926 322 926 325 930 c
        327 933 329 933 331 932 c
        334 930 341 932 344 934 c
        347 937 352 936 358 932 c
        363 928 363 928 369 931 c
        376 934 383 934 389 929 c
        391 927 395 927 401 927 c
        410 928 411 927 411 924 c
        412 922 413 919 415 919 c
        419 918 433 921 440 925 c
        443 927 445 927 446 925 c
        447 924 450 923 453 923 c
        455 923 459 922 461 920 c
        465 915 475 913 479 916 c
        481 918 481 917 481 913 c
        481 906 486 902 493 902 c
        500 902 503 899 503 893 c
        503 887 507 883 516 879 c
        519 878 523 876 523 874 c
        524 873 521 869 518 865 c
        515 862 512 858 512 857 c
        512 857 515 853 519 849 c
        525 843 526 841 528 830 c
        529 822 531 814 533 811 c
        536 806 536 806 532 803 c
        528 800 528 799 529 789 c
        530 778 530 777 526 775 c
        524 774 521 772 520 770 c
        519 768 517 765 514 763 c
        512 762 509 756 507 752 c
        506 747 503 742 502 741 c
        500 738 504 735 509 735 c
        515 735 515 729 509 725 c
        506 722 503 719 503 717 c
        503 714 510 702 514 700 c
        516 698 517 696 517 695 c
        517 691 521 689 529 689 c
        535 689 536 689 538 685 c
        548 665 548 660 539 660 c
        536 661 532 661 531 662 c
        h f
        483 691 m
        483 692 482 692 481 693 c
        480 694 480 693 480 692 c
        480 691 480 690 481 690 c
        482 690 483 691 483 691 c
        h f
        423 703 m
        424 704 423 705 421 705 c
        418 705 417 704 418 703 c
        418 702 420 701 421 701 c
        421 701 423 702 423 703 c
        h f
        485 709 m
        485 711 484 712 482 712 c
        478 712 479 713 488 722 c
        498 731 497 734 487 726 c
        480 720 l
        472 726 l
        463 732 463 732 461 729 c
        459 727 458 723 459 721 c
        459 717 460 716 468 715 c
        476 713 478 712 478 709 c
        478 707 478 705 479 704 c
        481 702 485 706 485 709 c
        h f
        423 721 m
        422 722 420 726 417 730 c
        410 742 399 741 402 728 c
        403 720 405 719 416 719 c
        422 719 424 720 423 721 c
        h f
        393 728 m
        397 730 395 736 392 736 c
        388 737 385 731 387 728 c
        389 726 390 726 393 728 c
        h f
        335 897 m
        335 898 334 899 333 900 c
        332 900 331 899 331 897 c
        331 896 332 895 333 895 c
        334 895 335 896 335 897 c
        h f
        336 924 m
        336 925 336 927 335 928 c
        334 928 333 928 333 928 c
        333 927 332 926 332 924 c
        331 922 332 921 333 921 c
        335 921 336 923 336 924 c
        h f
        261 665 m
        259 666 259 667 261 669 c
        263 672 265 670 264 666 c
        264 664 263 664 261 665 c
        h f
        291 671 m
        291 672 293 672 293 671 c
        296 670 296 669 292 669 c
        291 669 290 670 291 671 c
        h f
        236 675 m
        236 676 238 677 241 678 c
        243 679 245 681 245 682 c
        245 683 247 685 249 685 c
        252 686 252 688 252 693 c
        251 698 252 700 255 702 c
        259 704 259 706 257 709 c
        255 714 256 719 259 719 c
        260 719 261 717 261 716 c
        261 714 263 711 266 709 c
        268 707 271 703 271 701 c
        271 699 273 694 275 690 c
        277 685 277 682 276 680 c
        274 676 264 675 258 678 c
        252 681 247 680 250 677 c
        252 675 251 674 244 675 c
        240 675 236 675 236 675 c
        h f
        232 713 m
        232 714 233 715 234 715 c
        235 715 236 715 236 714 c
        236 713 235 711 234 711 c
        233 710 232 711 232 713 c
        h f
        195 739 m
        192 741 194 746 198 745 c
        200 745 201 744 202 742 c
        202 739 198 737 195 739 c
        h f
        453 954 m
        451 957 451 957 454 957 c
        456 957 458 955 458 953 c
        459 949 456 949 453 954 c
        h f
        441 957 m
        441 958 442 959 445 959 c
        447 959 449 958 449 957 c
        449 956 448 955 446 955 c
        444 955 442 956 441 957 c
        h f
        356 979 m
        357 982 363 985 364 984 c
        365 983 365 982 364 980 c
        363 977 355 976 356 979 c
        h f
        0.82 0.30 0.07 rg
        474 616 m
        472 617 471 619 471 621 c
        471 625 469 627 462 630 c
        458 632 452 635 450 636 c
        447 639 433 659 433 661 c
        433 662 449 663 457 663 c
        461 663 465 664 467 665 c
        469 666 472 666 475 665 c
        478 664 480 664 483 665 c
        486 667 488 667 489 665 c
        492 663 504 663 506 665 c
        506 666 510 667 513 667 c
        518 667 520 666 521 663 c
        522 660 525 658 527 657 c
        529 656 535 653 539 651 c
        548 647 550 642 543 643 c
        540 643 537 642 534 641 c
        531 638 530 636 530 630 c
        530 619 528 618 507 620 c
        491 622 489 622 485 619 c
        480 615 479 615 474 616 c
        h f
        357 647 m
        356 648 357 651 360 651 c
        361 651 361 650 361 648 c
        361 645 359 645 357 647 c
        h f
        280 660 m
        274 663 274 666 280 664 c
        287 663 289 660 287 658 c
        286 657 283 658 280 660 c
        h f
        261 665 m
        259 666 259 667 261 669 c
        263 672 265 670 264 666 c
        264 664 263 664 261 665 c
        h f
        532 670 m
        529 675 529 680 532 683 c
        535 689 542 675 539 669 c
        536 664 534 664 532 670 c
        h f
        291 671 m
        291 672 293 672 293 671 c
        296 670 296 669 292 669 c
        291 669 290 670 291 671 c
        h f
        470 672 m
        469 672 469 676 469 680 c
        469 683 468 687 467 688 c
        466 689 466 689 467 690 c
        469 692 469 692 466 693 c
        457 696 453 698 453 699 c
        453 700 455 703 457 706 c
        460 710 462 710 467 709 c
        473 708 478 701 478 694 c
        478 689 481 688 484 690 c
        486 692 486 693 485 696 c
        483 698 483 700 486 703 c
        487 705 489 709 489 712 c
        489 717 502 731 507 731 c
        512 731 511 729 506 725 c
        503 724 501 721 501 719 c
        501 713 510 696 516 691 c
        520 688 523 685 523 684 c
        523 682 511 678 509 679 c
        507 679 504 678 502 675 c
        498 670 l
        497 674 l
        497 677 497 680 497 682 c
        497 685 495 686 490 684 c
        488 684 487 682 487 681 c
        487 679 485 678 483 678 c
        481 678 477 676 475 674 c
        473 672 471 671 470 672 c
        h f
        236 675 m
        236 676 238 677 241 678 c
        243 679 245 681 245 682 c
        245 683 247 685 249 685 c
        252 686 252 688 252 693 c
        251 698 252 700 255 702 c
        258 704 259 706 257 708 c
        255 713 255 714 258 714 c
        261 714 272 702 272 700 c
        271 699 273 695 274 691 c
        277 685 277 682 276 680 c
        274 676 264 675 258 678 c
        252 681 247 680 250 677 c
        252 675 251 674 244 675 c
        240 675 236 675 236 675 c
        h f
        413 693 m
        408 697 406 698 408 699 c
        410 699 412 701 413 703 c
        415 705 416 705 417 703 c
        418 701 420 701 422 702 c
        425 703 425 703 421 705 c
        416 708 413 708 407 704 c
        401 701 390 702 390 706 c
        390 709 386 709 385 705 c
        385 703 383 701 381 700 c
        379 698 377 698 375 701 c
        371 705 369 705 364 701 c
        360 698 345 697 336 700 c
        333 701 333 702 335 705 c
        337 708 337 709 335 709 c
        334 710 333 711 333 713 c
        333 717 328 718 324 714 c
        320 711 320 711 323 717 c
        325 723 326 723 332 723 c
        335 723 340 722 341 720 c
        343 718 345 717 346 717 c
        349 717 347 721 343 725 c
        339 729 338 730 340 734 c
        341 737 343 741 345 743 c
        347 745 347 746 346 748 c
        344 750 342 750 338 747 c
        334 745 332 745 328 746 c
        326 747 322 748 320 748 c
        318 748 314 749 311 751 c
        305 756 295 755 290 750 c
        286 747 285 747 285 752 c
        285 757 284 758 280 758 c
        276 759 275 760 275 763 c
        276 768 272 770 270 767 c
        268 766 268 765 270 762 c
        271 760 271 758 270 757 c
        269 757 268 755 268 753 c
        268 746 253 741 249 746 c
        248 748 247 753 246 758 c
        246 766 l
        239 766 l
        234 765 232 766 230 770 c
        227 774 226 774 217 773 c
        207 772 206 772 195 781 c
        189 785 183 789 182 789 c
        180 789 178 791 176 793 c
        173 796 173 799 173 812 c
        173 822 173 828 172 829 c
        171 830 170 832 170 835 c
        170 837 169 839 168 840 c
        167 841 166 846 167 855 c
        168 862 167 875 166 882 c
        164 898 164 900 172 905 c
        177 909 180 910 206 911 c
        207 911 210 918 212 925 c
        213 929 215 930 217 926 c
        219 920 236 918 236 924 c
        236 925 239 926 247 926 c
        253 926 258 927 260 928 c
        262 930 263 930 265 926 c
        266 924 269 921 272 921 c
        274 920 278 919 279 919 c
        281 918 283 919 283 922 c
        285 927 291 925 291 919 c
        292 916 293 911 295 909 c
        298 905 299 905 302 906 c
        306 909 313 920 312 922 c
        311 923 311 926 310 929 c
        310 933 310 934 313 932 c
        315 931 317 931 317 932 c
        317 932 318 931 320 929 c
        324 924 l
        325 929 l
        328 935 331 934 331 927 c
        331 922 332 921 334 922 c
        336 922 337 924 337 927 c
        337 931 337 932 340 932 c
        342 932 344 933 344 934 c
        346 937 352 936 358 932 c
        363 928 363 928 369 931 c
        376 934 383 934 389 929 c
        391 927 395 927 401 927 c
        410 928 411 927 411 924 c
        412 922 413 919 415 919 c
        419 918 433 921 440 925 c
        443 927 445 927 446 925 c
        447 924 450 923 453 923 c
        455 923 459 922 461 920 c
        465 915 475 913 479 916 c
        482 918 482 918 481 913 c
        480 908 480 907 484 905 c
        486 903 490 902 492 902 c
        498 902 501 900 501 895 c
        501 890 507 883 515 880 c
        525 876 526 873 516 863 c
        509 858 508 855 509 853 c
        510 852 512 852 513 852 c
        514 852 518 850 520 847 c
        525 844 526 840 528 830 c
        529 822 531 814 533 812 c
        534 809 535 806 535 806 c
        535 806 532 804 530 803 c
        527 801 526 800 527 795 c
        528 792 529 786 529 783 c
        530 778 530 776 528 776 c
        525 776 507 758 506 754 c
        504 747 496 742 488 741 c
        482 740 481 739 480 735 c
        479 729 474 728 468 732 c
        465 735 464 735 459 731 c
        456 729 452 728 450 729 c
        448 730 446 729 443 726 c
        439 721 l
        430 725 l
        425 728 420 732 419 734 c
        418 736 415 740 413 743 c
        409 747 408 748 405 746 c
        403 745 399 744 395 744 c
        392 744 387 742 384 740 c
        377 736 377 736 361 740 c
        358 741 356 740 355 736 c
        353 732 353 731 359 727 c
        364 724 368 723 371 723 c
        373 724 376 724 377 723 c
        377 722 380 721 382 722 c
        385 723 388 722 390 720 c
        393 717 397 716 409 716 c
        428 715 432 714 430 710 c
        430 709 429 705 428 702 c
        427 698 428 697 432 696 c
        439 695 438 690 431 690 c
        427 690 423 690 422 689 c
        421 689 416 691 413 693 c
        h f
        349 703 m
        349 704 348 705 347 705 c
        346 705 345 704 344 703 c
        344 702 345 701 346 701 c
        348 701 349 702 349 703 c
        h f
        338 897 m
        339 899 339 900 335 901 c
        331 903 331 903 331 899 c
        331 894 335 893 338 897 c
        h f
        471 909 m
        468 913 465 913 464 908 c
        463 906 465 905 468 905 c
        472 905 472 906 471 909 c
        h f
        232 712 m
        232 713 233 714 234 714 c
        235 714 236 713 235 712 c
        235 711 234 710 233 710 c
        233 710 232 711 232 712 c
        h f
        292 726 m
        292 727 293 728 294 728 c
        295 728 296 727 297 726 c
        297 725 296 724 294 724 c
        293 724 292 725 292 726 c
        h f
        453 954 m
        452 956 452 957 452 957 c
        453 957 455 956 456 954 c
        457 952 458 950 457 950 c
        456 950 454 952 453 954 c
        h f
        441 957 m
        441 958 442 959 445 959 c
        447 959 449 958 449 957 c
        449 956 448 955 446 955 c
        444 955 442 956 441 957 c
        h f
        357 979 m
        357 980 359 982 361 982 c
        364 982 365 982 364 980 c
        363 976 355 976 357 979 c
        h f
        0.90 0.18 0.04 rg
        473 623 m
        471 627 467 630 457 635 c
        451 637 434 657 437 659 c
        444 662 453 664 468 664 c
        483 663 486 663 488 660 c
        490 658 491 658 492 659 c
        494 661 497 662 499 662 c
        502 662 506 663 508 664 c
        514 669 523 666 522 659 c
        522 658 525 656 529 654 c
        537 650 538 646 534 641 c
        532 639 530 634 529 631 c
        527 620 523 619 505 620 c
        496 621 489 621 487 620 c
        481 617 474 618 473 623 c
        h f
        532 670 m
        529 675 529 680 532 683 c
        533 685 534 685 536 683 c
        538 681 539 679 538 674 c
        538 665 534 663 532 670 c
        h f
        471 677 m
        472 678 473 682 473 685 c
        473 690 473 690 476 688 c
        480 686 487 688 487 690 c
        487 691 486 694 485 695 c
        483 698 483 700 486 703 c
        487 705 489 709 489 711 c
        489 716 503 731 507 731 c
        511 731 511 729 506 725 c
        503 724 501 721 501 719 c
        501 713 510 697 515 691 c
        518 688 521 685 521 684 c
        521 681 515 679 506 679 c
        501 679 499 680 498 683 c
        496 687 490 687 487 682 c
        486 681 483 680 481 680 c
        479 680 476 678 474 676 c
        470 673 470 673 471 677 c
        h f
        415 694 m
        411 697 l
        418 697 l
        427 697 432 693 426 691 c
        420 690 420 690 415 694 c
        h f
        459 697 m
        454 700 454 703 458 707 c
        463 712 472 709 474 703 c
        475 701 476 698 475 697 c
        472 694 463 694 459 697 c
        h f
        376 702 m
        373 705 366 706 364 703 c
        362 699 354 701 352 706 c
        352 708 350 710 348 710 c
        345 710 341 705 343 703 c
        343 702 343 701 341 701 c
        340 701 338 703 337 706 c
        335 714 331 719 327 719 c
        325 719 324 720 324 721 c
        325 724 338 724 341 720 c
        343 718 345 717 346 717 c
        349 717 347 721 343 725 c
        339 729 338 730 340 734 c
        341 737 343 741 345 743 c
        347 745 347 746 346 748 c
        344 750 342 750 338 747 c
        334 745 332 745 328 746 c
        326 747 322 748 320 748 c
        318 748 314 749 312 751 c
        306 755 300 756 293 752 c
        286 749 287 749 286 753 c
        285 756 284 758 280 758 c
        276 759 275 760 275 763 c
        276 768 272 770 270 767 c
        268 766 268 765 270 762 c
        271 760 271 758 270 757 c
        269 757 268 755 268 753 c
        268 746 253 741 249 746 c
        248 748 247 753 246 758 c
        246 766 l
        239 766 l
        234 765 232 766 230 770 c
        227 774 226 774 217 773 c
        207 772 206 772 195 781 c
        189 785 183 789 182 789 c
        177 789 171 798 173 802 c
        175 809 173 830 169 838 c
        166 844 166 846 168 849 c
        170 852 170 858 169 873 c
        167 895 168 898 177 904 c
        184 908 199 908 202 904 c
        206 900 209 901 213 910 c
        217 918 l
        225 917 l
        235 915 239 913 241 909 c
        242 907 245 905 247 905 c
        250 905 254 913 253 918 c
        252 920 253 921 256 922 c
        261 924 264 921 261 917 c
        260 915 260 914 261 911 c
        265 906 269 906 266 911 c
        264 915 268 917 276 915 c
        280 915 283 915 284 916 c
        285 919 287 918 290 912 c
        293 903 299 899 304 902 c
        306 903 308 904 308 905 c
        308 906 310 909 313 912 c
        317 916 319 917 325 916 c
        332 916 333 916 329 913 c
        327 911 326 907 325 903 c
        325 896 325 895 330 893 c
        334 891 336 891 341 894 c
        346 896 348 898 348 902 c
        348 905 347 906 343 907 c
        337 907 337 909 342 913 c
        346 916 346 916 343 921 c
        341 924 340 927 340 928 c
        340 929 341 930 343 931 c
        345 932 347 933 348 933 c
        350 934 351 932 351 928 c
        352 922 356 919 356 925 c
        356 928 357 929 364 929 c
        368 929 372 930 374 931 c
        377 934 378 934 383 929 c
        388 925 389 925 399 926 c
        409 927 410 927 410 923 c
        411 916 415 916 439 923 c
        442 923 442 923 441 914 c
        440 905 l
        448 906 l
        452 907 456 908 458 909 c
        459 911 460 910 462 906 c
        464 900 468 899 474 903 c
        478 905 478 905 483 902 c
        486 901 490 899 493 898 c
        498 897 499 896 499 891 c
        500 887 499 886 495 885 c
        493 884 490 882 490 881 c
        490 879 492 878 494 879 c
        503 882 507 882 507 881 c
        508 873 507 860 506 859 c
        503 857 505 854 510 852 c
        520 847 525 840 528 824 c
        531 809 531 808 528 802 c
        525 797 525 795 526 788 c
        528 780 l
        518 770 l
        513 764 507 757 506 754 c
        502 746 497 742 488 741 c
        482 740 481 739 480 735 c
        479 729 474 728 468 732 c
        465 735 464 735 459 731 c
        456 729 452 728 450 729 c
        448 730 446 729 443 726 c
        439 721 l
        430 725 l
        425 728 420 732 419 734 c
        418 736 415 740 413 743 c
        409 747 408 748 405 746 c
        403 745 399 744 395 744 c
        392 744 387 742 384 740 c
        377 736 377 736 361 740 c
        358 741 356 740 355 736 c
        353 732 353 731 358 728 c
        364 724 367 723 382 722 c
        385 722 389 720 391 719 c
        393 717 396 716 399 716 c
        429 715 432 715 430 710 c
        430 707 427 707 419 707 c
        413 708 406 707 403 706 c
        399 704 396 704 393 706 c
        388 709 388 709 384 704 c
        380 698 379 698 376 702 c
        h f
        483 878 m
        483 882 480 886 477 886 c
        474 886 474 880 476 877 c
        480 874 483 874 483 878 c
        h f
        374 893 m
        374 894 374 895 373 895 c
        373 895 372 894 371 893 c
        371 892 371 891 372 891 c
        373 891 374 892 374 893 c
        h f
        438 896 m
        438 897 438 898 437 898 c
        436 898 435 897 435 896 c
        435 895 436 895 437 895 c
        438 895 438 895 438 896 c
        h f
        363 913 m
        364 915 359 920 356 920 c
        354 920 354 915 356 913 c
        359 910 361 910 363 913 c
        h f
        514 873 m
        514 874 515 875 516 875 c
        518 875 519 874 519 873 c
        519 872 518 871 516 871 c
        515 871 514 872 514 873 c
        h f
        473 911 m
        473 912 474 912 475 912 c
        477 912 478 912 478 911 c
        478 910 477 909 476 909 c
        475 909 474 910 473 911 c
        h f
        0.81 0.17 0.06 rg
        473 623 m
        471 627 467 630 457 635 c
        451 637 434 657 437 659 c
        444 662 453 664 468 664 c
        483 663 486 663 488 660 c
        490 658 491 658 492 659 c
        494 661 496 662 498 662 c
        500 662 505 663 509 664 c
        517 667 519 667 521 661 c
        522 658 525 655 529 654 c
        535 651 539 647 536 644 c
        536 644 533 643 530 643 c
        528 643 526 642 526 641 c
        529 628 528 625 525 622 c
        520 618 510 619 501 623 c
        494 627 491 626 486 622 c
        479 617 475 618 473 623 c
        h f
        533 673 m
        531 681 531 682 535 678 c
        538 676 539 673 538 672 c
        537 668 534 668 533 673 c
        h f
        471 677 m
        472 678 473 682 473 685 c
        473 690 473 690 477 688 c
        486 684 488 680 481 680 c
        479 680 476 678 474 676 c
        470 673 470 673 471 677 c
        h f
        500 680 m
        501 683 505 684 505 682 c
        505 681 503 679 502 679 c
        500 678 499 679 500 680 c
        h f
        492 693 m
        494 696 495 697 497 695 c
        501 692 498 688 493 688 c
        489 688 489 689 492 693 c
        h f
        415 694 m
        411 697 l
        418 697 l
        427 697 432 693 426 691 c
        420 690 420 690 415 694 c
        h f
        459 698 m
        455 701 458 705 465 705 c
        472 705 474 704 474 699 c
        474 696 473 696 468 696 c
        465 696 461 697 459 698 c
        h f
        376 702 m
        373 705 366 706 364 703 c
        362 699 354 701 352 706 c
        352 708 350 710 348 710 c
        345 710 341 705 343 703 c
        343 702 342 701 341 701 c
        339 701 338 702 338 703 c
        338 706 345 712 349 712 c
        352 712 355 713 357 714 c
        363 719 368 718 369 713 c
        370 710 372 708 375 707 c
        378 706 379 705 379 703 c
        379 699 379 698 376 702 c
        h f
        493 703 m
        492 704 493 706 493 708 c
        494 710 493 713 492 714 c
        491 716 491 717 493 719 c
        495 721 496 720 498 717 c
        499 715 500 713 499 713 c
        499 713 499 711 500 708 c
        502 703 496 698 493 703 c
        h f
        393 706 m
        391 709 401 715 404 714 c
        411 712 412 709 406 707 c
        400 704 394 704 393 706 c
        h f
        379 714 m
        379 720 382 722 386 719 c
        388 717 388 716 386 713 c
        384 708 379 709 379 714 c
        h f
        333 715 m
        332 717 330 719 328 719 c
        326 719 325 719 326 720 c
        328 722 340 717 339 715 c
        338 711 335 712 333 715 c
        h f
        346 739 m
        347 742 351 743 351 740 c
        351 738 349 737 348 737 c
        346 737 345 738 346 739 c
        h f
        422 749 m
        419 755 418 757 414 757 c
        409 757 400 753 399 751 c
        397 749 383 745 374 745 c
        363 746 357 748 354 753 c
        351 759 348 759 338 757 c
        335 756 328 756 322 756 c
        317 757 313 758 314 758 c
        317 758 315 767 312 767 c
        311 767 307 765 304 763 c
        296 758 293 757 293 762 c
        293 766 278 774 267 775 c
        261 776 259 776 259 774 c
        259 770 260 770 251 774 c
        244 776 243 776 240 772 c
        236 768 232 769 236 774 c
        239 777 239 777 232 780 c
        227 783 226 783 208 779 c
        200 777 197 780 197 791 c
        197 792 195 795 192 796 c
        188 798 187 798 183 795 c
        179 792 179 792 177 797 c
        176 800 175 808 174 816 c
        174 823 173 832 173 835 c
        172 837 172 840 172 841 c
        172 842 170 843 169 843 c
        166 843 165 847 168 849 c
        170 850 170 856 169 871 c
        168 895 168 898 177 904 c
        184 908 199 908 202 904 c
        206 900 209 901 213 910 c
        217 918 l
        225 917 l
        235 915 239 913 241 909 c
        242 907 245 905 247 905 c
        250 905 254 913 253 918 c
        252 920 253 921 256 922 c
        261 924 264 921 261 917 c
        260 915 260 914 261 911 c
        265 906 269 906 266 911 c
        264 915 268 917 276 915 c
        280 915 283 915 284 916 c
        285 919 287 918 290 912 c
        293 903 299 899 304 902 c
        306 903 308 904 308 905 c
        308 906 310 909 313 912 c
        317 916 319 917 325 916 c
        332 916 333 916 329 913 c
        327 911 326 907 325 903 c
        325 896 325 895 330 893 c
        334 891 336 891 341 894 c
        346 896 348 898 348 902 c
        348 905 347 906 343 907 c
        337 907 337 909 342 913 c
        346 916 346 916 343 921 c
        341 924 340 927 340 928 c
        340 929 341 930 343 931 c
        345 932 347 933 348 933 c
        350 934 351 932 351 928 c
        352 922 356 919 356 925 c
        356 928 357 929 364 929 c
        368 929 372 930 374 931 c
        377 934 378 934 383 929 c
        388 925 389 925 399 926 c
        409 927 410 927 410 923 c
        411 916 415 916 439 923 c
        442 923 442 923 441 914 c
        440 905 l
        448 906 l
        452 907 456 908 458 909 c
        459 911 460 910 462 906 c
        464 900 468 899 474 903 c
        478 905 478 905 483 902 c
        486 901 490 899 493 898 c
        498 897 499 896 499 891 c
        500 887 499 886 495 885 c
        493 884 490 882 490 881 c
        490 879 492 878 494 879 c
        503 882 507 882 507 881 c
        508 870 507 860 505 859 c
        502 858 505 854 510 852 c
        520 847 525 840 528 823 c
        532 808 l
        528 804 l
        524 801 524 801 526 790 c
        528 779 l
        521 773 l
        517 769 510 765 507 764 c
        500 762 500 762 501 756 c
        501 750 501 749 497 748 c
        493 747 491 748 490 750 c
        488 752 486 753 485 753 c
        482 753 471 747 471 745 c
        471 742 457 740 452 743 c
        450 744 445 746 441 747 c
        436 748 434 747 431 744 c
        426 739 425 739 422 749 c
        h f
        483 878 m
        483 882 480 886 477 886 c
        474 886 474 880 476 877 c
        480 874 483 874 483 878 c
        h f
        374 893 m
        374 894 374 895 373 895 c
        373 895 372 894 371 893 c
        371 892 371 891 372 891 c
        373 891 374 892 374 893 c
        h f
        438 896 m
        438 897 438 898 437 898 c
        436 898 435 897 435 896 c
        435 895 436 895 437 895 c
        438 895 438 895 438 896 c
        h f
        363 913 m
        364 915 359 920 356 920 c
        354 920 354 915 356 913 c
        359 910 361 910 363 913 c
        h f
        514 873 m
        514 874 515 875 516 875 c
        518 875 519 874 519 873 c
        519 872 518 871 516 871 c
        515 871 514 872 514 873 c
        h f
        473 911 m
        473 912 474 912 475 912 c
        477 912 478 912 478 911 c
        478 910 477 909 476 909 c
        475 909 474 910 473 911 c
        h f
        0.80 0.09 0.02 rg
        474 627 m
        473 629 470 631 469 631 c
        467 631 465 632 464 634 c
        464 635 462 637 460 637 c
        458 638 456 640 456 642 c
        455 644 453 645 450 645 c
        447 645 445 647 443 651 c
        441 654 440 657 441 657 c
        441 657 444 659 447 660 c
        450 662 454 663 454 663 c
        455 662 462 662 469 661 c
        479 661 483 660 484 659 c
        485 656 491 656 497 658 c
        500 660 502 660 507 657 c
        511 656 515 655 516 656 c
        518 656 521 654 525 651 c
        530 646 530 645 528 644 c
        525 642 525 640 525 636 c
        526 633 526 631 524 631 c
        523 631 523 630 524 628 c
        525 623 520 621 513 624 c
        507 626 486 626 484 624 c
        482 621 477 622 474 627 c
        h f
        462 703 m
        462 704 463 705 464 705 c
        465 705 465 704 464 703 c
        464 702 463 701 463 701 c
        462 701 462 702 462 703 c
        h f
        357 708 m
        358 709 359 712 359 714 c
        360 716 361 717 362 717 c
        363 717 363 716 362 715 c
        362 714 363 713 366 712 c
        371 709 372 707 366 706 c
        359 704 356 704 357 708 c
        h f
        495 709 m
        494 711 494 713 495 713 c
        496 715 499 712 499 709 c
        499 705 496 706 495 709 c
        h f
        452 743 m
        450 744 445 746 441 747 c
        435 748 433 747 432 745 c
        431 742 426 741 424 743 c
        423 744 423 746 425 747 c
        426 750 426 750 425 749 c
        423 748 422 749 420 752 c
        419 755 417 757 414 757 c
        409 757 399 753 399 751 c
        399 750 398 750 396 750 c
        394 751 388 750 382 747 c
        374 744 371 744 367 746 c
        365 747 362 748 360 748 c
        358 748 356 750 355 752 c
        352 759 345 761 337 757 c
        330 754 323 755 325 759 c
        326 762 324 763 321 760 c
        318 757 315 759 315 764 c
        315 768 312 768 305 764 c
        300 760 293 760 293 764 c
        293 766 275 774 267 775 c
        261 776 259 776 259 774 c
        259 772 258 771 256 772 c
        255 772 253 773 252 773 c
        251 773 251 774 252 775 c
        253 779 247 779 244 776 c
        242 775 240 775 236 778 c
        230 782 215 785 215 782 c
        215 779 210 780 205 784 c
        202 786 199 788 198 787 c
        198 787 197 788 197 790 c
        197 791 197 794 199 794 c
        200 795 200 796 200 797 c
        199 798 198 797 197 796 c
        195 794 195 794 192 796 c
        191 797 188 798 186 797 c
        180 795 178 800 177 812 c
        177 828 176 834 172 840 c
        170 843 170 845 171 848 c
        174 852 174 872 172 874 c
        171 874 171 879 171 884 c
        171 893 171 894 176 894 c
        179 895 180 896 180 898 c
        180 900 181 902 183 903 c
        186 906 195 906 198 903 c
        203 899 212 896 215 898 c
        217 899 217 901 217 904 c
        216 907 217 909 220 911 c
        224 916 232 914 241 907 c
        248 901 252 902 255 908 c
        256 911 256 911 260 907 c
        265 903 270 902 271 907 c
        272 911 275 912 277 908 c
        277 906 279 906 282 906 c
        284 907 287 906 291 903 c
        299 896 305 897 310 905 c
        318 917 325 914 322 900 c
        319 891 322 889 336 890 c
        343 891 349 892 350 893 c
        351 894 351 898 352 902 c
        352 910 352 910 358 909 c
        361 909 364 910 365 911 c
        367 914 367 915 363 920 c
        358 926 359 927 367 926 c
        391 923 395 922 402 924 c
        406 925 407 924 408 920 c
        410 916 411 916 417 917 c
        436 920 438 920 438 917 c
        438 916 436 912 433 910 c
        428 905 427 905 428 895 c
        429 888 428 884 426 882 c
        423 879 424 877 428 877 c
        431 877 432 877 431 880 c
        430 883 431 885 437 889 c
        441 891 444 895 444 896 c
        444 901 450 906 455 905 c
        457 905 460 902 463 897 c
        465 893 467 889 468 887 c
        469 886 471 882 472 878 c
        474 872 475 872 482 871 c
        487 870 489 870 492 873 c
        496 877 503 878 503 874 c
        503 873 501 870 500 867 c
        494 861 498 853 510 848 c
        515 846 520 843 522 841 c
        523 839 526 831 527 823 c
        529 811 529 808 527 807 c
        523 804 521 792 523 790 c
        528 784 520 772 506 765 c
        502 763 500 760 500 759 c
        501 755 498 753 490 753 c
        483 753 474 749 469 745 c
        468 743 465 742 462 743 c
        460 744 458 743 458 742 c
        458 740 456 740 452 743 c
        h f
        415 875 m
        418 879 414 882 407 883 c
        401 883 400 880 404 876 c
        407 872 412 872 415 875 c
        h f
        377 890 m
        379 892 378 894 374 898 c
        368 904 l
        367 898 l
        367 895 367 891 368 890 c
        369 887 375 887 377 890 c
        h f
        480 889 m
        474 892 473 896 476 899 c
        478 901 480 900 486 897 c
        497 891 498 891 496 889 c
        493 886 486 886 480 889 c
        h f
        0.68 0.08 0.02 rg
        516 626 m
        512 628 515 631 519 631 c
        522 631 523 630 522 628 c
        522 625 519 624 516 626 c
        h f
        473 633 m
        460 637 456 640 458 643 c
        459 644 459 647 459 649 c
        457 656 460 657 480 657 c
        494 657 501 656 503 654 c
        506 653 509 651 511 650 c
        514 649 515 647 511 644 c
        510 642 508 640 508 639 c
        508 635 500 633 493 634 c
        489 634 487 634 485 632 c
        482 629 483 629 473 633 c
        h f
        448 659 m
        448 662 452 664 455 662 c
        458 661 457 658 452 658 c
        450 657 448 658 448 659 c
        h f
        457 755 m
        457 756 455 757 453 757 c
        451 757 449 758 449 759 c
        449 762 442 763 417 765 c
        406 766 403 767 399 770 c
        396 773 394 774 393 773 c
        393 773 391 773 389 774 c
        386 776 385 775 382 773 c
        381 771 380 768 380 767 c
        381 765 379 766 373 770 c
        364 775 363 775 353 775 c
        347 774 340 774 338 775 c
        334 777 332 777 331 775 c
        330 772 329 772 328 775 c
        326 780 317 783 313 780 c
        310 778 308 778 305 782 c
        303 784 299 785 296 785 c
        290 785 284 788 284 792 c
        284 794 283 796 282 797 c
        280 800 274 800 274 798 c
        274 795 266 789 263 789 c
        262 789 261 791 260 793 c
        260 795 256 799 251 802 c
        242 808 234 816 234 820 c
        234 822 232 823 224 823 c
        216 822 213 821 210 818 c
        208 816 204 814 203 814 c
        199 814 198 819 201 822 c
        205 824 205 824 201 826 c
        199 826 193 829 188 831 c
        179 835 178 836 179 841 c
        180 845 178 855 174 860 c
        173 862 174 864 176 865 c
        186 872 186 875 177 877 c
        173 878 171 880 171 885 c
        169 892 171 896 174 893 c
        177 891 178 892 181 899 c
        184 906 191 907 199 902 c
        202 900 206 898 208 898 c
        209 898 213 895 216 892 c
        219 888 222 886 223 886 c
        227 886 227 888 224 893 c
        219 899 218 905 220 909 c
        222 913 235 912 241 906 c
        248 900 251 900 254 906 c
        256 911 l
        260 907 l
        265 903 270 902 271 907 c
        273 912 274 912 279 905 c
        283 900 283 900 285 902 c
        287 905 288 905 291 901 c
        299 896 305 897 310 905 c
        313 909 315 912 316 912 c
        316 912 318 911 320 909 c
        323 906 323 906 321 901 c
        318 896 317 884 320 884 c
        320 884 322 885 324 886 c
        325 888 331 889 336 890 c
        343 891 345 891 345 889 c
        345 886 349 884 354 884 c
        356 884 356 884 354 888 c
        352 890 351 895 352 901 c
        352 910 l
        358 909 l
        361 909 364 910 365 911 c
        367 914 365 922 362 925 c
        359 927 359 927 363 927 c
        365 927 368 926 371 925 c
        373 924 377 923 380 922 c
        382 922 387 919 390 916 c
        393 913 397 911 397 911 c
        400 911 401 918 399 920 c
        398 921 397 921 399 922 c
        400 923 401 922 401 921 c
        401 920 402 920 404 920 c
        406 921 408 920 409 918 c
        410 916 416 916 429 919 c
        432 920 433 919 433 916 c
        432 914 430 911 428 908 c
        425 905 424 903 425 896 c
        426 890 426 888 423 885 c
        419 882 418 882 413 884 c
        405 886 385 884 384 881 c
        383 879 385 878 388 876 c
        390 875 394 873 395 871 c
        397 868 399 868 405 868 c
        409 868 414 868 417 867 c
        420 866 423 867 427 869 c
        432 871 433 873 432 878 c
        432 884 433 885 438 886 c
        443 888 444 889 444 894 c
        444 904 455 909 461 901 c
        463 897 463 895 462 893 c
        459 888 460 887 466 886 c
        469 886 470 885 470 880 c
        471 875 472 873 476 871 c
        479 869 483 868 484 868 c
        486 868 489 864 492 861 c
        495 855 500 852 508 849 c
        522 842 525 839 525 828 c
        525 823 526 819 527 818 c
        530 815 528 810 519 806 c
        511 802 510 801 512 798 c
        513 796 515 794 516 794 c
        517 793 517 792 515 788 c
        513 786 512 782 513 781 c
        514 776 508 774 498 777 c
        493 778 488 780 487 781 c
        484 784 480 784 479 782 c
        478 780 477 775 477 770 c
        477 760 477 759 471 758 c
        468 757 465 756 464 755 c
        463 752 459 752 457 755 c
        h f
        376 889 m
        380 892 381 900 378 900 c
        377 900 375 902 373 904 c
        371 907 370 908 369 905 c
        368 904 367 900 367 898 c
        367 895 365 891 363 888 c
        359 883 l
        366 885 l
        370 885 374 887 376 889 c
        h f
        237 889 m
        236 890 235 891 234 891 c
        233 891 233 890 234 888 c
        237 885 239 886 237 889 c
        h f
        363 899 m
        363 902 360 904 358 903 c
        357 901 359 896 361 896 c
        362 896 363 897 363 899 c
        h f
        394 903 m
        394 905 388 911 386 911 c
        385 911 383 905 384 903 c
        385 901 394 901 394 903 c
        h f
        493 868 m
        489 871 491 874 497 875 c
        502 876 502 876 500 870 c
        499 865 498 865 493 868 c
        h f
        0.56 0.05 0.02 rg
        466 787 m
        465 791 468 791 427 788 c
        423 788 414 789 406 791 c
        399 793 391 794 388 794 c
        386 794 381 795 378 797 c
        376 798 373 799 372 799 c
        372 798 363 798 353 797 c
        338 796 336 796 333 799 c
        328 804 317 806 313 802 c
        310 799 309 799 303 806 c
        298 813 297 813 293 811 c
        287 807 282 809 281 814 c
        281 818 280 818 267 817 c
        250 815 245 817 245 826 c
        244 832 243 833 238 834 c
        232 836 231 838 234 840 c
        237 842 236 844 232 845 c
        225 847 218 855 220 858 c
        222 860 221 862 220 866 c
        216 876 217 878 225 877 c
        231 876 233 874 236 871 c
        238 866 238 866 234 863 c
        232 861 231 859 232 858 c
        233 857 236 859 239 861 c
        242 863 243 866 243 869 c
        242 874 245 875 253 872 c
        256 871 258 872 261 874 c
        265 877 265 878 263 882 c
        262 884 262 886 263 886 c
        266 888 265 892 262 893 c
        260 894 261 894 265 894 c
        273 895 284 890 287 886 c
        288 884 291 882 293 882 c
        297 882 297 881 296 877 c
        295 873 296 871 298 870 c
        300 869 300 868 299 863 c
        296 858 296 858 303 856 c
        307 855 308 856 311 862 c
        313 868 315 870 320 871 c
        325 873 327 874 328 877 c
        328 879 330 882 333 884 c
        337 886 338 886 341 883 c
        343 882 347 880 350 879 c
        353 879 357 876 359 874 c
        361 871 363 870 370 870 c
        377 870 380 869 384 866 c
        388 862 390 861 402 861 c
        410 861 419 860 423 860 c
        432 858 439 858 448 860 c
        453 861 455 862 457 866 c
        459 869 461 871 462 871 c
        463 871 465 867 466 862 c
        469 855 469 852 467 849 c
        466 845 466 843 467 841 c
        470 836 470 836 478 843 c
        484 848 484 848 481 851 c
        478 854 478 855 481 857 c
        484 860 489 856 489 851 c
        489 850 490 848 492 847 c
        496 846 498 839 496 836 c
        492 832 495 825 500 825 c
        503 825 505 824 506 823 c
        506 822 508 821 510 821 c
        515 821 515 817 511 816 c
        507 815 501 806 501 801 c
        501 797 498 797 495 800 c
        488 807 476 802 476 793 c
        476 790 471 783 468 783 c
        467 783 466 785 466 787 c
        h f
        276 879 m
        276 882 275 883 272 883 c
        269 884 268 883 268 880 c
        268 877 271 875 275 876 c
        276 876 276 878 276 879 c
        h f
        329 791 m
        329 792 330 793 332 793 c
        334 794 334 793 334 792 c
        333 791 329 790 329 791 c
        h f
        503 836 m
        501 839 501 840 503 841 c
        507 843 507 843 511 840 c
        514 838 514 837 512 835 c
        508 833 506 833 503 836 c
        h f
        197 861 m
        197 863 197 864 199 864 c
        200 863 200 862 200 861 c
        200 860 200 859 199 859 c
        197 859 197 860 197 861 c
        h f
        197 871 m
        196 872 197 874 198 874 c
        202 876 200 878 197 877 c
        192 876 190 881 193 885 c
        196 888 205 887 210 882 c
        212 881 214 879 214 879 c
        215 879 215 871 214 870 c
        214 870 210 869 206 869 c
        200 868 198 869 197 871 c
        h f
        235 901 m
        232 902 232 905 235 905 c
        238 905 240 902 238 901 c
        238 901 236 901 235 901 c
        h f
        0.47 0.05 0.02 rg
        402 796 m
        394 797 386 799 383 801 c
        381 802 375 804 370 804 c
        365 804 360 805 358 807 c
        356 809 352 811 348 812 c
        345 813 341 816 340 818 c
        338 820 337 821 333 819 c
        328 817 316 819 310 822 c
        308 824 304 825 301 825 c
        299 825 295 826 293 829 c
        290 833 288 833 283 831 c
        279 830 276 830 273 832 c
        270 833 265 834 261 834 c
        253 834 247 838 250 841 c
        251 842 252 844 252 846 c
        252 847 254 850 256 853 c
        258 857 261 858 266 858 c
        271 859 272 858 276 852 c
        279 848 281 844 281 843 c
        281 842 283 840 285 838 c
        288 836 290 836 293 837 c
        295 838 300 839 305 840 c
        314 841 315 841 316 837 c
        317 835 318 834 320 834 c
        328 836 328 836 329 832 c
        330 827 337 822 341 824 c
        343 825 349 827 354 827 c
        362 828 362 828 366 822 c
        369 819 370 815 370 814 c
        370 813 372 811 375 809 c
        378 807 380 807 383 808 c
        384 809 390 811 395 811 c
        403 812 404 812 405 809 c
        406 806 408 805 410 805 c
        411 806 414 807 415 807 c
        416 807 418 806 419 802 c
        420 797 420 794 417 794 c
        417 795 410 795 402 796 c
        h f
        442 798 m
        442 799 444 800 447 800 c
        449 800 451 799 451 798 c
        451 797 449 796 447 796 c
        444 796 442 797 442 798 c
        h f
        455 809 m
        453 812 452 812 451 810 c
        450 809 450 810 450 813 c
        451 816 451 817 448 817 c
        447 817 446 818 446 820 c
        447 822 460 826 471 827 c
        478 827 478 827 475 814 c
        474 810 473 809 469 809 c
        467 808 463 808 461 807 c
        459 807 456 808 455 809 c
        h f
        409 830 m
        405 836 398 838 382 837 c
        370 836 366 837 365 838 c
        363 840 362 841 361 839 c
        360 837 360 838 361 841 c
        362 845 361 846 359 846 c
        356 846 355 850 358 851 c
        360 851 365 852 370 854 c
        377 856 379 856 383 854 c
        385 852 387 852 390 853 c
        392 854 395 855 398 855 c
        400 855 403 856 404 857 c
        405 860 413 859 416 855 c
        418 853 421 851 424 850 c
        432 847 430 844 420 843 c
        415 843 411 843 410 842 c
        408 840 414 835 418 835 c
        421 835 422 835 422 832 c
        421 828 411 827 409 830 c
        h f
        454 846 m
        454 847 456 848 457 848 c
        457 848 458 847 458 846 c
        458 845 457 844 455 844 c
        454 844 453 845 454 846 c
        h f
        318 858 m
        315 861 317 862 325 862 c
        329 862 333 861 333 860 c
        333 860 333 859 332 859 c
        322 857 319 857 318 858 c
        h f
        346 870 m
        344 872 344 873 346 875 c
        350 878 354 876 354 871 c
        354 867 349 866 346 870 c
        h f
        Q
    }}
\usepackage{transparent}
\newcount\picnum
\newcount\tmpnum
\def\coffeehbox#1{\hbox to0pt{\kern\the\tmpnum mm #1 \hss}}
\def\coffeevbox#1{\vbox to0pt{\kern\the\tmpnum mm #1 \vss}}
\def\coffeerotbox#1{\rotatebox[origin=c]{\the\tmpnum}{#1}}
\makeatletter
\def\coffeepic#1{
    \ifcase#1\transparent{\strip@pt\dimexpr \the\tmpnum pt / 300}\coffeeA%
    \or \transparent{\strip@pt\dimexpr \the\tmpnum pt / 300}\coffeeB%
    \or \transparent{.\the\tmpnum}\coffeeC%
    \or \transparent{\strip@pt\dimexpr \the\tmpnum pt / 600}\coffeeD \fi
}
\makeatother
\newcommand\randomcoffee{
    \setrannum\picnum{0}{3}
    \setrannum\tmpnum{99}{99}\edef\coffeescale{.\the\tmpnum}
    \setrannum\tmpnum{-50}{-0}\coffeehbox{ 
        \setrannum\tmpnum{-10}{0}\coffeevbox{
            \setrannum\tmpnum{-90}{90}\coffeerotbox{ 
                \setrannum\tmpnum{10}{60}\coffeepic\picnum}
        }%
    }%
}

\usepackage{soul}
\usepackage{xcolor}
\colorlet{lightyellow}{yellow!20}
\sethlcolor{lightyellow}
\soulregister\cite7
\soulregister\ref7
\soulregister\eqref7
\soulregister\sout7
\soulregister\ac7
\soulregister\acp7
\soulregister\emph7
\soulregister\textit7
\soulregister\textbf7

\usepackage[inline, nomargin, marginclue, multiuser, draft]{fixme}
\fxsetup{theme=color}
\usepackage{mdframed}
\mdfdefinestyle{hl}{innerleftmargin=0, innerrightmargin=0, innertopmargin=0, innerbottommargin=0, backgroundcolor=lightyellow, linewidth=0}
\makeatletter
\renewcommand*\FXLayoutInline[3]{%
    \@fxdocolon{#3}{%
        \@fxuseface{inline}%
        \begingroup
        \color{fx#1}\ignorespaces\hl{#3\mbox{\@fxcolon}#2}%
        \ifcoffee \randomcoffee \fi%
        \endgroup}}
\renewcommand*\FXEnvLayoutColorBegin[2]{%
    \@fxdocolon{#2}%
    \@fxuseface{env}%
    \begin{mdframed}[style=hl, fontcolor=fx#1]
        \ignorespaces#2\@fxcolon\ignorespaces}
        \renewcommand*\FXEnvLayoutColorEnd[2]{\end{mdframed}}
\makeatother

\usepackage{layouts}

\usepackage[normalem]{ulem}

\usepackage{xspace}

\usepackage{kantlipsum}

\usepackage{cite}

\usepackage{flushend}

\usepackage{tikz}
\usetikzlibrary{patterns}
\usepackage[precision=2, unit=mm]{lengthconvert}

\FXRegisterAuthor{rp}{rp3020}{R.P.}
\FXRegisterAuthor{ar}{ajr2227}{A.R.}

\title{ Highly Uniform \color{black} Thermally Undercut Silicon Photonic Devices in a 300 mm CMOS Foundry Process}

\author[1]{Robert Parsons}
\author[1]{Kaylx Jang}
\author[1]{Yuyang Wang}
\author[1]{Asher Novick}
\author[2]{A. Matthew Smith}
\author[2]{Christopher C. Tison}
\author[3]{Yonas Gebregiorgis}
\author[3]{Venkatesh Deenadayalan}
\author[3]{Matthew van Niekerk}
\author[4]{Lewis Carpenter}
\author[4]{Tat Ngai}
\author[4]{Gerald Leake}
\author[4]{Daniel Coleman}
\author[1]{Xiang Meng}
\author[3]{Stefan Preble}
\author[2]{Michael L. Fanto}
\author[1]{Keren Bergman}
\author[2,5,*]{Anthony Rizzo}
\affil[1]{Department of Electrical Engineering, Columbia University, New York, NY 10027, USA}
\affil[2]{Air Force Research Laboratory Information Directorate, Rome, NY 13441, USA}
\affil[3]{Electrical and Microelectronic Engineering, Rochester Institute of Technology, Rochester, NY 14623, USA}
\affil[4]{American Institute for Manufacturing Integrated Photonics (AIM Photonics), Albany, NY 12203, USA}
\affil[5]{Thayer School of Engineering, Dartmouth College, Hanover, NH 03755, USA}

\affil[*]{anthony.j.rizzo@dartmouth.edu}

\begin{abstract}
Silicon photonic devices fundamental to high-density wavelength-division multiplexed (DWDM) optical links and photonic switching networks, such as resonant modulators and Mach-Zehnder interferometers (MZIs), are highly sensitive to fabrication variations and operational temperature swings. However, thermal tuning to compensate for fabrication and operational temperature variations can result in prohibitive power consumption, challenging the scalability of energy-efficient photonic integrated circuits (PICs). In this work, we develop and demonstrate a wafer-scale thermal undercut process in a 300 mm complementary metal oxide semiconductor (CMOS) foundry that dramatically improves the thermal isolation of thermo-optic devices by selectively removing substrate material beneath the waveguides and resonators. This approach significantly reduces the power required for thermal tuning across multiple device architectures, achieving almost a 5$\times$ improvement in tuning efficiency in a state-of-the-art 4.5 \si{\micro\meter} radius microdisk modulator and a 40$\times$ improvement in efficiency for a MZI phase shifter. To the best of the authors' knowledge, we demonstrate the first wafer-scale   comparison of non-undercut and undercut silicon photonic devices \color{black} using comprehensive wafer-scale measurements across 64 reticles of a 300 mm silicon-on-insulator (SOI) wafer.   Further, we demonstrate a comprehensive wafer-scale analysis of the influence of undercut trench opening geometry on device tuning efficiency. Notably, we observe highly uniform performance across the full 300 mm wafer for multiple device types, emphasizing that our process can be scaled to large-scale photonic circuits with high yield. These results open new opportunities for large-scale integrated photonic circuits using thermo-optic devices, paving the way for scalable, low-power silicon photonic systems.
\end{abstract}

\begin{document}

\flushbottom
\maketitle

\thispagestyle{empty}

\section*{Introduction}
Silicon photonic devices, such as resonant modulators, resonant filters, and Mach-Zehnder interferometers (MZIs), are foundational elements in dense wavelength-division multiplexed (DWDM) optical links and photonic switching systems due to their compact size, high-speed operation, and low energy consumption \cite{bogaertsSiliconMicroringResonators2012, chenCombLaserdrivenDWDM2015, novick2024ultra, parsons2023efficient, 2ndOrderRing, luoHighBandwidthOnchip2010, horstCascadedMachZehnderWavelength2013, osgood2003fast, lee2018silicon, huang2020push}. These devices leverage silicon’s strong thermo-optic effect for efficient tuning and configurability, making them highly attractive for large-scale photonic integrated circuits (PICs) in data centers and high-performance computing (HPC) environments \cite{rizzo2023massively}. Static fabrication variations and dynamic temperature variations can significantly shift the optical properties of these devices, necessitating the use of integrated micro-heaters to stabilize and tune their performance \cite{rizzo2023fabrication}. As silicon photonics scales towards dense integration with electronic components in co-packaged optical interconnects, the demand for precise thermal tuning increases. While some resonant modulators with high electro-optic tuning abilities \cite{timurdogan2014ultralow, gevorgyan2022miniature} have been proposed to reduce the reliance on thermal tuning, their tuning range is below what is required for realistic temperature swings in co-packaged interconnects. Trimming the refractive index of devices post-fabrication has also been proposed \cite{jayatilleka2021post}, which compensates for fabrication variations, but does not solve the challenge of large temperature swings within the package. Therefore, thermal tuning of these devices is still required. However, conventional thermal tuning approaches, which require constant heating to counter localized temperature swings, impose a high power consumption burden. For instance, micro-heaters in resonant modulators can consume up to 25 mW P\textsubscript{$\pi$}, which is unsustainable for energy-efficient DWDM links and switching networks, where minimizing energy per bit is a critical requirement \cite{masood2013comparison, jacques2019optimization, liu2022thermo}. Reducing the energy consumption of these thermal tuning elements is essential for enabling scalable, low-power photonic systems \cite{zortman2010silicon, wang2024co}.

In this work, we address the challenge of thermal management and energy reduction by introducing a selective thermal substrate undercut technique across a range of key silicon photonic devices, including microdisk modulators, microring \& racetrack modulators, and linear thermo-optic phase shifters in MZIs. These devices, widely used for modulation, wavelength multiplexing, and switching, all rely on thermal tuning for stable operation \cite{padmaraju2014resolving}. By selectively removing the substrate material beneath the waveguides and resonators, the thermal undercut creates a region of enhanced thermal isolation, which significantly reduces the power required for thermo-optic tuning. This approach improves thermal efficiency while maintaining the performance and compactness of the devices. Previous demonstrations of thermal undercut show ample improvements in thermo-optic tuning efficiency, however these demonstrations have the disadvantages of requiring backside etching \cite{cunningham2010highly}, low-volume electron-beam lithography \cite{dong2010thermally, sun2010submilliwatt}, or not achieving full release from the substrate \cite{fang2011ultralow}. Recent foundry-supported thermal undercut demonstrations yield moderate improvement factors, limited by the sealing of the undercut \cite{giewont2019300, fang2024comparison}. Substrate undercut has also been applied to other contexts, including enabling integrated photonic MEMS \cite{jo2022wafer}.

Resonant modulators, which are crucial for high-speed modulation in DWDM systems due to their inherent wavelength selectivity, benefit from reduced tuning power with the thermal undercut technique. We apply the wafer-scale thermal undercut to microring modulators, racetrack-style ring modulators, and state-of-the-art vertical junction microdisk modulators. Our experimental results demonstrate significant gains in tuning efficiency, achieving almost a 5$\times$ improvement for the microdisk modulators, while maintaining CMOS-compatible drive voltages. Linear thermo-optic phase shifters in MZIs, crucial for interleaving, switching, and routing in photonic networks, exhibit even greater reductions in power consumption, with over 40$\times$ improvement in tuning efficiency with optimized trench geometries.   We show large-scale uniformity in performance for each device type, exhibiting the robustness of the substrate removal process across full 300 mm wafers. \color{black} Furthermore, we demonstrate the capability of arbitrarily defined undercut openings, providing room for future improvement in all designs with optimized opening geometry.


  To the best of the authors' knowledge we present, for the first time, comprehensive wafer-scale comparisons of non-undercut and undercut silicon photonic devices across multiple device types and heater geometries. Additionally, we illustrate tuning efficiency trends of undercut microdisk modulators based on systematically varied undercut geometric parameters. \color{black} By sweeping the dimensions and shapes of the undercut openings beneath these microdisk modulators, we observe that wider and more elongated trenches yield significantly improved thermal isolation and reduced power consumption. These measurements reveal the critical role of trench geometry in optimizing thermal performance, providing key insights for design strategies that maximize tuning efficiency across photonic systems. The reduction in power consumption achieved through this optimized undercut geometry has broad implications for DWDM links and other high-density photonic systems. In DWDM links, where minimizing power per channel is essential for scalability \cite{wang2024silicon}, the improved thermal efficiency allows for higher channel counts and lower energy-per-bit metrics, critical for data center and telecom applications. Similarly, in efficient large-scale photonic switching systems \cite{tu2019state}, the power savings gained through thermal undercut can alleviate the thermal load, enabling increased switching density with minimized power consumption.

\section*{Results}

\begin{figure}[ht]
    \centering
    \includegraphics[scale=0.62]{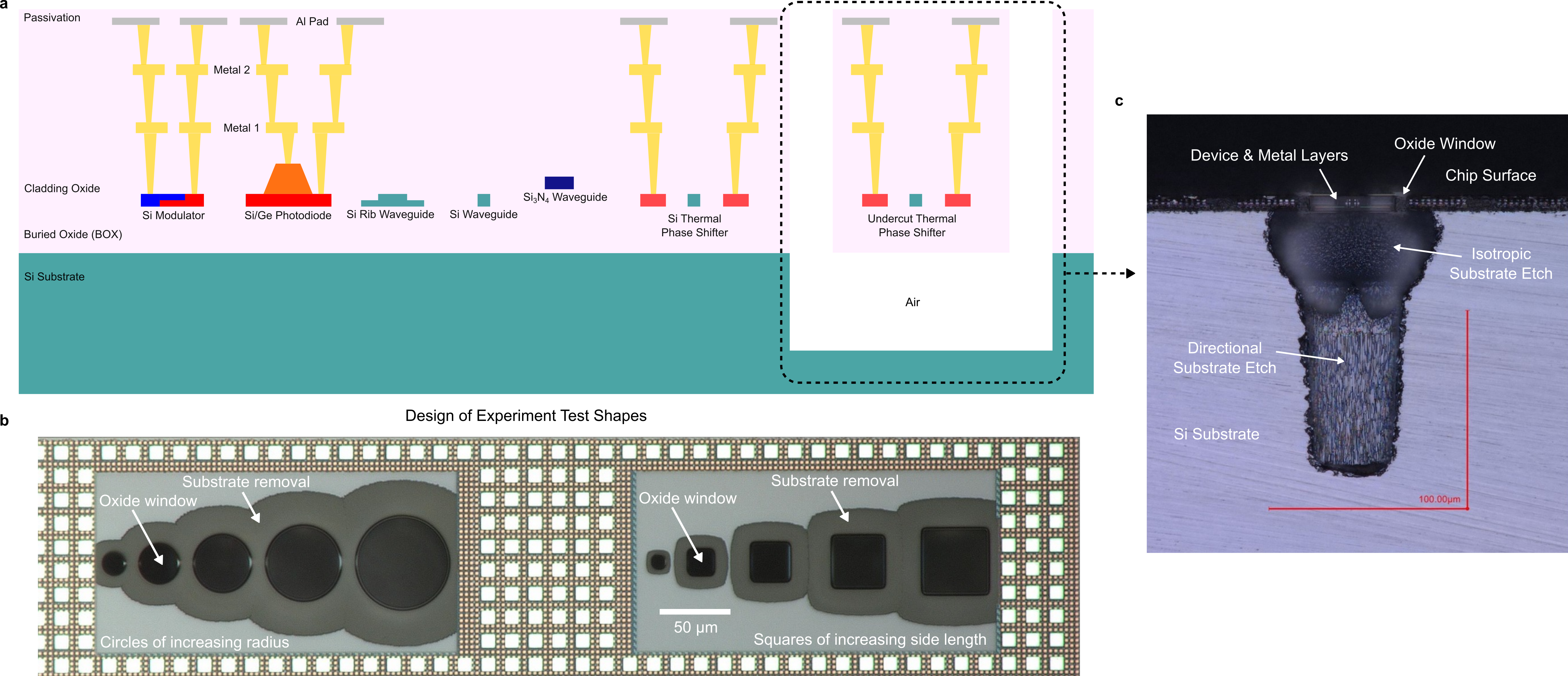}
    \caption{ \textbf{Fabrication process cross section and inspection of undercut design-of-experiment (DoE) structures.}
    \textbf{a,} Complete process cross-section of the AIM Photonics Actives Process including silicon and silicon nitride waveguides, free carrier-based high-speed modulators, germanium photodiodes, and doped-silicon thermal phase shifters. This work extends the process to include undercut thermal phase shifters for higher energy efficiency (far right).
    \textbf{b,} Optical microscope image of design-of-experiment test structures to evaluate the lateral extent of the isotropic substrate removal etch. These shapes include circles of increasing radius (left) and squares of increasing side length (right). The dark black regions indicate the extent of the oxide window etch through the cladding and buried oxide (BOX), and the lighter gray regions indicate the lateral extent of the silicon substrate removal beyond the extent of the oxide window. These DoE structures additionally demonstrate the ability of our undercut fabrication process to extend to arbitrary trench shapes.
    \textbf{c,} Optical microscope image of the chip cross-section after mechanically dicing through the center of an undercut device. The key features of the undercut fabrication are visible in the micrograph including the cladding, device, and metal layers (top), the oxide windows exposing access to the substrate from the chip surface, the first isotropic substrate removal etch to form the undercut, and a final directional substrate etch that arises from the Bosch etch used to define deep trench facets for edge coupling. }
    \label{fig:cross_section}
\end{figure}

We design and demonstrate robust wafer-scale thermal undercut in collaboration with a 300 mm CMOS foundry \cite{AIMPDK}, ensuring that the process can extend to high-volume manufacturing.   This thermal undercut process was designed to maintain full compatibility with the standard AIM Photonics Actives Process, as highlighted by the process cross section in Fig.~\ref{fig:cross_section}a. \color{black} As shown in Fig.~\ref{fig:cross_section}b, design of experiment test structures with both circular and square trench geometries validate the ability to process a range of arbitrary shapes in a standard CMOS fabrication process. This flexibility is essential for optimizing thermal isolation across different device architectures and confirms the robustness of the undercut technique for varied applications. Comprehensive design rules ensure devices retain structural integrity when surrounded by undercut trenches. The thermal undercut process employed in this work involves selectively removing silicon substrate material \cite{van2022wafer}. This creates an air-filled trench that significantly enhances thermal isolation by lowering the effective thermal conductivity surrounding the active photonic device. Silicon, with a thermal conductivity of approximately 150 W/m·K\cite{shanks1963thermal, slack1964thermal}, provides a highly conductive pathway for heat dissipation, causing significant thermal leakage from the micro-heater to the bulk substrate. Further, the buried oxide has a thermal conductivity of approximately 1.4 W/m·K\cite{cahill1990thermal, grove1967physics}, representing reduced thermal confinement. By introducing an undercut trench filled with air, which has a thermal conductivity of roughly 0.025 W/m·K\cite{stephan1985thermal}, we reduce both heat dissipation pathways, effectively creating an insulating barrier around the device. This drastic reduction in thermal conductivity between the heater and the substrate allows the device to reach higher temperatures with lower input electrical power, significantly improving thermal tuning efficiency.   Since the foundry process does not include a metal heater layer such as titanium nitride, we implement our resistive heaters using doped silicon in the silicon device layer (Fig. \ref{fig:cross_section}a). We use a suite of standard n-type and p-type ion-implanted doping layers \cite{AIMPDK} to modulate the resistance of the silicon microheater and form ohmic contacts between the metal vias and silicon. Depending on the heater geometry (length, width, etc.) and the target resistance for the heater, we employ a mix of low, medium, and high doping levels ($\approx 1e18$ cm\textsuperscript{-3}, $1e19$ cm\textsuperscript{-3}, and $1e20$ cm\textsuperscript{-3}, respectively). The metal vias and routing layers are fabricated using standard copper and aluminum back end of line (BEOL) processing \cite{AIMPDK}. \color{black}

\begin{figure}[ht]
    \centering
    \includegraphics[scale=0.99]{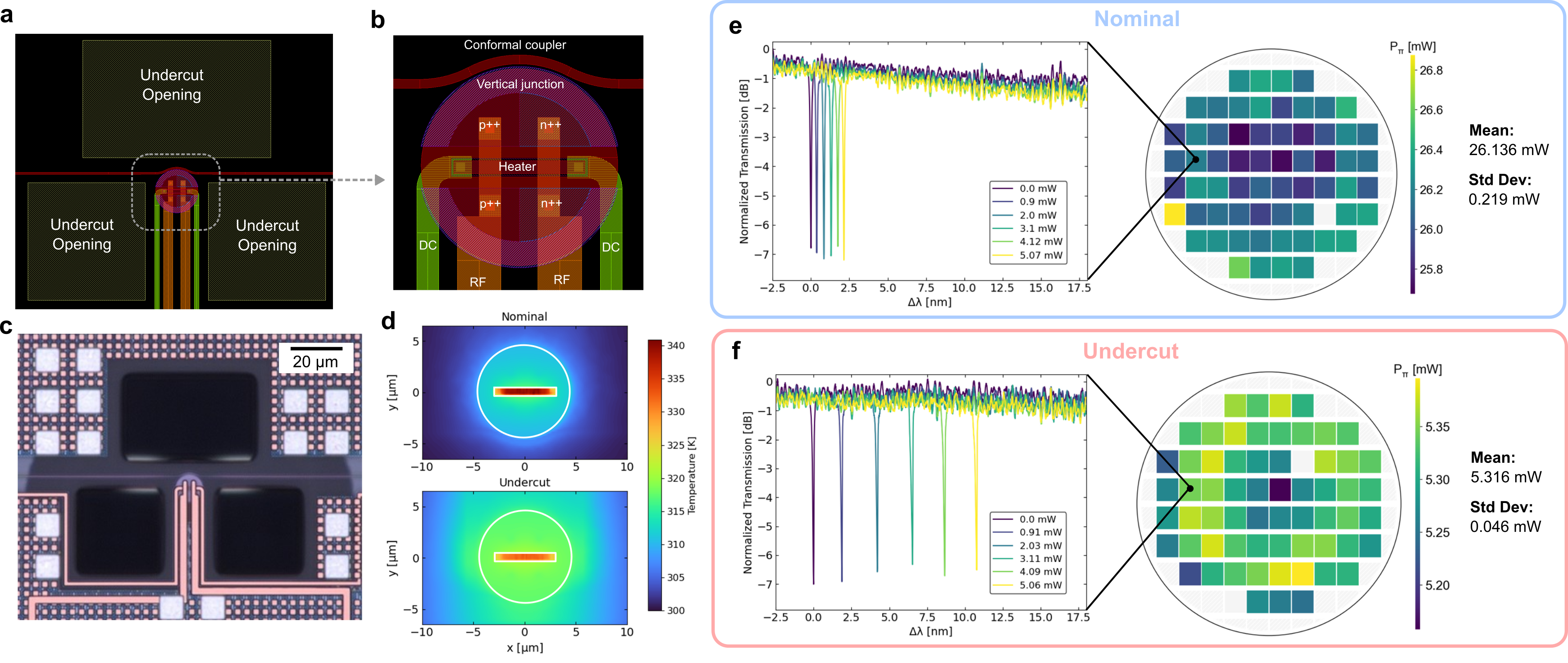}
    \caption{ \textbf{Undercut microdisk modulator wafer-scale thermo-optic measurements.}
    \textbf{a,} Mask layout of the undercut microdisk modulator highlighting the layers for the directional oxide etch (`undercut opening'). \textbf{b,} Mask layout of the microdisk modulator highlighting the device geometry including the vertical junction doping around the disk perimeter, conformal phase-matched coupler for selective excitation of the fundamental mode, heavily doped contact regions with vias, and DC/RF metal traces. \textbf{c,} Micrograph of the fabricated device corresponding to the mask layout in \textbf{a}. \textbf{d,} Simulated temperature profile of the microdisk with (bottom) and without (top) undercut given 1 mW of power dissipated from the heater, showing a dramatic increase in the temperature throughout the disk for the same amount of applied power. \textbf{e,} Wafer map and representative thermal tuning spectrum for a nominal non-undercut device, showing the inter-reticle statistics across the full 300 mm wafer. \textbf{f,} Wafer map and representative thermal tuning spectrum for the undercut device, showing a dramatic improvement in $P_{\pi}$ while maintaining consistent performance across the entire wafer.}
    \label{fig:disk_mod}
\end{figure}

Thermal simulations \cite{coenen2022thermal} were performed for a microdisk modulator both with and without undercut to compare their transient thermal responses \cite{lumerical}. The mask layout for a microdisk modulator surrounded by thermal undercut is shown in Fig.~\ref{fig:disk_mod}a. The comprehensive device design is detailed in ref. \cite{rizzo2022petabit}, and preliminary results of the undercut device are shown in ref. \cite{rizzo2023ultra}.   Although outside the scope of this work, recent improvements to the microdisk modulator RF contact geometry have yielded devices capable of 32 Gb/s on-off-keying non-return-to-zero (OOK-NRZ) transmission with low peak-to-peak driving voltages \cite{novick2024ultra}. \color{black} Simulations in which 1 mW of power was applied through the integrated micro-heater show a significantly higher temperature increase in the undercut microdisk compared to the nominal device, underscoring the enhanced thermal isolation achieved by the trench (Fig.~\ref{fig:disk_mod}d). This heightened temperature response for the same input power demonstrates the potential for lower power consumption, as less energy is required to reach the desired tuning temperatures in devices with thermal undercut. Additionally, transient thermal simulations were performed to provide insight into the dynamic response of the undercut device, revealing longer rise and fall times compared to the nominal device. Specifically, the nominal device had a rise time of 11.0 $\mu$s and a fall time of 11.3 $\mu$s, while introducing the undercut resulted in a rise time of 96.7 $\mu$s and a fall time of 95.7 $\mu$s. The approximately 8-9$\times$ slower heating and cooling transients are a result of the reduced thermal conductivity around the microdisk due to the air gap created by the undercut. This effect can be advantageous in applications where slower thermal responses improve stability by reducing sensitivity to temperature fluctuations.

Microdisk modulator test structures were measured across two 300 mm wafers. Two variants of the device were measured on each reticle: modulators both with and without thermal undercut. A micrograph of a microdisk modulator with thermal undercut is shown in Fig.~\ref{fig:disk_mod}c, which confirms that the device is fully released from the substrate. Further demonstrating the full release of the device, a micrograph of the side profile is shown in Fig.~\ref{fig:cross_section}c. A dicing saw was used to make a precise cut close to the undercut modulator on a singulated die to enable capture of the cross-section. The optical spectrum of a nominal microdisk modulator without thermal undercut at different applied thermo-optic phase shifter powers showing the shift in resonance is displayed in Fig.~\ref{fig:disk_mod}e. A drastic increase in the resonance shift for similar applied powers is shown in Fig.~\ref{fig:disk_mod}f, where the optical spectrum of a representative thermally undercut microdisk modulator is displayed. This large increase in tuning efficiency is additionally observed at the wafer-scale; the wafer maps are shown in Figs.~\ref{fig:disk_mod}e and \ref{fig:disk_mod}f, showing the \text{$P_{\pi}$} of the nominal non-undercut and undercut microdisks, respectively.   The \text{$P_{\pi}$} for the undercut microdisks is on average 5.316 mW with standard deviation 0.046 mW, which is an improvement of 4.92 as compared to the \text{$P_{\pi}$} of 26.136 mW with standard deviation 0.219 mW exhibited by the devices without undercut. Furthermore, these wafer-scale measurements confirm the structural integrity and high yield of the undercut devices. \color{black}

\begin{figure}[ht]
    \centering
    \includegraphics[scale=0.98]{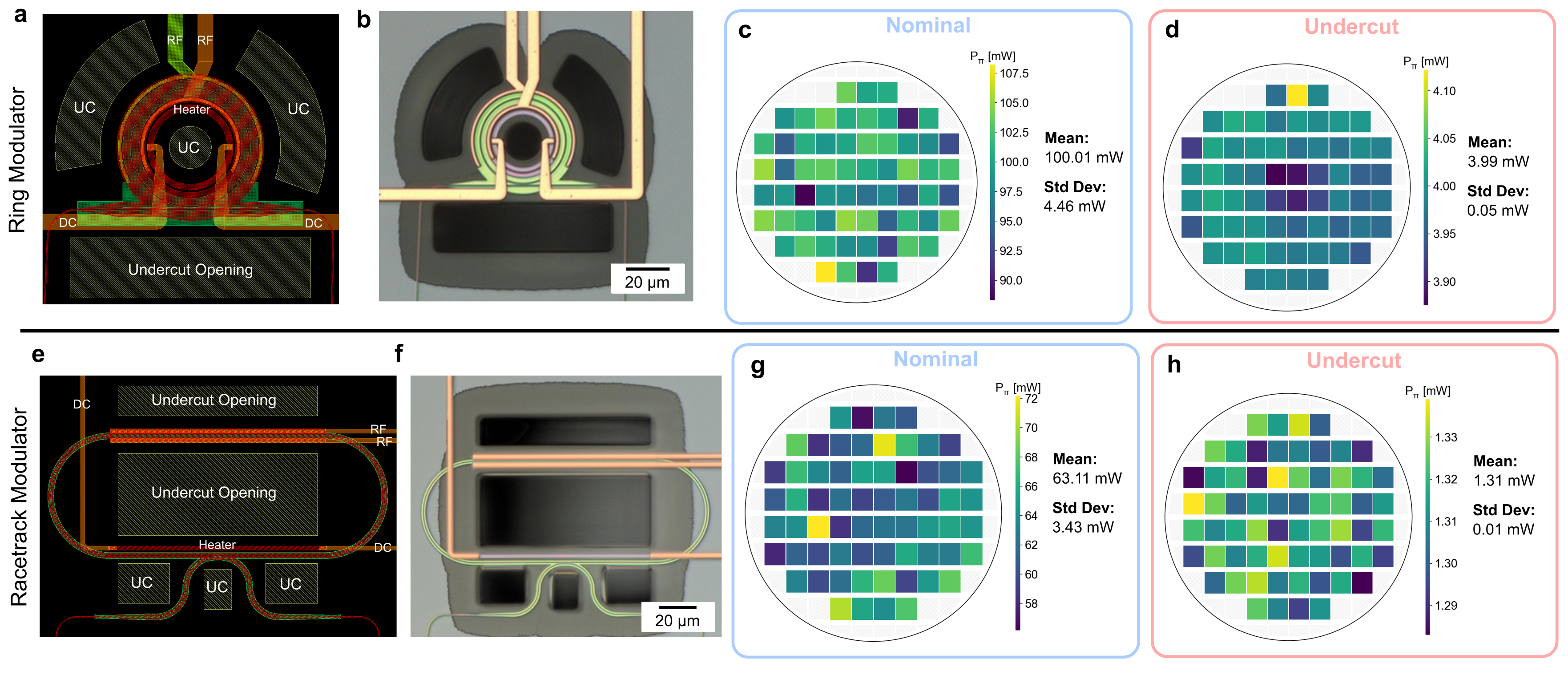}
    \caption{ \textbf{Undercut microring and racetrack modulator wafer-scale thermo-optic measurements.}
    \textbf{a,} Mask layout of the microring modulator highlighting the location of the undercut trench openings, interior doped silicon heater, high speed RF traces, and DC heater traces.
    \textbf{b,} Corresponding micrograph of undercut microring modulator.
    \textbf{c,} Wafer map of thermal tuning power required for a $\pi$ phase shift ($P_{\pi}$) of the nominal microring modulator.
    \textbf{d,} Wafer map of $P_{\pi}$ for the undercut microring modulator, showing a dramatic decrease relative to the nominal design.
    \textbf{e,} Mask layout of the racetrack modulator highlighting the location of the undercut trench openings, interior doped silicon heater, high speed RF traces, and DC heater traces.
    \textbf{f,} Corresponding micrograph of undercut racetrack modulator. 
    \textbf{g,} Wafer map of $P_{\pi}$ for the nominal racetrack modulator.
    \textbf{h,} Wafer map of $P_{\pi}$ for the undercut racetrack modulator, showing a dramatic decrease relative to the nominal design.}
    \label{fig:rings}
\end{figure}

To demonstrate the wide applicability of the undercut process, two other types of resonant modulators were fabricated, both with and without thermal undercut. The first device is a lateral junction microring modulator with a radius of 20 \si{\micro\meter}. The micrograph of this microring modulator with thermal undercut is shown in Fig.~\ref{fig:rings}b. The thermo-optic phase shifter is located concentrically inside the ring. Trench openings were placed both inside and outside the microring and alongside the access waveguide, as shown in the schematic in Fig.~\ref{fig:rings}a. The second device is a racetrack-style ring modulator; a micrograph of a representative device is shown in Fig.~\ref{fig:rings}f. A linear thermo-optic phase shifter is located internal to the ring along the straight access section. The PN-junction for high-speed modulation is located along the opposite straight section. For this modulator design, trench openings were placed both internally and externally along the straight sections of the ring and alongside the access waveguide, as shown in the schematic in Fig.~\ref{fig:rings}e. The \text{$P_{\pi}$} measured across a wafer is shown for the non-undercut microring modulator and racetrack ring modulator in Fig.~\ref{fig:rings}c and Fig.~\ref{fig:rings}g, respectively. The \text{$P_{\pi}$} wafer maps of the undercut microring and racetrack modulators are displayed in Fig.~\ref{fig:rings}d and Fig.~\ref{fig:rings}h, respectively.   We observe an average tuning efficiency improvement of 25.1, from average \text{$P_{\pi}$} of 100.01 mW and standard deviation 4.46 mW without undercut to average \text{$P_{\pi}$} of 3.99 mW and standard deviation 0.05 mW with undercut for the microring modulator. In the case of the racetrack-style ring modulator, the tuning efficiency improvement on average is 48.1, from average \text{$P_{\pi}$} of 63.11 mW and standard deviation 3.43 mW without undercut to average \text{$P_{\pi}$} of 1.31 mW and standard deviation 0.01 mW with undercut. \color{black} Both undercut devices exhibit drastic increases in tuning efficiency, indicating full release from the substrate.

While all the fully-released undercut resonant devices show substantial improvements in tuning efficiency, the improvement factor is more pronounced for less efficient thermo-optic phase shifter designs. The baseline tuning efficiencies of the microring and racetrack modulators without undercut have \text{$P_{\pi}$} values of 100.01 mW and 63.11 mW, respectively. In comparison, the already relatively efficient microdisk modulator exhibits a \text{$P_{\pi}$} of 26.136 mW without undercut. The higher baseline efficiency of the microdisk can be attributed to its integrated thermo-optic phase shifter, which is surrounded by only a thin 100 nm oxide layer for electrical isolation \cite{murthy2022mitigation}. The proximity of the phase shifter to the optical mode, with silicon as the dominant material between them, ensures effective thermal conduction due to silicon's higher thermal conductivity compared to oxide. In contrast, the thermo-optic phase shifters in the microring and racetrack modulators are positioned farther from the optical mode, separated by a larger oxide gap. This separation significantly reduces thermal efficiency and confinement in these devices compared to the microdisk. As a result, the microring and racetrack modulators benefit more, relatively speaking, from the enhanced thermal isolation provided by the substrate undercut, achieving greater improvement factors in tuning efficiency.   Consequently, more conservative phase shifter designs, such as those designed to incur negligible optical loss, can still achieve state-of-the-art tuning efficiency once undercut. \color{black}

\begin{figure}[ht]
    \centering
    \includegraphics[scale=0.99]{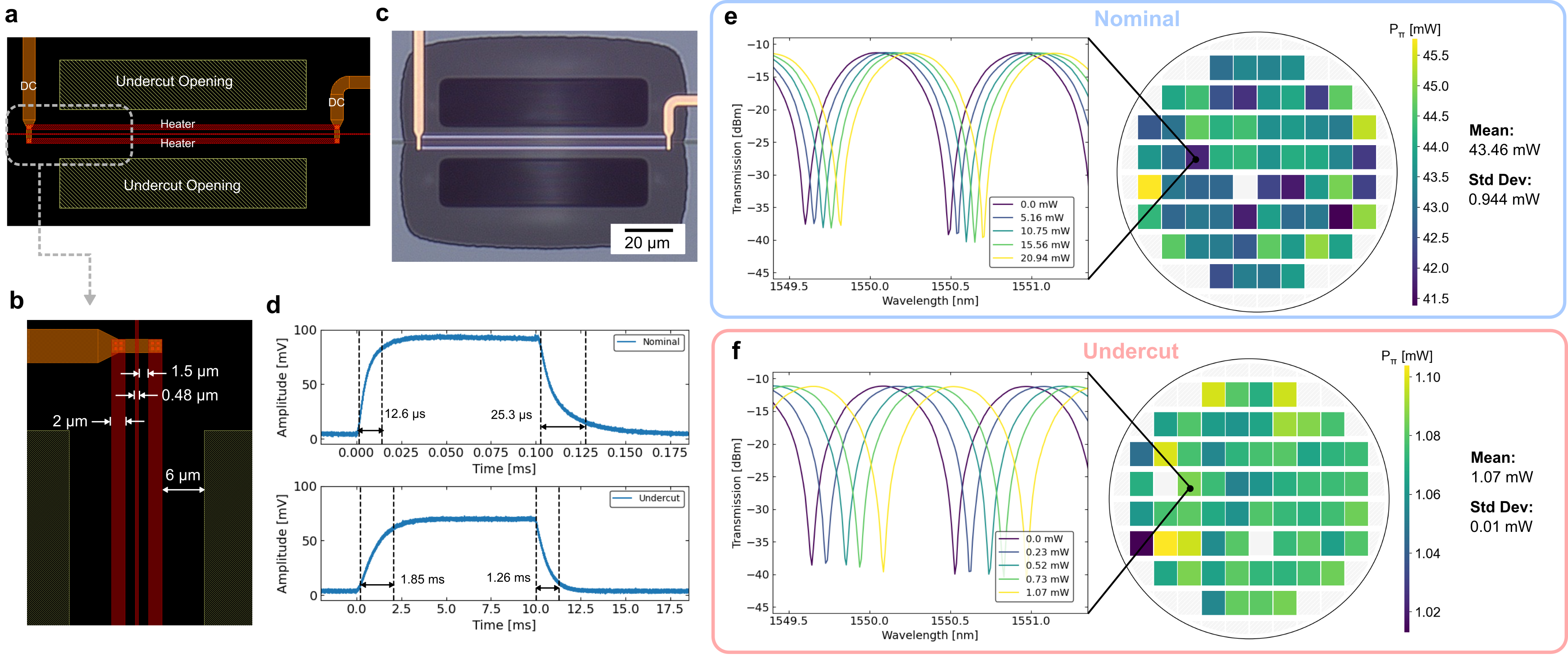}
    \caption{ \textbf{Undercut linear thermo-optic phase shifter wafer-scale measurements.}
    \textbf{a,} Mask layout of the undercut linear thermo-optic phase shifter highlighting the location of the undercut trench openings, doped silicon heaters on both sides of the waveguide, and DC heater traces.
    \textbf{b,} Zoomed view of the mask layout showing the dimensions of the waveguide width (0.48 \si{\micro\meter}), heater separation from the waveguide (1.5 \si{\micro\meter}), heater width (2 \si{\micro\meter}), and heater separation from the undercut trench openings (6 \si{\micro\meter}).
    \textbf{c,} Corresponding micrograph of the undercut thermo-optic phase shifter.
    \textbf{d,} Heater response to a square wave drive signal applied to the heater anode and cathode terminals, highlighting the rise and fall times for the nominal (top) and undercut (bottom) devices. The extreme thermal isolation of the undercut devices results in dramatically longer thermal time constants.
    \textbf{e,} Wafer map of $P_{\pi}$ and representative thermal tuning spectrum for the nominal linear thermo-optic phase shifter.
    \textbf{f,} Wafer map of $P_{\pi}$ and representative thermal tuning spectrum for the undercut linear thermo-optic phase shifter.}
    \label{fig:lin_ps}
\end{figure}

In addition to resonant devices, we designed and fabricated linear thermo-optic phase shifters both with and without undercut. The mask layout and a representative micrograph of the phase shifter with undercut are shown in Figs.~\ref{fig:lin_ps}a and \ref{fig:lin_ps}b and Fig.~\ref{fig:lin_ps}c, respectively. The phase shifter consists of two doped silicon strips parallel to the waveguide.   These doped silicon strips are placed at a distance away from the optical mode to avoid excess optical losses, resulting in the insertion loss being dominated by the intrinsic waveguide propagation loss. A 125 \si{\micro\meter} device length and 2 dB/cm propagation loss yields a total insertion loss of 0.025 dB. \color{black} To extract the tuning efficiency of the phase shifter, it was placed within an imbalanced MZI so the interference fringes could be tracked at varying heater powers. The transmission spectrum of the MZI without undercut and with undercut under different applied heater powers is shown in Fig.~\ref{fig:lin_ps}e and Fig.~\ref{fig:lin_ps}f, respectively. The undercut phase shifter requires significantly less power to induce a larger shift in interference fringe wavelength. This is demonstrated in the wafer maps showing \text{$P_{\pi}$} for the phase shifters without undercut and phase shifters with undercut in Fig.~\ref{fig:lin_ps}e and Fig.~\ref{fig:lin_ps}f, respectively.   An improvement factor of 40.5 is shown, moving from a non-undercut average \text{$P_{\pi}$} of 43.46 mW and standard deviation 0.94 mW to an average \text{$P_{\pi}$} of 1.07 mW and standard deviation 0.01 mW for the undercut device. \color{black} To characterize the time-dependent transient effects of substrate undercut on thermal modulation, we apply a square wave to the phase shifter. The transient is shown for both the nominal and undercut phase shifter in Fig.~\ref{fig:lin_ps}d (top and bottom, respectively). A pulse width of 100 \si{\micro\second} was applied to the non-undercut device, resulting in a rise time of 12.6 \si{\micro\second} and a fall time of 25.3 \si{\micro\second}. For the undercut device, a pulse width of 10 ms was used, resulting in a rise time of 1.853 ms and a fall time of 1.258 ms. The increase in rise/fall time has implications on the performance of these phase shifters in the context of optical switches and multiplexers. For instance, the increase in rise/fall time results in a slower optical switching speed. However, the reduced switching speed disadvantage is mostly negated if the network topology is configured before application runtime \cite{wu2024flexible, jouppi2023tpu}. Further, the increase in rise/fall time can be leveraged when considering time-multiplexed driving signals for many thermo-optic phase shifters to reduce electrical I/O \cite{ribeiro2020column}.

\begin{figure}[ht]
    \centering
    \includegraphics[scale=0.99]{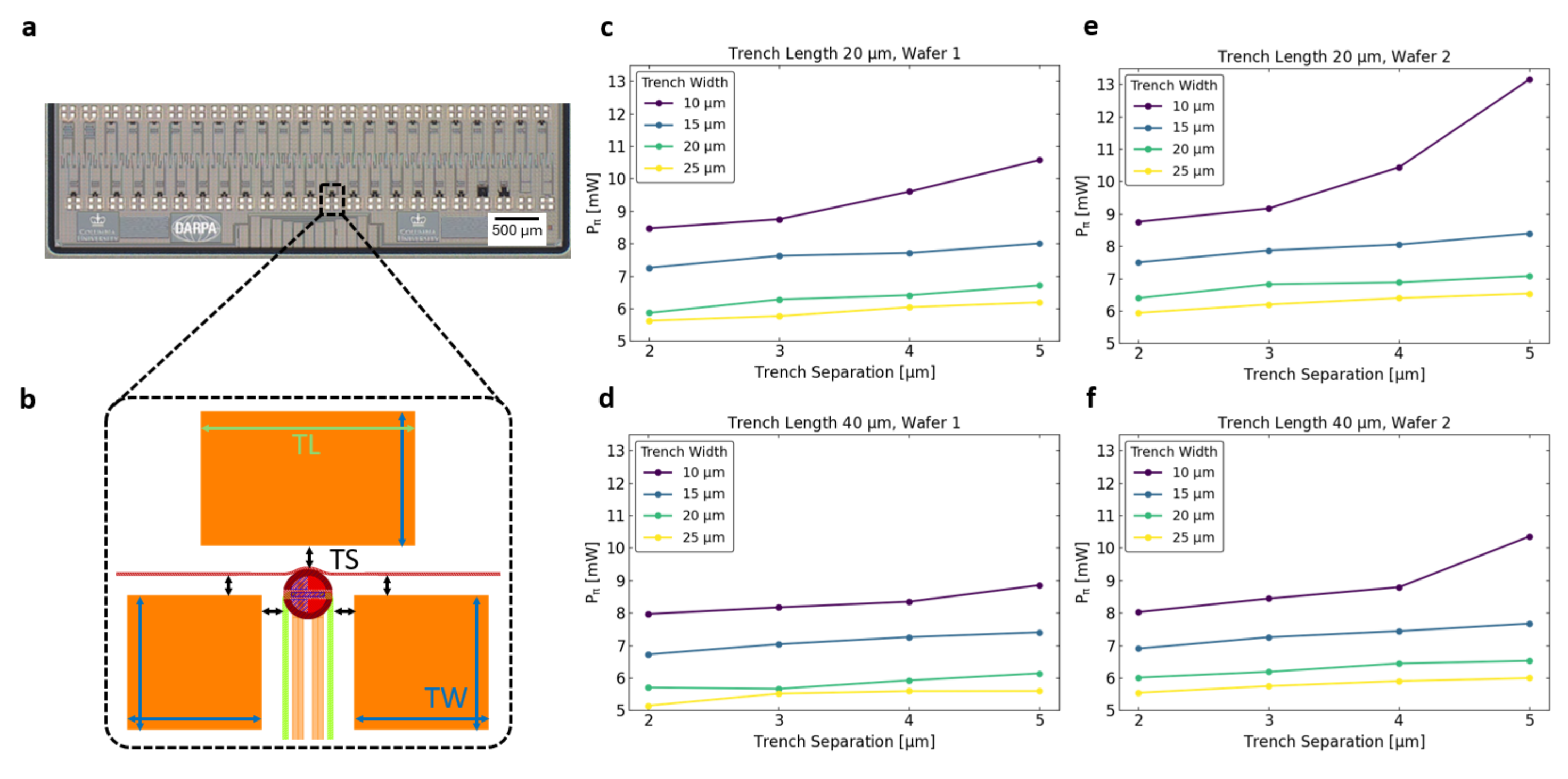}
    \caption{\textbf{Influence of undercut geometry on device thermo-optic performance.} 
    \textbf{a,} Optical microscope image of the undercut microdisk modulator test structures with parameter sweeps of opening geometry. \textbf{b,} Schematic layout of the microdisk modulator with undercut openings illustrating the swept design parameters. TS: trench separation; TL: trench length; TW: trench width. \textbf{c-f,} Average tuning power required for $\pi$ phase shift over varying trench width and trench separation with trench length of 20 \si{\micro\meter} on first wafer \textbf{(c)}, trench length of 40 \si{\micro\meter} on first wafer \textbf{(d)}, trench length of 20 \si{\micro\meter} on second wafer \textbf{(e)}, trench length of 40 \si{\micro\meter} on second wafer \textbf{(f)}.}
    \label{fig:trend}
\end{figure}

To elucidate the impact of undercut geometry on the tuning efficiency of devices, we swept the undercut geometric parameters across a number of microdisk modulator devices. A micrograph of the test structure sweep of undercut microdisk modulators is shown in Fig.~\ref{fig:trend}a. While keeping the microdisk modulator design constant for all cases, both the undercut opening size and separation were varied according to the schematic in Fig.~\ref{fig:trend}b. The tuning efficiency of each microdisk modulator in the undercut parameter sweep was extracted across two full wafers. The tuning efficiency of each modulator in the parameter sweep, averaged across the first wafer, is shown in Figs.~\ref{fig:trend}c and ~\ref{fig:trend}d, where the top trench length is 20 and 40 \si{\micro\meter}, respectively. Similarly, the tuning efficiencies of the modulators from the second full wafer are shown in Figs.~\ref{fig:trend}e and ~\ref{fig:trend}f. First, there is a clear trend in increased tuning efficiency (reduced \text{$P_{\pi}$}) in all cases as the trench width is increased from 10 \si{\micro\meter} to 25 \si{\micro\meter}. The larger trench opening increases the lateral undercut, ensuring the devices are fully released from the substrate. However, we see diminishing returns when increasing the trench width from 20 \si{\micro\meter} to 25 \si{\micro\meter}, suggesting the tuning efficiency of the devices is dominated by the volume of remaining buried oxide and substrate contacting the bottom of the device. Additionally, there is a trend in increased \text{$P_{\pi}$} when the trench separation is increased, which corresponds to wider oxide bridges between the trench openings. This trend explains the limited gains in tuning efficiency seen in processes using a sealed substrate undercut \cite{giewont2019300, pal2022low, fang2024comparison}; heat from the phase shifter is less confined to the device without the trench openings. Another notable trend is the marked consistency and uniformity in tuning efficiency of the undercut modulators between both wafers, demonstrating the mechanical robustness and reliability of the substrate removal process.   Only when aggressively pushing the boundaries of trench separation far beyond the foundry design rules do we see mechanical damage to the bus waveguide and metal traces. Using Fig.~\ref{fig:trend}b as a reference, we observed mechanical damage at a trench separation (TS) of 1 \si{\micro\meter}, but observed no damage and high mechanical stability at a trench separation of at least 2 \si{\micro\meter}. \color{black}

\begin{figure}[ht]
    \centering
    \includegraphics[scale=0.99]{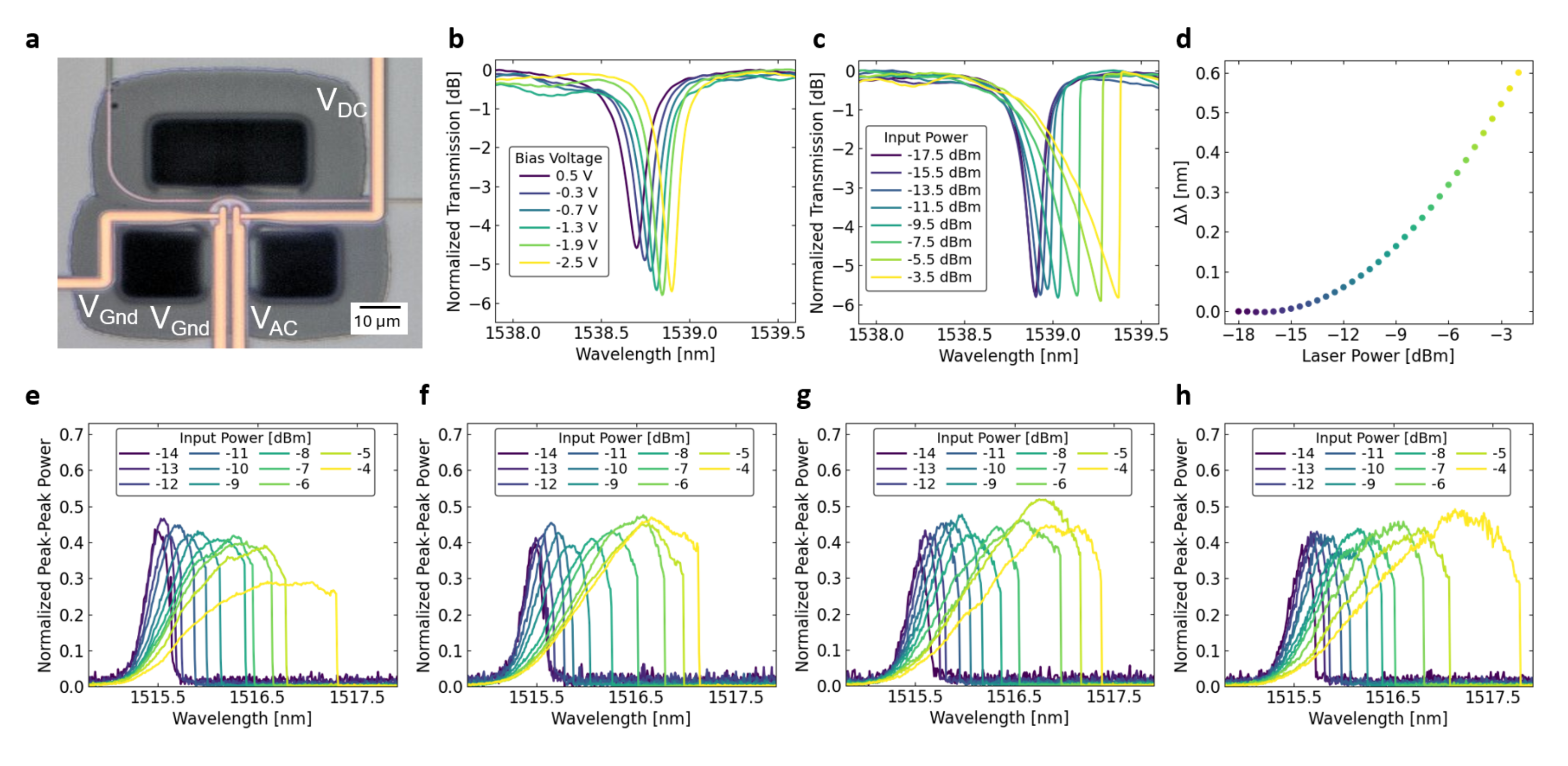}
    \caption{\textbf{Nonlinear spectral shifts in undercut microdisk modulator.} 
    \textbf{a,} Optical microscope image of the undercut resonant device with heater and RF electrical inputs annotated. \textbf{b,} Electro-optic depletion response of the device, swept from 0.5 V to -2.5 V at low optical power (-17.5 dBm). \textbf{c,} Transmission spectrum of the device with fixed -2.5 V DC bias and increasing laser powers. \textbf{d,} Induced spectral shift due to input optical power. \textbf{e-h,} Modulation at different input optical powers and modulation frequencies of \textbf{(e)} 1 MHz, \textbf{(f)} 50 MHz, \textbf{(g)} 300 MHz, and \textbf{(h)} 600 MHz.}
    \label{fig:nonlinear}
\end{figure}

The small size of microdisk modulators causes strong light confinement, leading to high optical power density and nonlinear effects like thermal bistability \cite{Almeida:04}. Thermal isolation exacerbates this by trapping heat in the microdisk, complicating high-power management in resonant modulators \cite{zheng2012enhanced}. We studied the effects of high optical power in thermally undercut microdisk modulators, and thermal instability even at -14 dBm input power is observed \cite{gebregiorgis2024wide}. The modulator is thermally isolated using wafer-scale undercut, shown in Fig.~\ref{fig:nonlinear}a. We examined electro-optic modulation of the PN-junction at varying laser powers. At low input laser power (-17.5 dBm), the DC spectral response showed a modulation efficiency of 65 pm/V under bias voltages from 0.5 V to -2.5 V (Fig.~\ref{fig:nonlinear}b). However, increasing laser power caused nonlinear spectral shifts with resonance red-shifting and broadening due to thermal instability, as shown in Fig.~\ref{fig:nonlinear}c \cite{de2019power}. Near-resonance laser light is partially absorbed via two-photon and free-carrier absorption, generating heat. This effect is enhanced by the thermal isolation of the undercut structure, leading to significant thermal instability at laser optical powers as low as -14 dBm (Fig.~\ref{fig:nonlinear}d).

Despite this, robust high-speed modulation was achievable over a broad spectral range at higher powers, up to -4 dBm. Thermally-induced resonance red-shifting enabled modulation over a range 6$\times$ wider than the resonator’s passive linewidth. We present an experimental investigation of optical modulation in thermally undercut microdisk modulators, focusing on the advantages of optically induced thermal nonlinearity. As laser power increases, the spectral shift of the resonator becomes nonlinear, accompanied by a redshift and resonance broadening due to thermal instability. We also explored the modulation behavior across varying operating frequencies and input optical powers. A 2 V peak-to-peak square wave was applied at different frequencies, and the peak-to-peak optical modulation was measured and normalized to compare modulation depths. Across all modulation frequencies, the spectral range broadens with increased laser power. However, at low frequencies (<1 MHz, Fig.~\ref{fig:nonlinear}e), although the spectral range expands, the peak-to-peak modulation decreases at higher laser powers, likely due to the device’s thermal response. In contrast, at higher data frequencies (>25 MHz, Figs.~\ref{fig:nonlinear}f-h), the modulation depth remains stable as laser power increases, achieving modulation over a spectral range six times broader than the resonator's linewidth at an optical power of -4 dBm. Therefore, at high data rates, the device's thermal sensitivity becomes a beneficial nonlinear effect, enabling robust modulation across a wide spectral range. Thermal isolation thus enhances modulation efficiency and reduces micro-heater tuning requirements.

\section*{Discussion}

In conclusion, we developed and experimentally verified a wafer-scale thermal undercut process in a 300 mm CMOS foundry applied to a range of fundamental silicon photonic devices with state-of-the-art performance and yield. We show an excellent 40$\times$ improvement in tuning efficiency of linear thermo-optic phase shifters with undercut, enabling a drastic reduction in power consumption of large-scale silicon photonic circuits. Undercut microring and racetrack-style modulators also exhibit high tuning efficiency, owing to their ability to accommodate both larger and elongated external trench openings along with an internal trench opening. Further, we apply the developed undercut etch to state-of-the-art microdisk modulators, achieving an almost 5$\times$ improvement in tuning efficiency. These efficiently tunable compact vertical-junction microdisk modulators with high modulation efficiency are fully compatible with CMOS driving voltages, enabling scalable, ultra efficient DWDM links.   Through comprehensive wafer-scale measurements comparing nominal non-undercut and undercut devices, we show high uniformity in $P_{\pi}$ across entire 300 mm wafers, indicating high yield and reproducibility for the developed process. Notably, we observe that high nominal power consumption (large $P_{\pi}$) can be largely offset by thermally undercutting the device, greatly relaxing the heater design requirements and allowing the heater to be placed far from the optical mode. \color{black}

Furthermore, by systematically varying the undercut trench geometry and measuring tuning efficiencies across multiple full wafers of microdisk modulators, we identified key trends that reveal how trench dimensions impact thermal tuning efficiency. Wider and more elongated trenches consistently provide improved thermal isolation, resulting in significant reductions in the power required for thermal tuning. Further, reduced trench separation also resulted in higher tuning efficiency. This finding suggests that trench geometry optimization can serve as a powerful design parameter for further minimizing power consumption in future device designs. By establishing predictable efficiency gains based on trench geometry, we provide a robust framework for integrating thermal undercut as a standardized process across different CMOS fabrication lines, ensuring reproducibility and consistency at the scale required for high-volume production. This level of scalability is critical for enabling low-cost, high-throughput manufacturing of photonic integrated circuits where consistent thermal performance is essential.

\section*{Methods}

\subsection*{Thermal Undercut Fabrication Process}

The thermal undercut fabrication process is run on a 300 mm, CMOS compatible, silicon photonic process \cite{AIMPDK}. The thermal undercut is run within the back-end-of-line (BEOL) giving it flexibility to be used with photonic, electronic and optoelectronic devices. The undercut can be combined with optical facets and waveguide access via trenches to both silicon and silicon nitride layers. The process consists of two primary steps: 1) silica and 2) silicon etching to ensure efficient and controlled reactive ion etching (RIE) of the different materials. Within each step photolithography is used to define the thermal undercut opening, a $\sim$10 \si{\micro\meter} photoresist is spun, baked, and exposed using i-line photolithography. Careful optimization of the thermal undercut opening sizes (Fig.~\ref{fig:trend}b TW and TL) and the proximity to the device (Fig.~\ref{fig:trend}b TS) is used to ensure RIE does not damage waveguides or metallization. The RIE etches of both the silica and silicon portion are closely monitored to ensure undercutting of optical facets does not happen and optical facets are produced with smooth straight surfaces.

\subsection*{Wafer-scale Measurements}

All optical measurements were taken using a ficonTEC TL1200 Wafer-Level and Component-Level Tester \cite{ficonTEC}. The wafer-level tester, with a temperature-controlled chuck, two six-axis stages with optical probes, and two five-axis stages with electrical probes, enables comprehensive wafer-scale measurements. Three dies per reticle, across 64 reticles, were measured to characterize a variety of silicon photonic devices both with and without thermal undercut. The first die contains many grating-coupled microdisk modulators with a sweep of undercut opening geometric parameters. An angled fiber array was coupled to each microdisk modulator with an automated alignment process, and simultaneously multi-contact DC electrical probes were landed to control the integrated thermo-optic phase shifter. A high-precision DC power supply (Keithley 2280S-32-6) was used to bias the thermo-optic phase shifter. After tuning the polarization to select for the fundamental transverse-electric (TE) mode with an internal polarization controller (Thorlabs MPC320), a tunable laser source (Keysight 81608A), lightwave measurement system (Keysight 8164B), and optical power meter (Keysight N7744A) were used to sweep the optical spectrum of each device at each phase shifter bias point.

A lidless periscope fiber array was used to edge-couple to the microring modulators, racetrack modulators, and imbalanced Mach Zehnder interferometers (MZIs) on the other two dies on each reticle. The periscope fiber arrays, manufactured by Keystone Photonics, enable wafer-scale low-loss edge-coupling by extending 3D-printed optics into the dicing trench \cite{dietrich2018situ}. Following edge-coupling, the optical spectra of each device at different thermo-optic phase shifter bias points were measured using the same methodology as the microdisk modulators. In total, two wafers without thermal undercut were measured, and two wafers with thermal undercut were measured. 

The tuning efficiency of each device was calculated by finding the resonance or interference fringe shift at each thermo-optic phase shifter bias point. A peak-finding algorithm was used to find the resonance or fringe wavelength from each optical spectrum sweep. Together with the recorded electrical power dissipated in the thermo-optic phase shifter and the measured free spectral range, the \text{$P_{\pi}$}, i.e. the power required for a phase shift of \text{$\pi$}, was calculated for each device.

The transient responses of the microdisk modulator and imbalanced MZI were measured both with and without thermal undercut. A tunable laser was aligned to the wavelength of the resonance and interference fringe of the microdisk and MZI, respectively. Next, the optical output was switched to a photodiode (Thorlabs PDA10CS), connected to an oscilloscope (Tektronix MSO5204B) set to trigger on the rising edge. An arbitrary function generator (Keithley 3390) was used to send a square wave to the thermo-optic phase shifter, with the resultant transient captured by the oscilloscope. The rise and fall times of each device were extracted by finding the 10\% and 90\% thresholds of each optical transient.

\bibliography{bibliography}

\section*{Acknowledgements}

This work was supported in part by the U.S. Advanced Research Projects Agency--Energy under ENLITENED Grant DE-AR000843, in part by the U.S. Defense Advanced Research Projects Agency under PIPES Grant HR00111920014, and in part by the Center for Ubiquitous Connectivity (CUbiC), sponsored by the Semiconductor Research Corporation (SRC) and DARPA under the JUMP 2.0 program. This material is based on research sponsored by the United States Air Force (FA8750-23-C-1001) and Air Force Research Laboratory under AIM Photonics (agreement number FA8650-21-2-1000) and also FA8750-21-2-0004. The U.S. Government is authorized to reproduce and distribute reprints for Governmental purposes notwithstanding any copyright notation thereon. The views and conclusions contained herein are those of the authors and should not be interpreted as necessarily representing the official policies or endorsements, either expressed or implied, of the United States Air Force, the Air Force Research Laboratory or the U.S. Government. Approved for Public Release; Distribution A Unlimited: AFRL-2025-1217.

\section*{Author contributions statement}

A.R. and A.N. conceived the device designs, performed initial simulations, and completed the mask layout of the undercut geometry sweeps. V.D. and M.vN. designed and completed the mask layout of the microring, racetrack, and linear thermal phase shifters. L.C., G.L., and D.C. developed the undercut process and fabricated the wafers. C.C.T., A.M.S, and M.L.F. contributed to earlier versions of the undercut design, measurements, and data analysis. Y.G. performed the wide spectral modulation experiments. R.P. led the wafer-scale measurements with assistance from K.J. and the data analysis with assistance from Y.W. and X.M. All authors reviewed the manuscript and contributed to the writing. A.R., K.B., M.L.F, and S.P. supervised the project.

\section*{Additional information}

\subsection*{Competing interests} 

The authors declare no competing interests. 

\subsection*{Data Availability}

The data that support the plots within this paper and other findings of this study are
available from the corresponding author upon reasonable request.

\subsection*{Correspondence}

Correspondence and requests for materials should be addressed to A.R. (email: anthony.j.rizzo@dartmouth.edu).



\end{document}